\documentclass[useAMS,usenatbib,onecolumn]{mn2e}

\usepackage{graphicx}

\title[Completely analytical families of anisotropic $\gamma$-models]
{Completely analytical families of anisotropic $\gamma$-models}
\author[Pieter Buyle, Chris Hunter, and Herwig Dejonghe]
{P. Buyle$^{1}$\thanks{Post-doctoral Fellow of the Fund for
Scientific Research - Flanders, Belgium (F.W.O.), e-mail:
Pieter.Buyle@UGent.be}, C. Hunter$^{2}$\thanks{E-mail:
hunter@math.fsu.edu}, and H. Dejonghe$^{1}$\thanks{E-mail:
Herwig.Dejonghe@UGent.be} \\ $^{1}$Astronomical Observatory, Ghent
University, Krijgslaan 289 - S9, B9000 Ghent, Belgium\\
$^{2}$Department of Mathematics, Florida State University,
Tallahassee, FL 32306-4510, USA}
\begin{document}

\date{Accepted 1988 December 15. Received 1988 December 14; in original form 1988 October 11}

\pagerange{\pageref{firstpage}--\pageref{lastpage}} \pubyear{2002}

\maketitle

\label{firstpage}

\begin{abstract}
We present new analytical distribution functions for anisotropic
spherical galaxies. They have the density profiles of the
$\gamma$-models, which allow a wide range of central density slopes,
and are widely used to fit elliptical galaxies and the bulges of
spiral galaxies.  Most of our models belong to two two-parameter
families. One of these parameters is the slope $\gamma$ of the central
density cusp. The other allows a wide range of varying radial and
tangential anisotropies, at either small or large radii. We give
analytical formulas for their distribution functions, velocity
dispersions, and the manner in which energy and transverse velocity
are distributed between orbits. We also give some of their observable
properties, including line-of-sight velocity profiles which have
been computed numerically. Our models can be used to provide a useful
tool for creating initial conditions for N-body and Monte Carlo
simulations.
\end{abstract}


\section{Introduction}
\label{sec:intro}

A galaxy is fully described by a distribution function ${\cal
F}(\bf{r},\bf{v})$ which specifies the positions and velocities of all
of its constituent stars and gas. Eddington (1916) showed how to
determine the distribution function for a known spherical mass
provided that the galaxy is dynamically isotropic, that is with the
binding energy $E$ being the only isolating integral. Spherical form
is of course an idealisation, and at best an approximation to the true
shape. Isotropy is generally also an approximation to the true dynamics
\citep{illing,binney}. Our anisotropic models allow us to
explore a range of possible dynamical behaviour while still
retaining the other simplifying approximation of sphericity.

The step from finding isotropic distribution functions for spherical
galaxies to that of finding anisotropic distribution functions for
them has turned out to be surprisingly steep \citep{herwig1,hunter}.
Anisotropic distribution functions depend also on the modulus of the
angular momentum $L$. Because of the limited supply of analytical models,
many workers have followed Hernquist (1993) in finding approximate
distribution functions by first solving the Jeans' equations for the
velocity dispersions and then using Gaussians to provide local
velocity distributions. Recently Kazantzidis, Magorrian \& Moore
(2004) have shown that this procedure can be hazardous when applied to
generate initial conditions for galaxies that are strongly
non-Gaussian.  In such cases, the Jeans plus Gaussian approximation
can lead to an initial state that is far from equilibrium, so that
much of its subsequent development is an artifact of that start.  As
the importance of numerical simulations increases, the issue of
controlling numerical errors in them becomes equally important. One
can avoid invalid initial conditions for N-body and Monte-Carlo
simulations simply by using a three-dimensional Monte-Carlo simulator
on an exact distribution function to derive the initial conditions
\citep{buyle}.

Some of the currently known anisotropic models have constant
anisotropies \citep{marel,wilkinson,treu}. Others are of the
Osipkov-Merritt type \citep{osipkov,merritt,hern} and are limited to
depending solely on an isolating integral which is a linear combination
of $E$ and $L^2$.  Models of this type are isotropic at small radii
but dominated by radial orbits at large radii.  Cuddeford (1991)
showed how to modify the Osipkov-Merritt algorithm to allow for
anisotropy at small radii and a central cusp, while Ciotti \& Pellegrini (1992)
constructed composite Osipkov-Merritt models.  Models with a greater
variety of anisotropic behaviour are known for special models; the
Plummer model \citep{plummer,herwig2} and the Hernquist model
\citep{hern,baes2}.  A greater variety of analytical systems with
different and varying kinematics is clearly desirable. Some have
recently been provided by An \& Evans (2006a). Theirs are for two
families of densities; the generalised isochrone family which includes
the Hernquist and isochrone models as special cases, and
a generalised Plummer family which includes the Plummer
sphere.  Ours differ because they are for the densities given by the
$\gamma$-models \citep{dehn,tremaine}.

The $\gamma$-models include the models of Hernquist \citep{hern},
Dehnen \citep{dehn} and Jaffe \citep{jaffe} as special cases.
Isotropic distribution functions are known for all $\gamma$-models
\citep{tremaine,baes}. The two new families of analytical anisotropic
spherical models, which we present in this paper, introduce an
additional parameter $q$ which allows the anisotropy of the system to
be varied.  One family is isotropic at large radii, and $q/2$ gives
Binney's anisotropy parameter $\beta$ \citep{BT} at its center.  The
second family is isotropic at small radii, and $-q/2$ gives the value
of $\beta$ at large radii. A greater variety of behaviour can be
obtained by combining the two families, though we do not pursue that
possibility here.

In Section \ref{sec:basics} we review the basic equations needed for
our study, and introduce the $\gamma$-models and their isotropic
distribution functions. Before introducing our first new family in
Section \ref{sec:anisomodels}, we first give some surprisingly simple
constant anisotropy $\beta=1/2$ models of which only the $\gamma=1$
case was known previously. We then show how to construct the first
family of variable anisotropy models. We give analytical expressions
for their distribution functions, velocity dispersions, and energy and
transverse velocity distributions.  Section \ref{sec:anisomodels}
concludes with a numerical investigation of some line-of-sight
velocity profiles. Section \ref{sec:moremodels} gives the same
properties for our second family of models with its different
anisotropic behaviour. The analysis here is compact because it makes
extensive use of that done in Section \ref{sec:anisomodels} for the
first family.  Section \ref{sec:moremodels} concludes with another set
of constant anisotropy models; ones composed entirely of radial
orbits.  They exist for all $\gamma$-models with $\gamma \ge 2$. We
summarize our conclusions in Section \ref{sec:conclusions}. The
Appendices give mathematical details of our derivations, and collect
formulas.

\section{Basic properties}
\label{sec:basics}

\subsection{General formulae}

The isotropic motion of stars and gas in a spherical galaxy is
described by a mass distribution function ${\cal F}(E)$ which is
solely a function of the binding energy $E$. The mass density $\rho$
is obtained from the distribution function by the integration
\begin{equation}
\rho(\psi)=4\pi \int_0^{\psi}{\cal F}(E)\sqrt{2(\psi-E)}dE
\label{rhoiso}
\end{equation}
where $\psi$ is the gravitational potential. Eddington (1916) solved
the integral equation (\ref{rhoiso}) to get
\begin{equation}
{\cal F}(E)=\frac{1}{2\pi^2}D_E\int_0^E\frac{d\rho(\psi)}{d\psi}
\frac{d\psi}{\sqrt{2(E-\psi)}},
\label{Fiso}
\end{equation}
where $D_E$ denotes differentiation with respect to $E$. Eddington's
solution makes use of the mass density $\rho(\psi)$ expressed as a 
function of the potential. Further kinematic information about the system can
be obtained once the density is expressed in this form. The velocity
dispersions for example are given by
\begin{equation}
\sigma^2(\psi)=\sigma^2_r(\psi)=\sigma^2_{\varphi}(\psi)
=\sigma^2_{\theta}(\psi)=\frac{1}{\rho(\psi)}\int_0^{\psi}\rho(\psi ')d\psi '.
\end{equation}

The distribution function of an anisotropic model of a spherical 
galaxy function depends also on the modulus of the angular momentum vector
\begin{equation}
L=r\sqrt{v_{\varphi}^2+v^2_{\theta}}=rv_t.
\end{equation}
Then a double integration is needed to derive the mass
density from the distribution function ${\cal F}(E,L)$ as
\begin{equation}
\rho(\psi,r)=2\pi \int_0^{\psi}dE\int_0^{2(\psi-E)}
             \frac{{\cal F}(E,L)}{\sqrt{2(\psi-E)-v_t^2}}dv_t^2.
\label{rhoani}
\end{equation}
It gives what Dejonghe (1986,1987) has named the augmented mass 
density $\rho(\psi,r)$. Unlike the $\rho(\psi)$ of the isotropic
case in equation (\ref{rhoiso}), this augmented mass density is not 
determined by the spatial dependence of $\rho(r)$ and $\psi(r)$.
Instead different feasible distribution functions
${\cal F}(E,L)$ give rise to different augmented densities for the 
same mass density. The inversion of equation (\ref{rhoani}) to determine the
distribution function which corresponds to a known augmented mass
density is generally unstable if done numerically \citep{herwig1}.
Velocity dispersions can be derived  directly from the augmented density, 
without determining the corresponding distribution function, as
\begin{equation}
\sigma_r^2(\psi ,r)=\frac{1}{\rho(\psi ,r)}\int_0^{\psi}\rho(\psi ',r)d\psi '
\label{disprani}
\end{equation}
\begin{equation}
\sigma_{\varphi}^2(\psi,r)=\sigma_{\theta}^2(\psi,r)
=\frac{1}{2}\sigma_t^2(\psi,r)
=\frac{1}{\rho(\psi ,r)}\int_0^{\psi}D_{r^2}[r^2\rho(\psi ',r)]d\psi '.
\label{disptani}
\end{equation}
However, until the distribution function has been determined,
one must beware that these dispersions might correspond to an unphysical model.

\subsection{Isotropic $\gamma$-models}
\label{subsec:isotropicmodels}

The $\gamma$-models were independently introduced by Dehnen (1993) and
Tremaine et al. (1994) and are a generalisation of a series of spherical
models such as the Hernquist model \citep[$\gamma=1$]{hern}, the
Dehnen model \citep[$\gamma=\frac{3}{2}$]{dehn} and the Jaffe model
\citep[$\gamma=2$]{jaffe} that are famous for their analytical
description of the observed cuspy slopes in elliptical galaxies. The
dimensionless mass density of these models is given by
\begin{equation}
\rho(r)=\frac{3-\gamma}{4\pi}\frac{1}{r^{\gamma}(1+r)^{4-\gamma}},
\label{gammadens}
\end{equation}
where $\gamma$ can have a value between 0 and 3 and determines the
growth of the density at small radii while density decays as $r^{-4}$
at large radii. The potential of a $\gamma$-model is given by
\begin{equation}
\psi(r)=\frac{1}{2-\gamma}\left[1-\left(\frac{r}{1+r}\right)^{2-\gamma}\right],
\label{gammapot}
\end{equation}
for $\gamma\ne 2$. The combination of equations (\ref{gammadens}) and
(\ref{gammapot}) gives the density as the following function of the potential:
\begin{equation}
\label{augrho}
\rho(\psi)=\frac{3-\gamma}{4\pi}
\left[1-(2-\gamma)\psi\right]^{\frac{-\gamma}{2-\gamma}}
\left\{1-[1-(2-\gamma)\psi]^{\frac{1}{2-\gamma}}\right\}^{4}.
\end{equation}
Tremaine et al. (1994) used equations (\ref{Fiso}) and (\ref{augrho}) 
to derive isotropic distribution functions for all $\gamma$-models.
Baes et al. (2005) showed that they can all be expressed
in terms of hypergeometric functions as
\begin{eqnarray}
\label{isodf}
{\cal F}(E,\gamma)=\frac{3-\gamma}{4\pi^3}\sqrt{2E}\left[
\right. &-& \left. (\gamma-4)
\ _2F_1\left(1,\frac{-\gamma}{2-\gamma};\frac{3}{2};
(2-\gamma)E\right)+2(\gamma-3)\ _2F_1\left(1,\frac{1-\gamma}{2-\gamma};
\frac{3}{2};(2-\gamma)E\right)\right.\nonumber\\
&-& \left. 2(\gamma-1)\ _2F_1\left(1,\frac{3-\gamma}{2-\gamma};\frac{3}{2};
(2-\gamma)E\right)+\gamma\ _2F_1\left(1,\frac{4-\gamma}{2-\gamma};
\frac{3}{2};(2-\gamma)E\right) \right].
\end{eqnarray}
The distribution function for the $\gamma=2$ case of the Jaffe (1983)
model can be recovered by taking the $\gamma \to 2$ limit. Then the
hypergeometric functions become confluent ones \citep{abramowitz}
according to $ \lim_{\gamma \to
2}~_2F_1(1,a(\gamma)/(2-\gamma);3/2;(2-\gamma)E) = M( 1, 3/2,
a(2)E)$. The result is equivalent to that given by Jaffe in terms of
Dawson's integrals.

\section{Anisotropic $\gamma$-models}
\label{sec:anisomodels}

The first of our two-parameter families of anisotropic models is
introduced in \S\ref{subsec:firstmodels}.  A preliminary section
\S\ref{subsec:simpleaniso} gives a set of constant and radially biased
anisotropic models whose distribution functions are much simpler than
the isotropic ones (\ref{isodf}).  These models exist for all $\gamma
\ge 1$, and resemble components of the family which is the main topic
of this section. The distribution functions of this family are given
in \S\ref{subsec:firstmodels} as infinite series.  Convergence
requirements generally restrict this family to the range $0 < \gamma <
2$, though particular models for which the series can be summed
explicitly, such as those of \S\ref{subsec:whenqisone}, exist for
larger $\gamma$ values. Later sections give the velocity dispersions,
distributions of energy and transverse velocity, and line profiles of
this family of anisotropic models.

\subsection{Simple models for $\gamma \ge 1$}
\label{subsec:simpleaniso}

An \& Evans (2006a) give rules for deriving anisotropic distribution
functions which depend on angular momentum via a power law.  The case
of an inverse first power is particularly simple.  We first extract an
$r^{-1}$ power from the augmented density (\ref{augrho}), and then
apply their algorithm.  The result is the one-parameter family of
distribution functions
\begin{equation}
\label{wyndf}
{\cal F}(E,L,\gamma)=\frac{3-\gamma}{8\pi^3L}
\left\{1-\left[1-(2-\gamma)E\right]^{\frac{1}{2-\gamma}}\right\}^2
\left\{4-\gamma+ 
\frac{\gamma-1}{\left[1-(2-\gamma)E\right]^{\frac{1}{2-\gamma}}}\right\}.
\end{equation}
This family exists only for $\gamma \ge 1$ because of the second term
in the last brace. That term, which becomes large and dominant as
$(2-\gamma)E$ tends to its central value of 1, is negative when
$\gamma<1$. The restriction to $\gamma \ge 1$ is a simple instance of
An \& Evans's (2006b) cusp slope-central anisotropy theorem because
Binney's anisotropy parameter $\beta$ has the constant value of $1/2$
for the models (\ref{wyndf}).  Two examples are
\begin{equation}
{\cal F}(E,L,1)=\frac{3E^2}{4\pi^3L}, 
\quad {\cal F}(E,L,2)=\frac{e^E-3e^{-E}+2e^{-2E}}{8\pi^3L}.
\end{equation}
The first was given by Baes \& Dejonghe (2002), while the second can
either be obtained directly, or from the $\gamma \to 2$ limit of
equation (\ref{wyndf}).

\subsection{A two-parameter family of anisotropic models}
\label{subsec:firstmodels}

We now construct a family of anisotropic models for different $\gamma$-models
by writing the mass density (\ref{gammadens})
in the separable augmented form of $r^{-q}(1+r)^q$ times
a function of $\psi$ as follows:
\begin{eqnarray}
\rho(\psi,r)&=&\frac{3-\gamma}{4\pi}\frac{(1+r)^q}{r^q}
\left[1-(2-\gamma)\psi\right]^{\frac{q-\gamma}{2-\gamma}}
\left\{1-\left[1-(2-\gamma)\psi\right]^{\frac{1}{2-\gamma}}
\right\}^4,
\nonumber\\
&=&\frac{3-\gamma}{4\pi}\frac{(1+r)^q}{r^q}\sum_{k=0}^4{4 \choose k}
(-1)^{k}\left[1-(2-\gamma)\psi\right]^{\frac{k+q-\gamma}{2-\gamma}},\nonumber\\
&=&\frac{3-\gamma}{4\pi}\frac{(1+r)^q}{r^q}\sum_{p=4}^{\infty}
(-1)^p[(2-\gamma)\psi]^p
\sum_{k=0}^4(-1)^{k}{4 \choose k}{\frac{k+q-\gamma}{2-\gamma} \choose p}.
\label{gammanirho}
\end{eqnarray}
The parameter $q$ allows a range of models for each mass density
(\ref{gammadens}).  The binomial expansion in powers of
$(2-\gamma)\psi$ is valid provided that $2 > \gamma > 0$ because
then $(2-\gamma)\psi \in [0,1]$.  It begins with $p=4$ because the
augmented density (\ref{gammanirho}) varies as $\psi^4$ for small
$\psi$, i.e. at large distances.  Series converge rapidly except in
the central regions where the distribution function becomes infinite
because of the cuspiness of the mass density there.

\begin{figure}
\centering
\includegraphics[width=6cm,angle=0,clip=true]{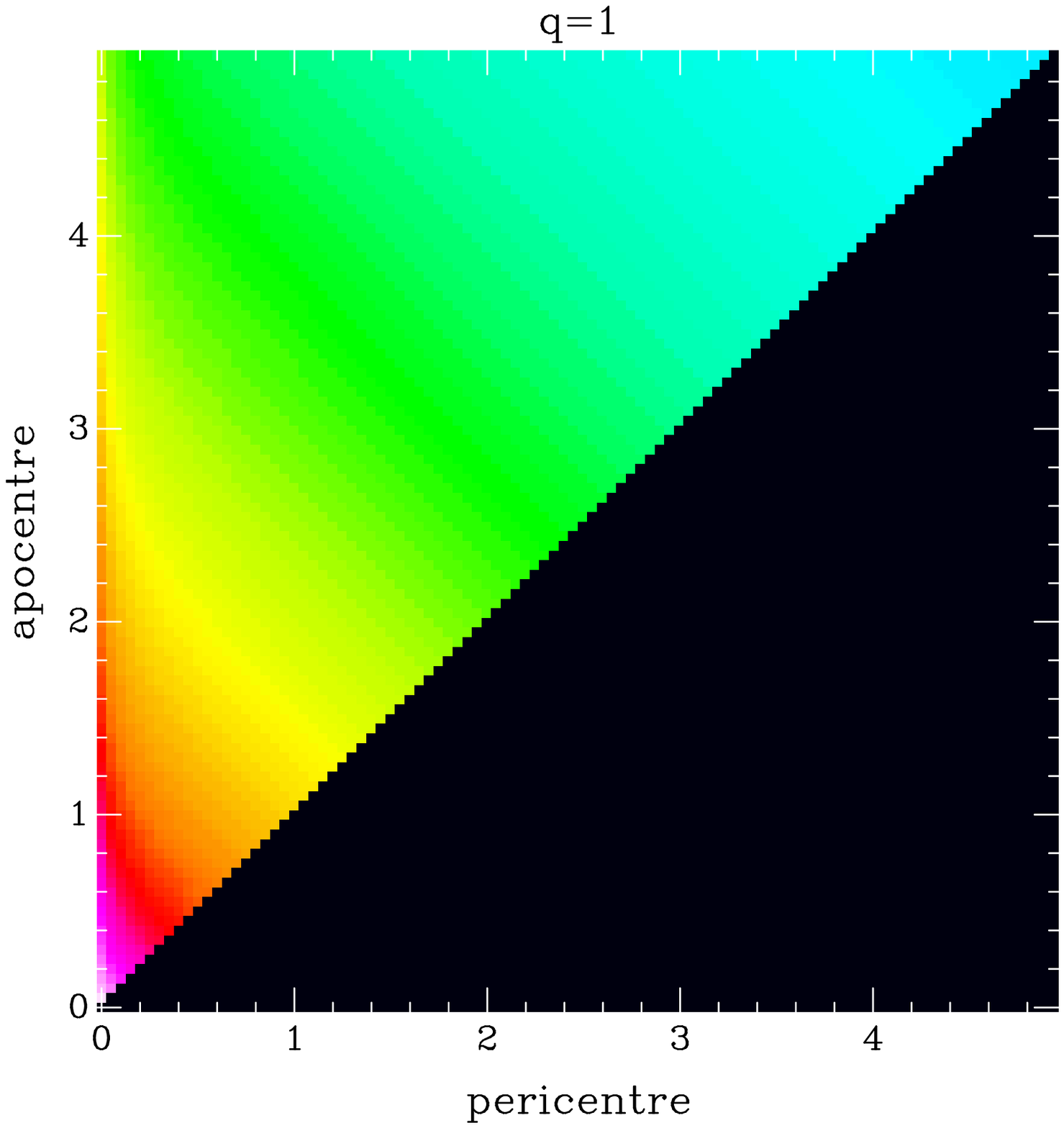}\hspace{1cm}
\includegraphics[width=6cm,angle=0,clip=true]{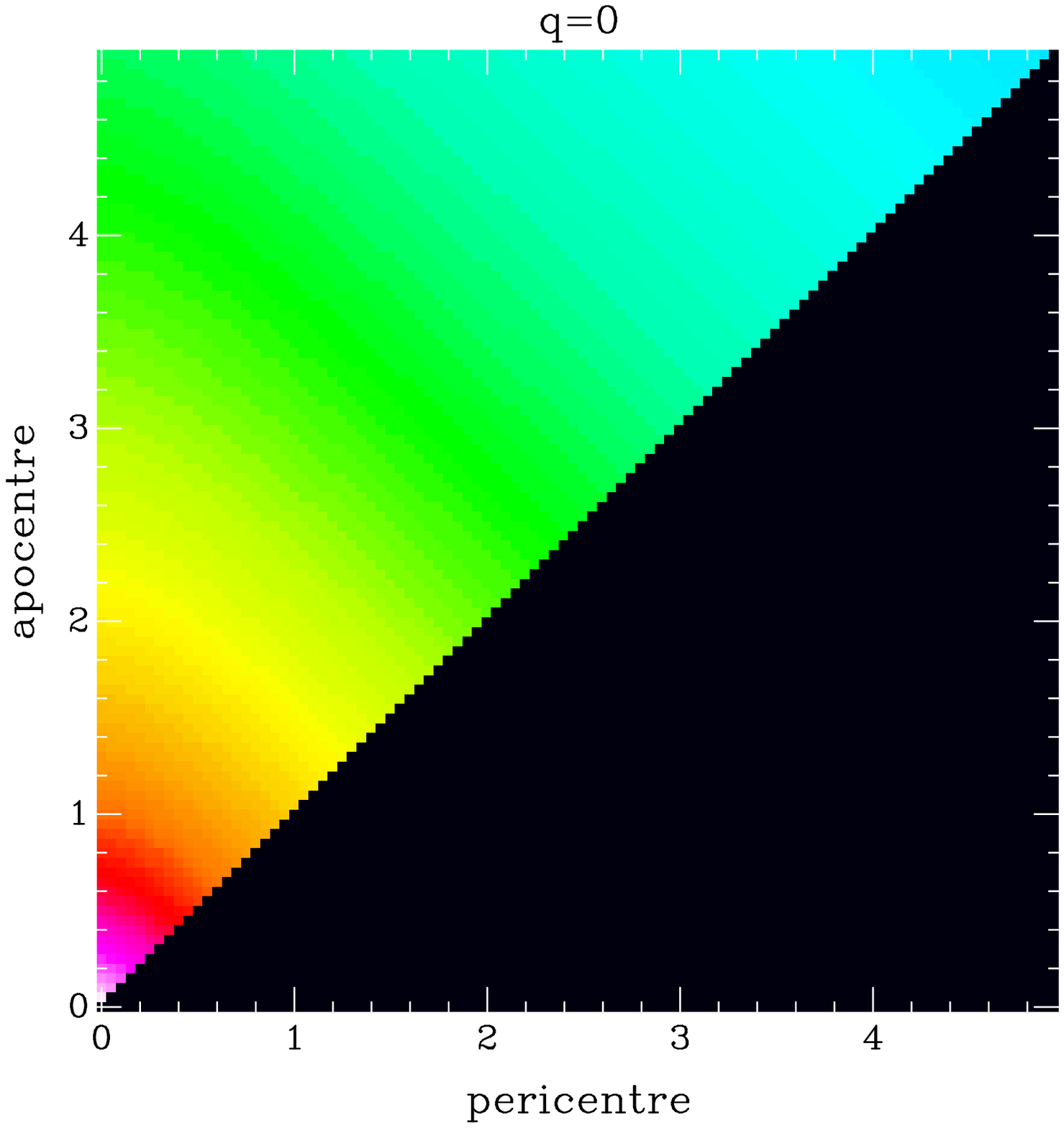}\vspace{1cm}
\includegraphics[width=6cm,angle=0,clip=true]{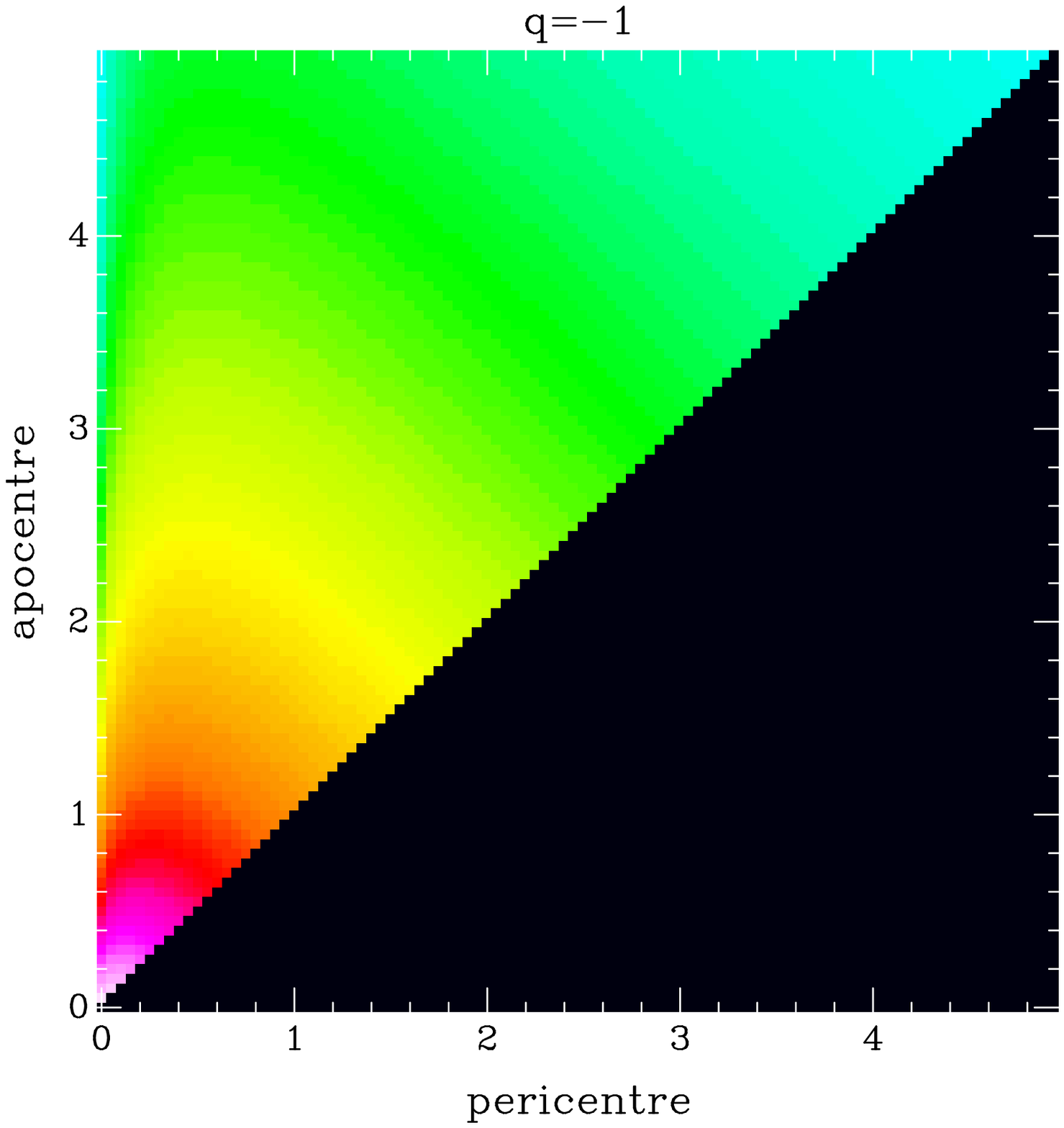}\hspace{1cm}
\includegraphics[width=6cm,angle=0,clip=true]{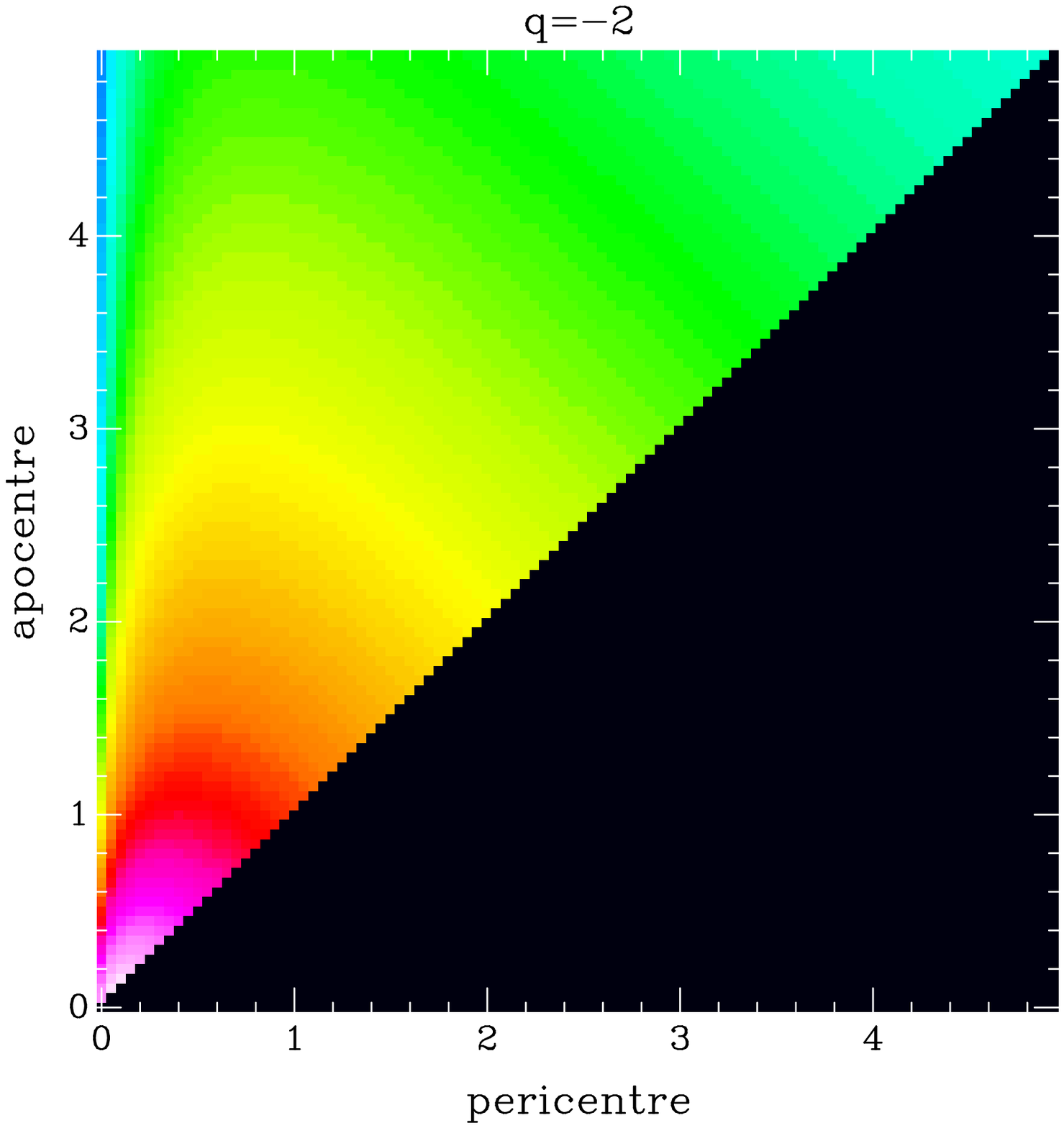}
\caption{Turning-point diagram for several Hernquist models of the
first family of \S\ref{subsec:firstmodels}. The colour shows how the
density of orbits varies with pericentre and apocentre. Density declines 
from a high of white at the very centre, and then through the spectrum
from red to a low of blue. Note the high red densities of elongated
orbits along the left boundary for the most radial $q=1$ model, and
their steady decline with increasing tangentiality as $q$
decreases.\label{figseruke1}}
\end{figure}

We define $F_q^p(E,L)$ to be the component of the distribution function which
corresponds to the component $r^{-q}(1+r)^q\psi^p$ of augmented density.  
It is given by formulas (\ref{FlargeL}) and (\ref{FsmallL}) in Appendix 
\ref{app:formulas}.
The full distribution function is then found by summing
\begin{equation}
{\cal F}(E,L,\gamma,q)=\frac{3-\gamma}{4\pi}\sum_{p=4}^{\infty}
C(\gamma,p,q)F_q^p(E,L),
\label{andistrart}
\end{equation}
where the coefficients $C(\gamma,p,q)$ are defined as
\begin{equation}
C(\gamma,p,q)=(-1)^p(2-\gamma)^p\sum_{k=0}^4 (-1)^{k} {4 \choose k}
{\frac{k+q-\gamma}{2-\gamma} \choose p}.
\label{Ccoeffs}
\end{equation}
The distribution function components $F_q^p(E,L)$ do not depend on the
parameter $\gamma$. This parameter influences the rapidity
with which the series (\ref{andistrart}) converges through the
coefficients $C(\gamma,p,q)$, and especially their dependence on the 
power $(2-\gamma)^p$. Hence the closer $\gamma$ is to 2, the more
rapidly does the series (\ref{andistrart}) converge.

The elementary distribution functions are simpler in certain special
cases which are featured in the figures. For $q\in[1,0,-1,-2]$, we
have 
\begin{eqnarray}
\label{simpleDFs}
F_q^p(E,L)=\frac{pE^p}{(2\pi E)^{3/2}} \left[
\frac{\Gamma(p)}{\Gamma(p-\frac{1}{2})}+\frac{q\sqrt{2E}}{\sqrt{\pi}L} \right.
&\times& \left.  \left\{ \begin{array}{lll}1, && q=0,1, \\
\ _2F_1\left(\frac{1}{2},-q;p,-\frac{2E}{L^2}\right), && q=-1,-2.
\end{array} \right\} \right].
\end{eqnarray}
As we see from equation (\ref{binneybeta}) below, $q/2$ gives the
central value of Binney's anisotropy parameter, and hence An \&
Evans's (2006b) cusp slope-central anisotropy theorem restricts $q$ to
the range $q \le \gamma$.  It is also necessary that $q <2$ because
${\cal F}$ grows as $L^{-q}$ as $L \to 0$ when $q>0$ (See equation
(\ref{FsmallL})), and the integration over ${\cal F}$ in equation
(\ref{rhoani}) diverges if $q \ge 2$.

The expansion of the augmented density in powers of $\psi$ 
is simple for the $\gamma=1$ Hernquist models, for which
\begin{equation}
C(1,p,q)= (-1)^{p}{q-1 \choose p-4}, \qquad p \ge 4.
\end{equation}
We display distribution functions for these models for different $q$ values
in Fig. \ref{figseruke1}. They are plotted there in peri- and apocentre
space. Circular orbits lie along the lower diagonal boundary, while
orbits become increasingly radial as the left vertical boundary is
approached. Hence densities are larger near there for the radially
anisotropic $q=1$ case relative to the isotropic $q=0$ case, but
relatively smaller for the tangentially anisotropic $q=-1$ and $q=-2$ cases.

\subsubsection{Explicit $q=1$ distribution functions for $\gamma \ge 1$}
\label{subsec:whenqisone}

Equations (\ref{andistrart}) and  (\ref{simpleDFs}) show that the
distribution functions for $q=1$ have the simple form
\begin{equation}
{\cal F}(E,L,\gamma,q=1)=f_0(E) + \frac{f_1(E)}{L}.
\end{equation}
The two terms correspond to the two components $\rho_0(\psi)$ 
and $\rho_1(\psi)/r$ of the augmented density (\ref{gammanirho}).
That allows us to avoid the expansion in powers of $\psi$, and to
calculate the two components directly combining the methods of
Baes et al. (2005) and An \& Evans (2006a) respectively as in
Sections \ref{subsec:isotropicmodels} and \ref{subsec:simpleaniso}.
The results, which are similar to but not the same as the earlier
(\ref{isodf}) and (\ref{wyndf}), are
\begin{eqnarray}
f_0(E)&=&\frac{3-\gamma}{8\pi^3}\sqrt{2E}
\left[ -3(\gamma-5)
\ _2F_1\left(1,\frac{-1-\gamma}{2-\gamma};\frac{3}{2};(2-\gamma)E\right)
+8(\gamma-4)\ _2F_1\left(1,\frac{-\gamma}{2-\gamma};
\frac{3}{2};(2-\gamma)E\right)\right.\nonumber\\
&&\left.-6(\gamma-3)\ _2F_1\left(1,\frac{1-\gamma}{2-\gamma};\frac{3}{2};
(2-\gamma)E\right)+(\gamma-1)\ _2F_1\left(1,\frac{3-\gamma}{2-\gamma};
\frac{3}{2};(2-\gamma)E\right) \right],
\end{eqnarray}
and 
\begin{equation}
f_1(E)=\frac{3-\gamma}{8\pi^3}
\left\{1-\left[1-(2-\gamma)E\right]^{\frac{1}{2-\gamma}}\right\}^3
\left\{5-\gamma + 
\frac{\gamma-1}{\left[1-(2-\gamma)E\right]^{\frac{1}{2-\gamma}}}\right\}.
\end{equation}
This solution is also restricted to $\gamma \ge 1$, like that of
Section \ref{subsec:simpleaniso}, because otherwise ${\cal F}$ becomes
negative in the physical range. However $\gamma$ may exceed $2$,
because the limitation imposed by the series expansion in $\psi$ used
to obtain the solutions in Section \ref{subsec:firstmodels} no longer
applies. The distribution functions are particularly simple for 
the $\gamma=1$ Hernquist case for which
\begin{equation}
\label{oneoneDF}
{\cal F}(E,L,1,1)=\frac{8E^2\sqrt{2E}}{5\pi^3} + \frac{E^3}{\pi^3L}.
\end{equation}

\subsection{Second-order moments}
\label{subsec:secondordermoments}

An analytical expression for the velocity dispersions of all models
can be derived by means of equations (\ref{disprani}) and
(\ref{disptani}).  Since the augmented density is separable in $r$ and
$\psi$ and depends on $r$ as $(1+r)^qr^{-q}$, the tangential velocity
dispersion can easily be derived from equation (\ref{disptani}) as
\begin{equation}
\sigma_{\theta}^2(r,q)=\left(1-\frac{q}{2}\frac{1}{1+r}\right)\sigma_r^2(r,q).
\label{disptone}
\end{equation}
Hence Binney's anisotropy parameter is the monotonic function
\begin{equation}
\label{binneybeta}
\beta(r)=1-\frac{\sigma_{\theta}^2(r)}{\sigma_r^2(r)}=\frac{q}{2(1+r)},
\end{equation}
for all $\gamma$. It has the same sign as $q$. Systems are more radial
than isotropic when $q>0$, and more tangential than isotropic when
$q<0$ as seen in Fig. \ref{fig:fig3}, though they all tend to
isotropy at large $r$ where $\beta \to 0$. 
The condition $\sigma_{\theta}^2(r,q) \ge 0$
restricts $q \le 2$. There is no lower limit on $q$, and the limit
$q \to -\infty$ gives a system with all orbits circular.

\begin{figure}
\centering
\includegraphics[width=5cm,angle=0,clip=true]{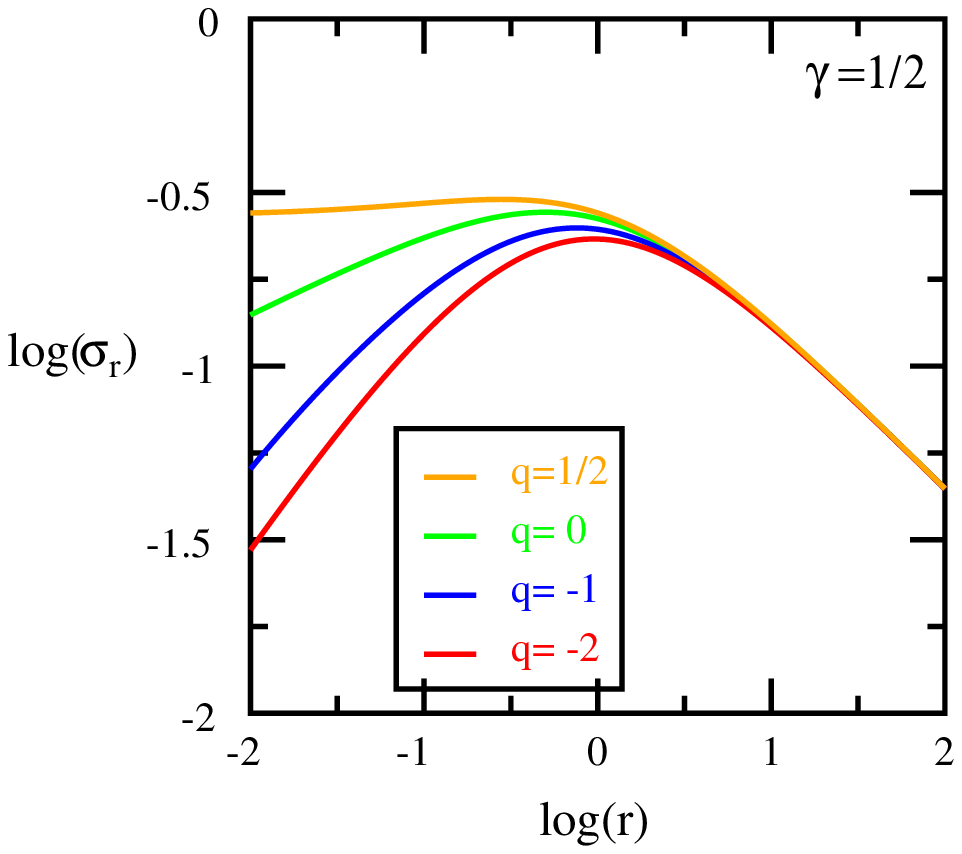}\hspace{1cm}
\includegraphics[width=5cm,angle=0,clip=true]{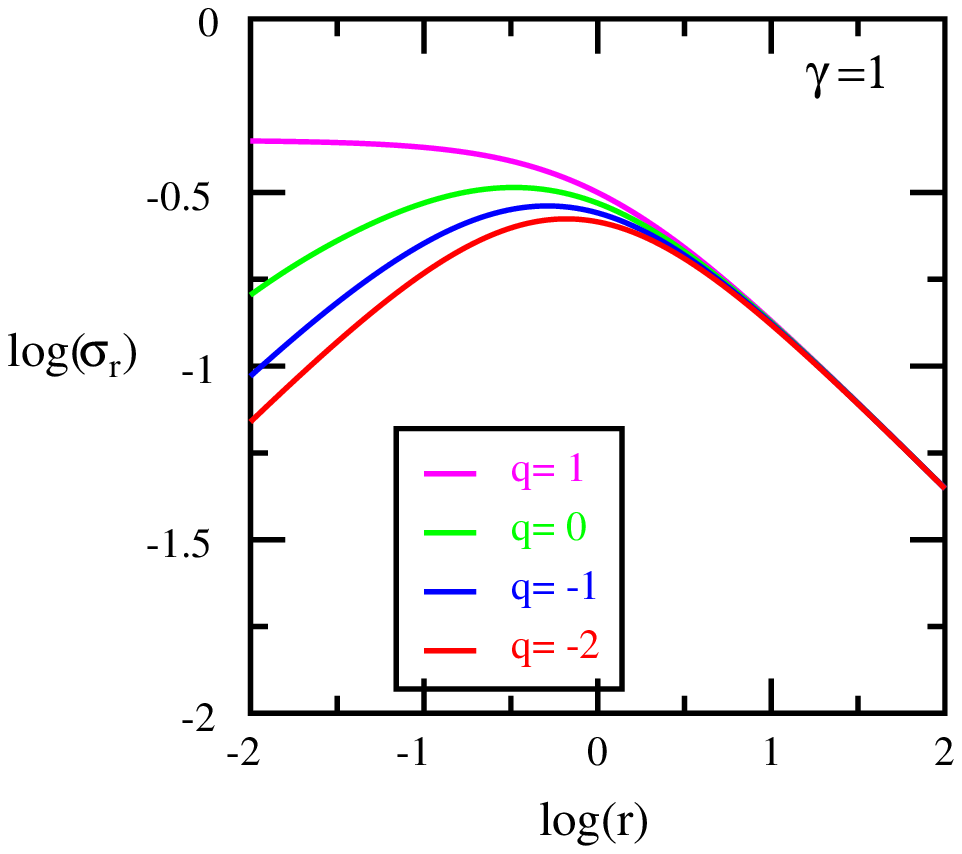}\hspace{1cm}
\includegraphics[width=5cm,angle=0,clip=true]{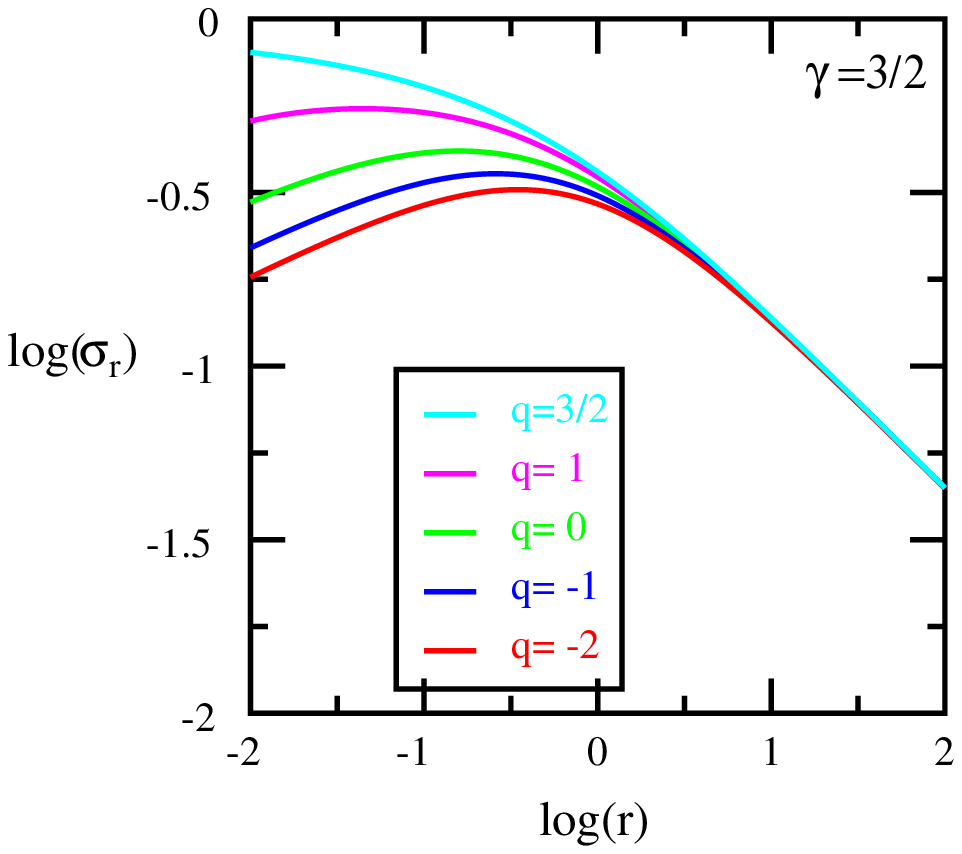}
\caption{Radial velocity dispersion profiles for models of the first family 
for three $\gamma$ values. Dispersions increase with $q$, and the top
curve corresponds to the highest allowed value $q=\gamma$.\label{fig:fig2}}
\end{figure}

\begin{figure}
\begin{center}
\includegraphics[width=6cm,angle=0,clip=true]{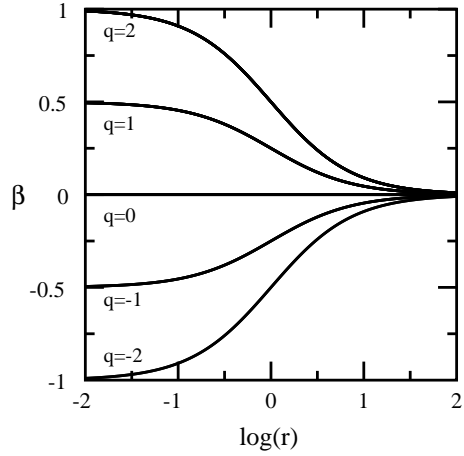}
\caption{The variation of the anisotropy parameter with radius $r$
for the models of \S\ref{subsec:firstmodels}. It is
independent of the model parameter $\gamma$.\label{fig:fig3}}
\end{center}
\end{figure}

The simple separable form of the 
augmented density in equation (\ref{gammanirho}), combined with equation
(\ref{disprani}), leads to a compact explicit expression for the
radial velocity dispersion for all $\gamma$ and $q$ in terms
of an incomplete Beta function \citep{abramowitz} as
\begin{equation}
\sigma_r^2(r,q)=r^{\gamma-q}(1+r)^{4+q-\gamma} 
B_{\frac{1}{1+r}} (5 , 2-2\gamma+q),
\label{disprone}
\end{equation}
cf \citep{baes2}. 
Fig. \ref{fig:fig2} shows the radial dispersion $\sigma_r(r,q)$ tending to
a common form at large $r$ where all models tend to isotropy.
Equation (\ref{disprone}) shows that common form to be $1/\sqrt{5r}$.
The figures also show that $\sigma_r(r,q)$
increases with increasing $q$ near the centre as the orbits there become
more strongly radial.


\subsection{Energy distribution}
\label{subsec:energydist}

The anisotropic models differ in way in which energy is distributed
among their orbits. We label the part of the mass density that is
contributed by the energy range $[E, E+dE]$ as the energy density
${\cal F}_E$. It is given by the inner integral of equation
(\ref{rhoani}) as
\begin{equation}
{\cal F}_E(\psi,r,E)=2\pi\int_0^{2(\psi-E)}
\frac{{\cal F}(E,L)}{\sqrt{2(\psi-E)-v_t^2}}dv_t^2.
\label{formfe}
\end{equation}
We define as $F_{E,q}^p(\psi,r,E)$ the components of energy density 
which are obtained with the elementary distribution functions $F_q^p(E,L)$ 
on the right hand side of this equation (\ref{formfe}). They are
evaluated in Appendix \ref{app:evaluations}. Then the energy density for the
full model is given by the sum
\begin{equation}
\label{energysum}
{\cal F}_E(\psi,r,E)=\frac{3-\gamma}{4\pi}\sum_{p=4}^{\infty}
C(\gamma,p,q) F_{E,q}^p(\psi,r,E),
\end{equation}
with the coefficients $C(\gamma,p,q)$ that were introduced 
in equation (\ref{Ccoeffs}). The energy density for the simple $\gamma=q=1$ 
distribution function (\ref{oneoneDF}) is
\begin{equation}
\label{oneoneFE}
{\cal F}_E(\psi,r,E)=\psi^3\left[\frac{64}{5\pi^2}
\left(\frac{E}{\psi}\right)^{5/2}\sqrt{1-\frac{E}{\psi}}
+\frac{2}{\pi r}\left(\frac{E}{\psi}\right)^3 \right],
\quad 0 \le \frac{E}{\psi} \le 1.
\end{equation}

\begin{figure}
\begin{center}
\includegraphics[width=5cm,angle=0,clip=true]{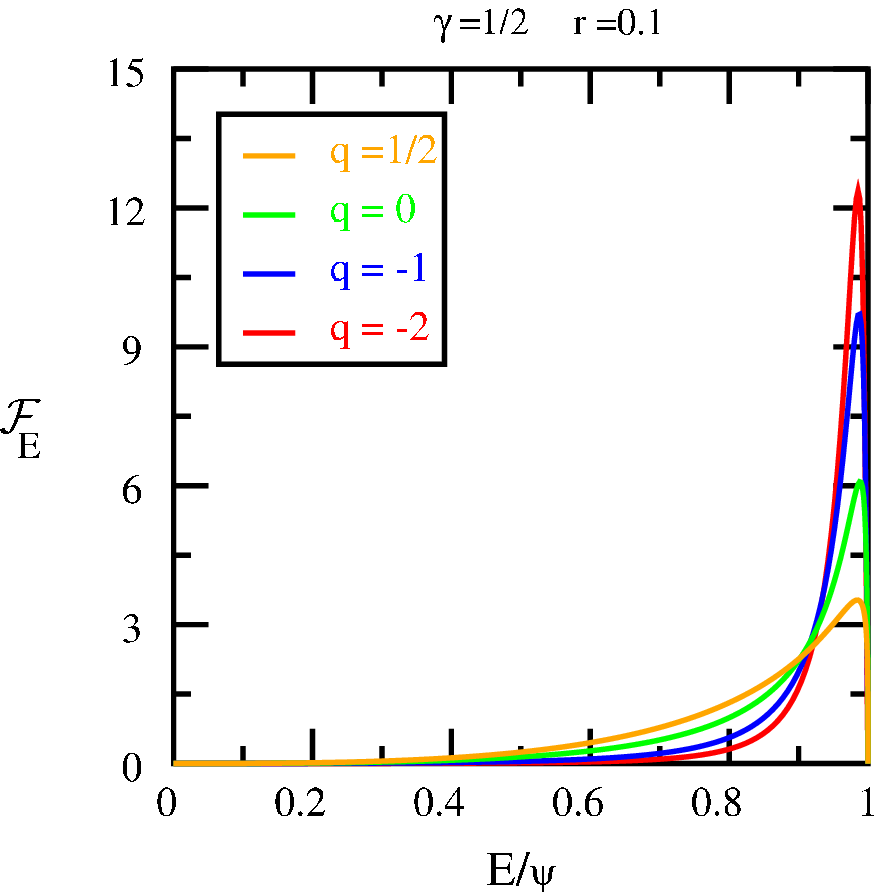}\hspace{1cm}
\includegraphics[width=5cm,angle=0,clip=true]{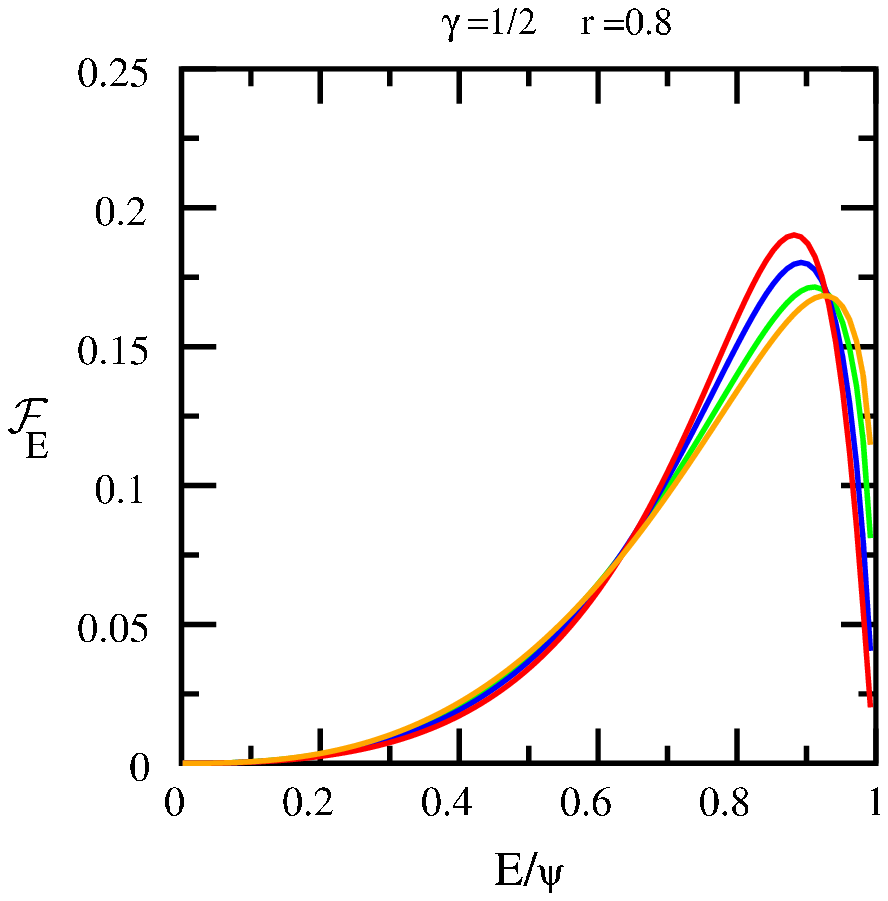}\hspace{1cm}
\includegraphics[width=5cm,angle=0,clip=true]{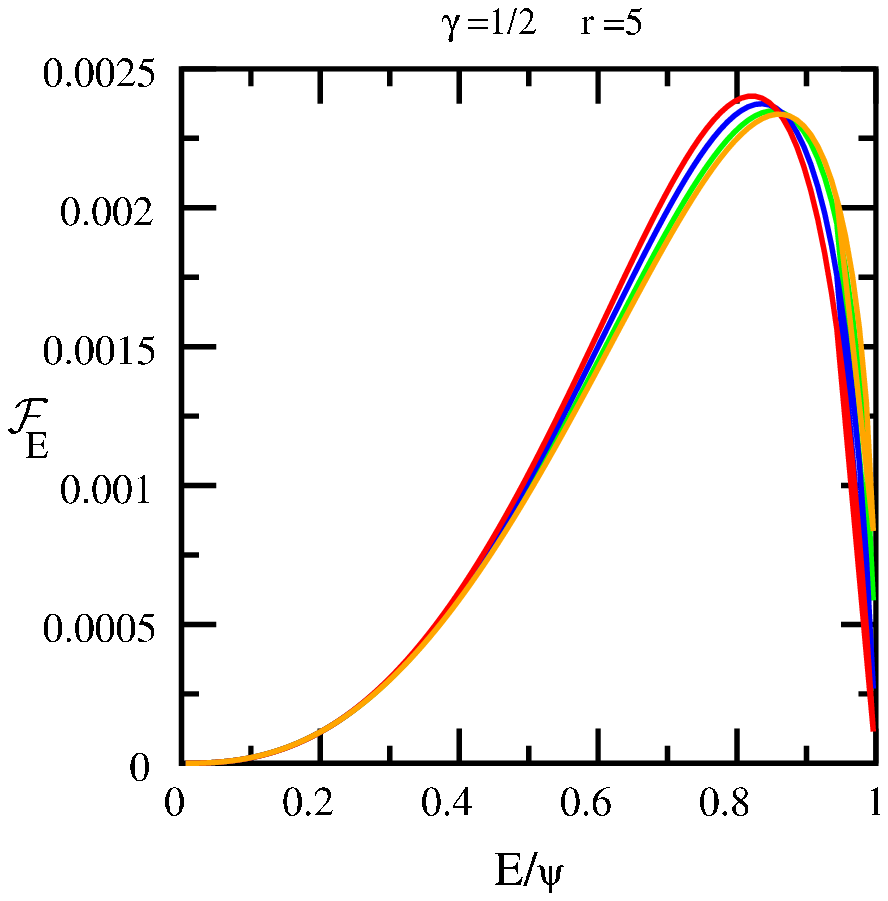}\vspace{1cm}
\includegraphics[width=5cm,angle=0,clip=true]{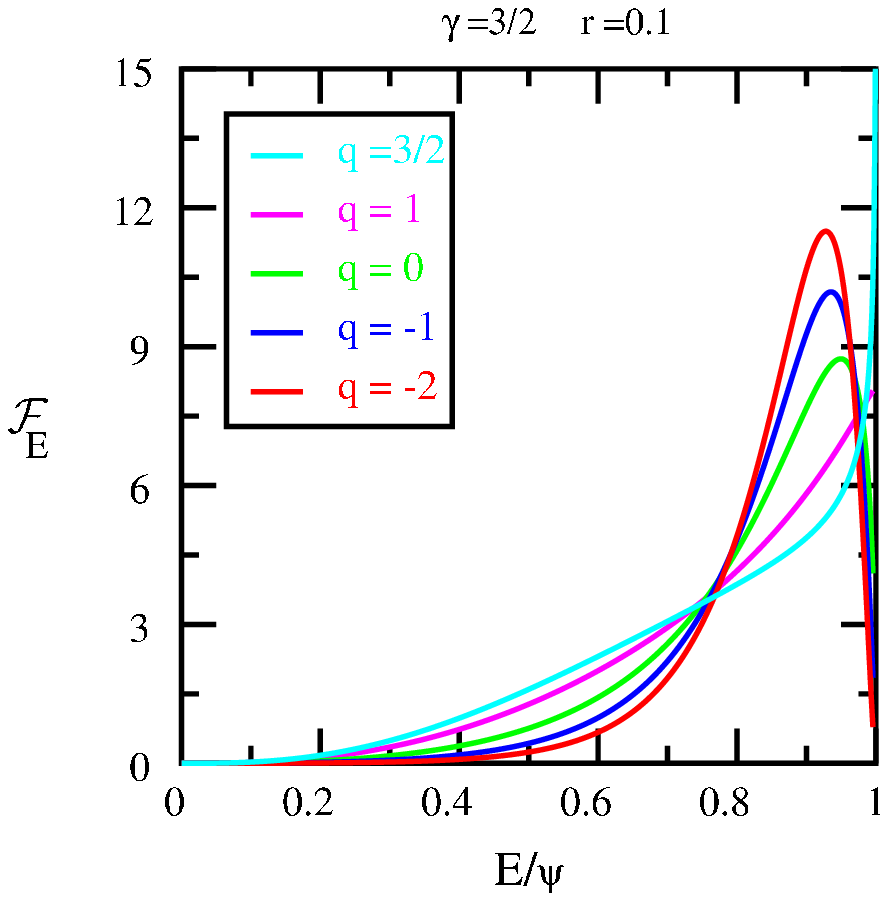}\hspace{1cm}
\includegraphics[width=5cm,angle=0,clip=true]{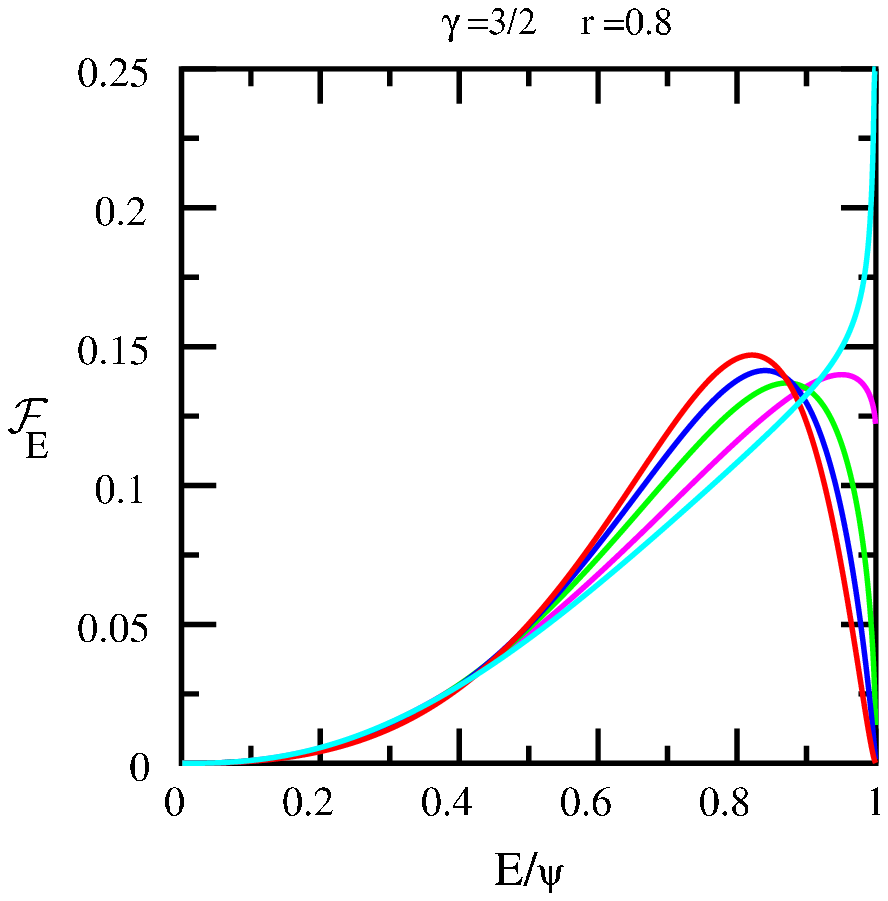}\hspace{1cm}
\includegraphics[width=5cm,angle=0,clip=true]{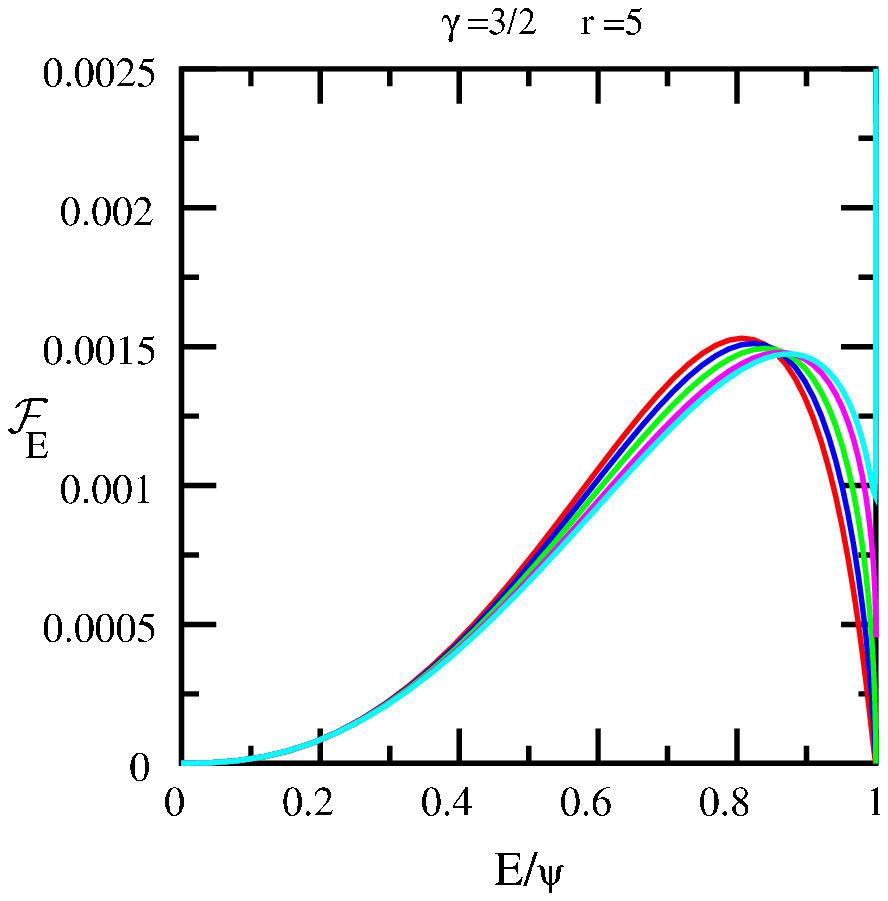}\hspace{1cm}
\caption{Energy densities for the $\gamma =1/2$ and 
$\gamma =3/2$ models of the first family at spatial radii $r=0.1$, 
$r=0.8$, and $r=5$. The same colour-codings of the values of $q$ are used
through each row. The energy $E$ is scaled with its maximum
value $\psi(r)$ at that radius. Each curve in a panel encloses the
same area. That area is the local mass density divided by $\psi$, and
varies from panel to panel with both $r$ and $\gamma$. The singular
growth of the $q=\gamma=3/2$ curve for $r=5$ does not occur until
$E/\psi> 0.999$. \label{fig:fig4}}
\end{center}
\end{figure}

Fig.~\ref{fig:fig4} shows the energy densities for two
values of $\gamma$ at three radii: one small, one intermediate, and
one large.  We chose the $\gamma$ values of $1/2$ and $3/2$ to
represent the range $0<\gamma<2$ for which our solutions apply.
${\cal F}_E$ is more concentrated at higher energy at small $r$ for the more
tangential models (smaller $q$) and less concentrated for
the more radial models (larger $q$). The situation is reversed at
large $r$. The reason is that the more radial orbits are less tightly
bound than the more tangential ones near the center, while the
opposite is the case at large $r$.  Fig.1 of Dejonghe (1987) for
anisotropic Plummer models shows the same phenomenon, though our
${\cal F}_E$ are more concentrated towards high energies than his because
our $\gamma$-models are more centrally concentrated than his Plummer
model.  Also, unlike his, the differences between our models diminish
at large $r$ where they become isotropic.

Whereas ${\cal F}_E$ drops to zero at the upper limit $E/\psi=1$ for all
of the upper row of $\gamma=1/2$ models, and also for the $q \le 1$,
$\gamma=3/2$ models of the lower row, it becomes infinite as $E/\psi \to 1$
for the most radial and most cuspy $q=\gamma=3/2$ model, and has a finite
limit for the $q=1$ model, as in the $\gamma=1$ case in 
equation (\ref{oneoneFE}). This behaviour is a consequence of the
$(1-E/\psi)^{(1-q)/2}$ dependence of $F^p_{E,q}$ for $E/\psi >r^2/(1+r^2)$
given by equation (\ref{FElargeE}). That dependence is a consequence of the
$L^{-q}$ dependence of the distribution function for $L^2<2E$ as given
by equation (\ref{FsmallL}). 


\begin{figure}
\begin{center}
\includegraphics[width=5cm,angle=0,clip=true]{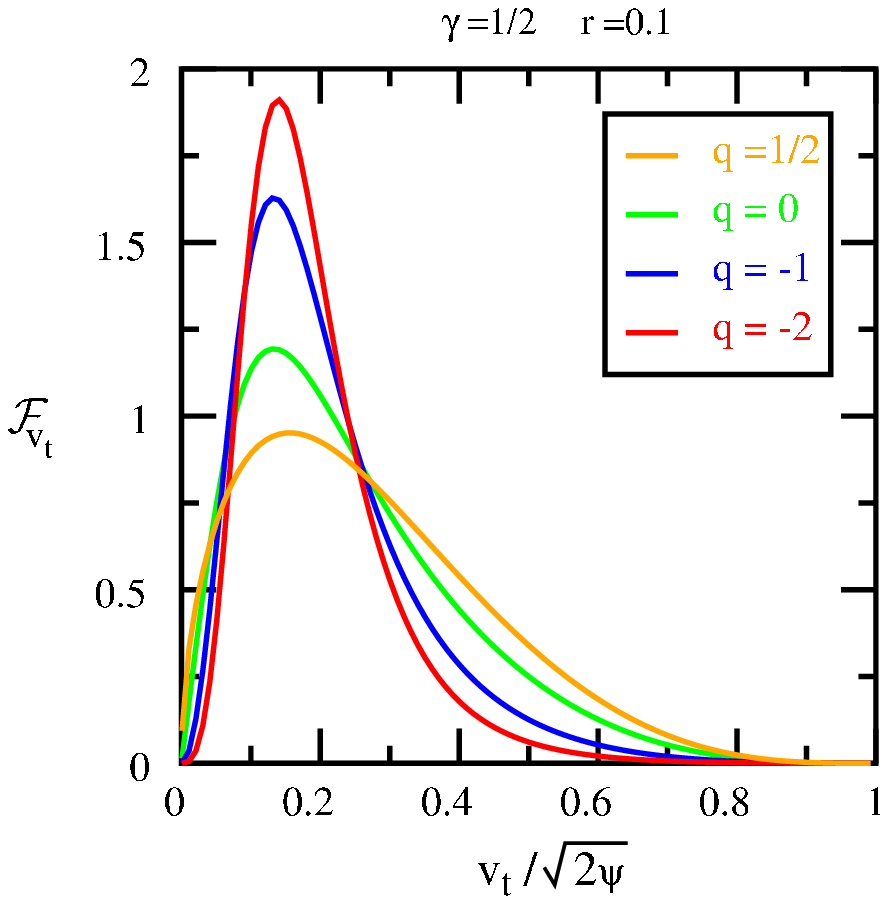}\hspace{1cm}
\includegraphics[width=5cm,angle=0,clip=true]{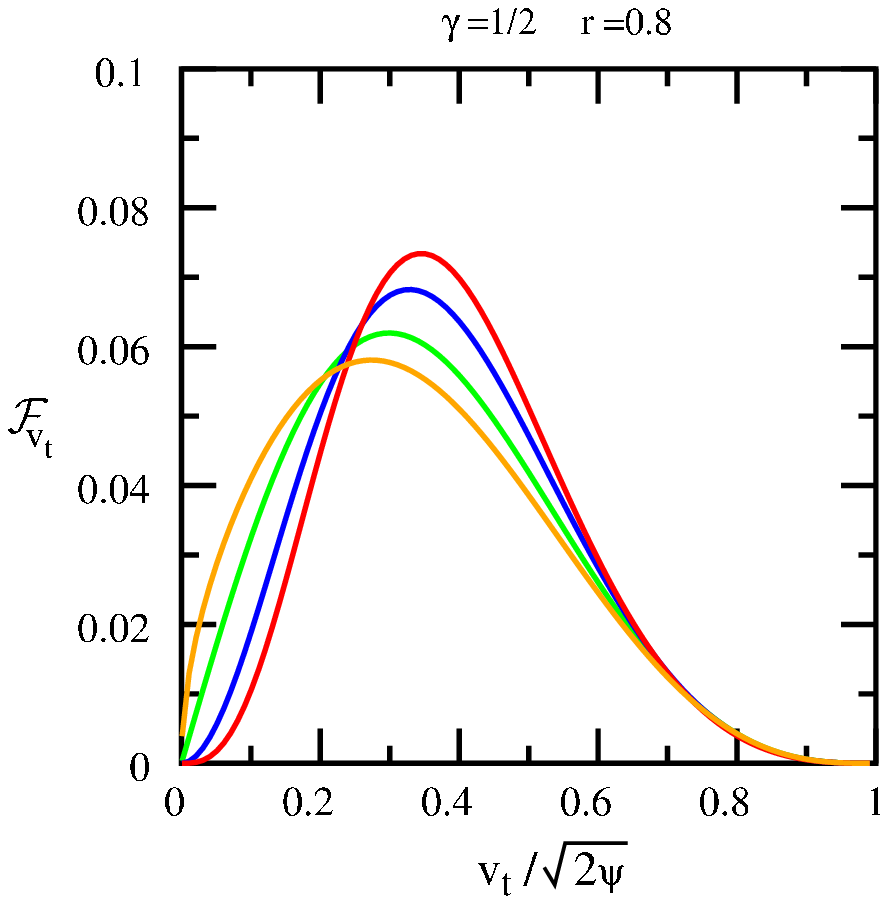}\hspace{1cm}
\includegraphics[width=5cm,angle=0,clip=true]{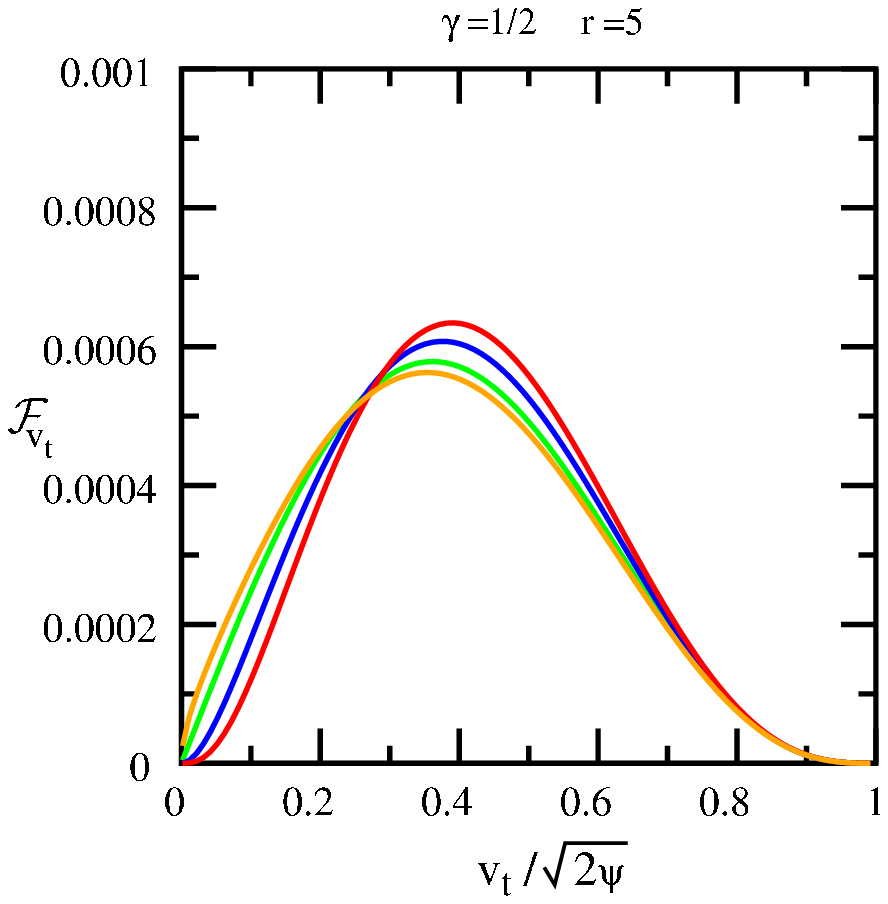}\vspace{1cm}
\includegraphics[width=5cm,angle=0,clip=true]{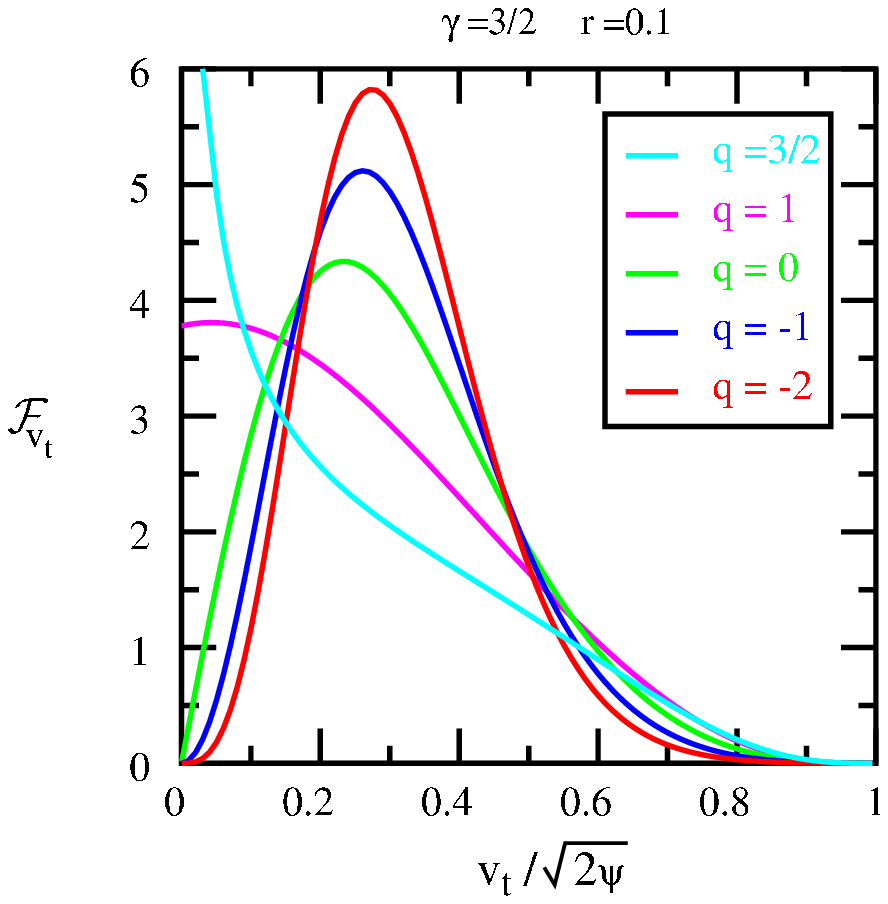}\hspace{1cm}
\includegraphics[width=5cm,angle=0,clip=true]{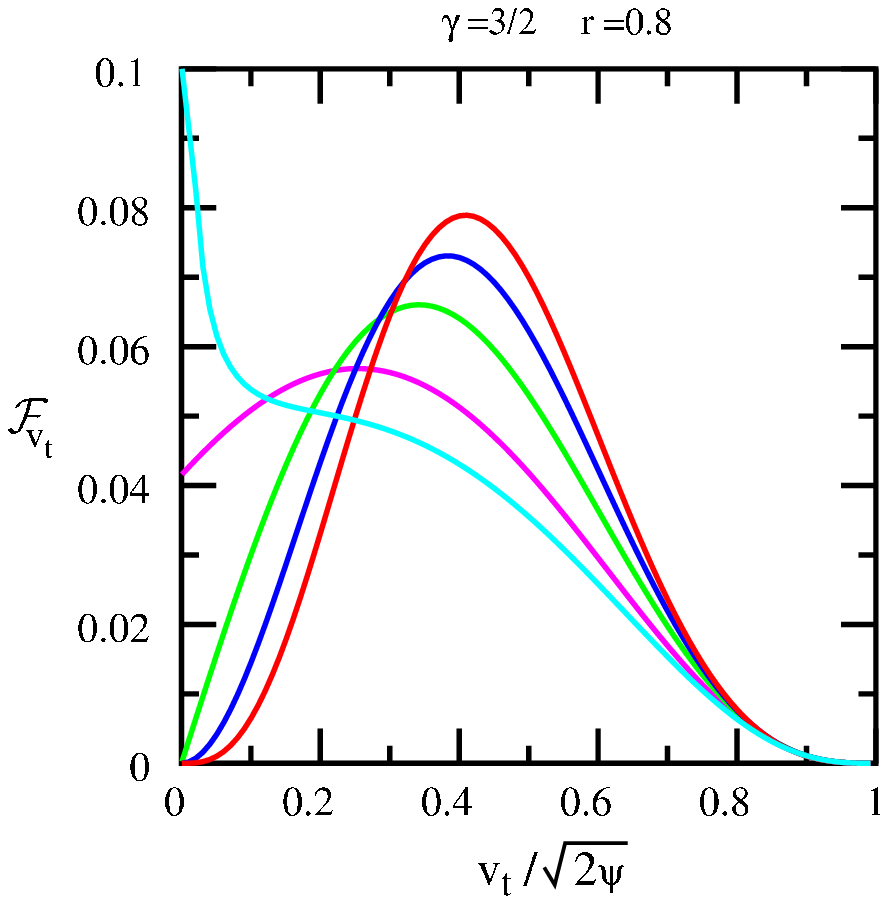}\hspace{1cm}
\includegraphics[width=5cm,angle=0,clip=true]{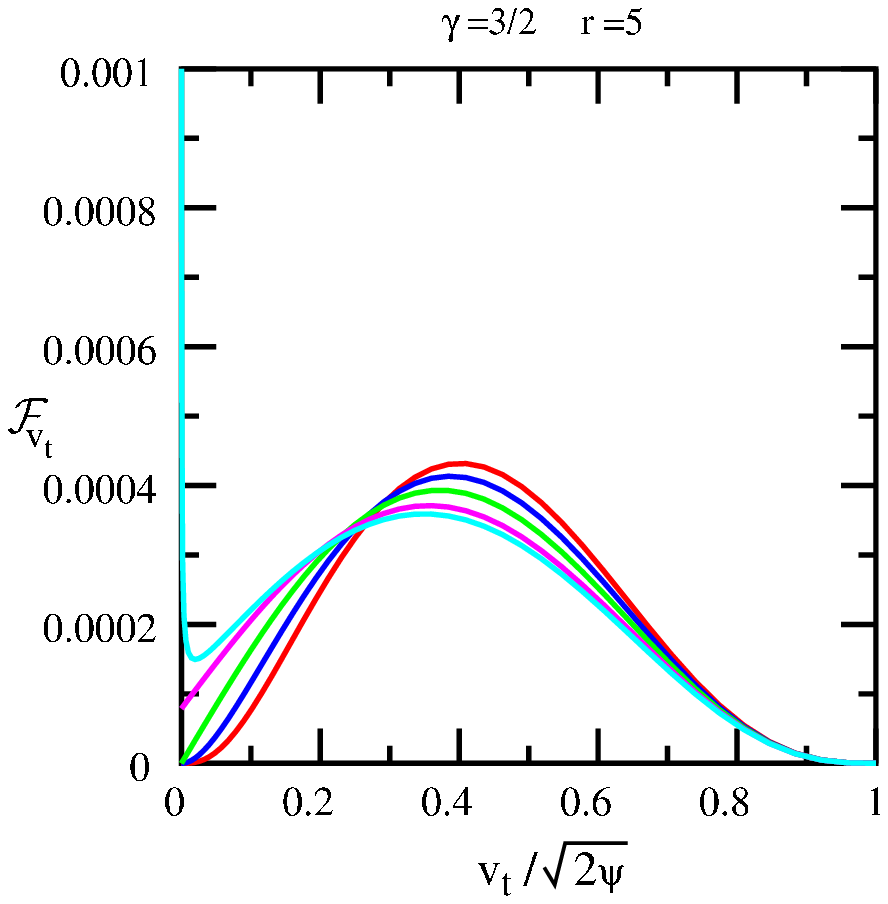}\hspace{1cm}
\caption{The transverse velocity densities for the $\gamma =1/2$ and
$\gamma =3/2$ models of the first family at spatial radii $r=0.1$,
$r=0.8$, and $r=5$. The same colour-codings of the values of $q$ are
used through each row. The transverse velocity is scaled with its maximum
value $\sqrt{2\psi}$ at that radius.  All the curves in a single panel
again enclose the same area; it is now the local mass density divided
by $\sqrt{2\psi}$.\label{fig:fig5}}
\end{center}
\end{figure}

\subsection{Distribution of the transverse motions}
\label{subsec:transveldist}

The anisotropic models also differ in way in which transverse velocity
is distributed among their orbits. We label the part of the mass
density which is contributed by the transverse velocity range $[v_t,
v_t+dv_t]$ as the transverse velocity density ${\cal F}_{v_t}$.  It is
given by the inner integral of equation (\ref{rhoani}), after the
order of its integrations has been changed, as
\begin{equation}
{\cal F}_{v_t}(\psi,r,v_t)=4\pi v_t\int_0^{\psi-v_t^2/2}
\frac{{\cal F}(E,L)}{\sqrt{2(\psi-E)-v_t^2}}dE.
\label{formfvt}
\end{equation}
We evaluate the transverse velocity densities 
$F_{v_t,q}^p(\psi,r,v_t)$ corresponding to the elementary distribution 
functions $F_q^p(E,L)$ in Appendix \ref{app:evaluations}. 
The transverse velocity density for the full model is 
\begin{equation}
\label{transvelsum}
{\cal F}_{v_t}(\psi,r,v_t)=\frac{3-\gamma}{4\pi}\sum_{p= 4}^{\infty}
C(\gamma,p,q) F_{v_t,q}^p(\psi,r,v_t),
\end{equation}
with the same coefficients $C(\gamma,p,q)$ as before.
The transverse velocity density for the simple $\gamma=q=1$ 
distribution function (\ref{oneoneDF}) is
\begin{equation}
\label{oneoneFvt}
{\cal F}_{v_t}(\psi,r,v_t)=\frac{(2\psi)^{7/2}}{4\pi}
\left[t(1-t^2)^3+\frac{32}{35\pi r}(1-t^2)^{7/2}\right],
\quad 0 \le t=\frac{v_t}{\sqrt{2\psi}} \le 1.
\end{equation}

Fig.~\ref{fig:fig5} shows the transverse velocity densities for the
same set of models as those whose energy densities were shown
in Fig.~\ref{fig:fig4}. Now ${\cal F}_{v_t}$ is generally more
concentrated toward low velocities at small $r$ for the more
tangential models (smaller $q$) and less so for the more radial models
(larger $q$). The situation is reversed at large $r$, though
differences there are small because our models tend to isotropy
there. Fig. 4 of Dejonghe (1987) for anisotropic Plummer models shows
similar behaviour at small radii.  The generally higher transverse
velocities at small $r$ for the more radial models is due to orbits
which are close to their turning points and so have transverse
velocities which are well in excess of the local circular velocity.

The $L^{-q}$ dependence of the distribution function for 
$L^2<2E$ again has an important effect. It is now seen as $v_t \to 0$.
Equation (\ref{Fvtsmallvt}) shows that $F^p_{v_t,q}$ varies as
$v_t^{1-q}$ for $v_t/\sqrt{2\psi} < 1/\sqrt{1+r^2}$. It tends to $0$ as
$v_t \to 0$ for $q < 1$, has a finite limit there for $q = 1$, and becomes
infinite for $q > 1$. The infinite growth is more prominent
for the $\gamma=3/2$ models of Fig.~\ref{fig:fig5} than for those of
Fig.~\ref{fig:fig4} because of its lower power (-1/2 rather than -1/4).

\subsection{Observable properties}

\begin{figure}
\begin{center}
\includegraphics[width=6cm,angle=0,clip=true]{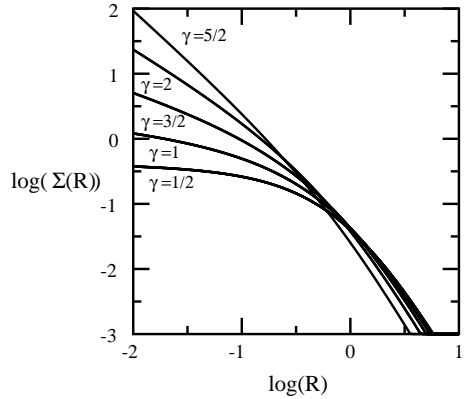}
\caption{The projected density distribution for different
$\gamma$-models.\label{fig:rhop}}
\end{center}
\end{figure}

\begin{figure}
\begin{center}
\includegraphics[width=5cm,angle=0,clip=true]{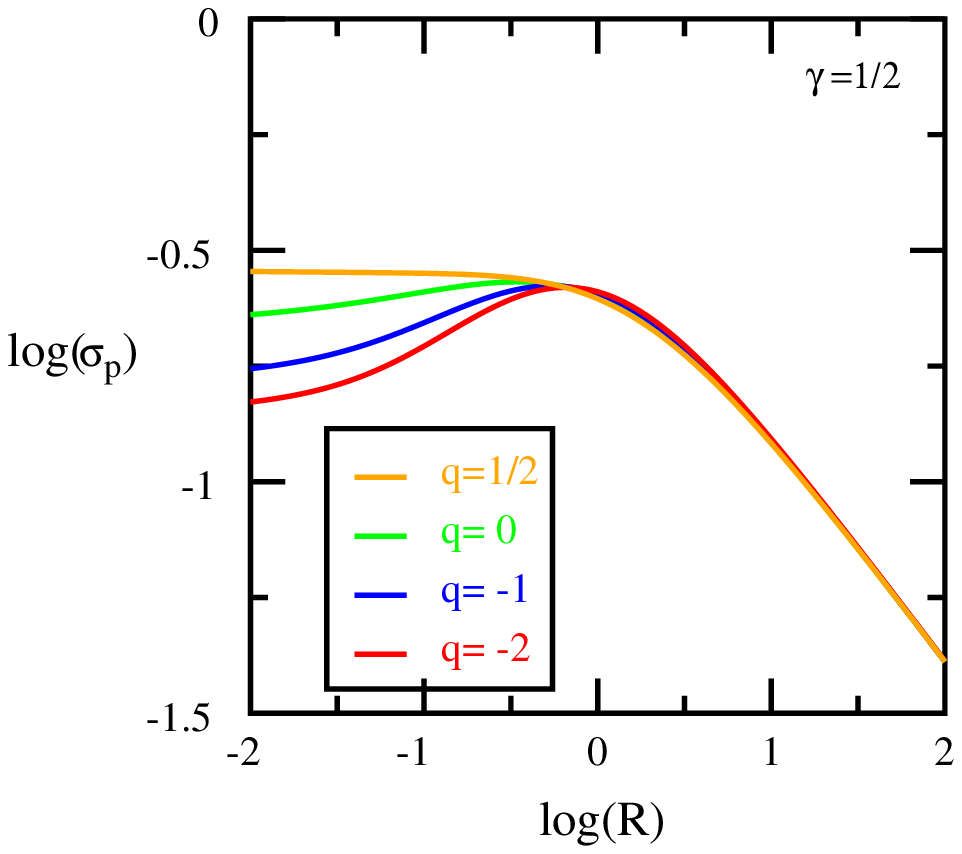}\hspace{1cm}
\includegraphics[width=5cm,angle=0,clip=true]{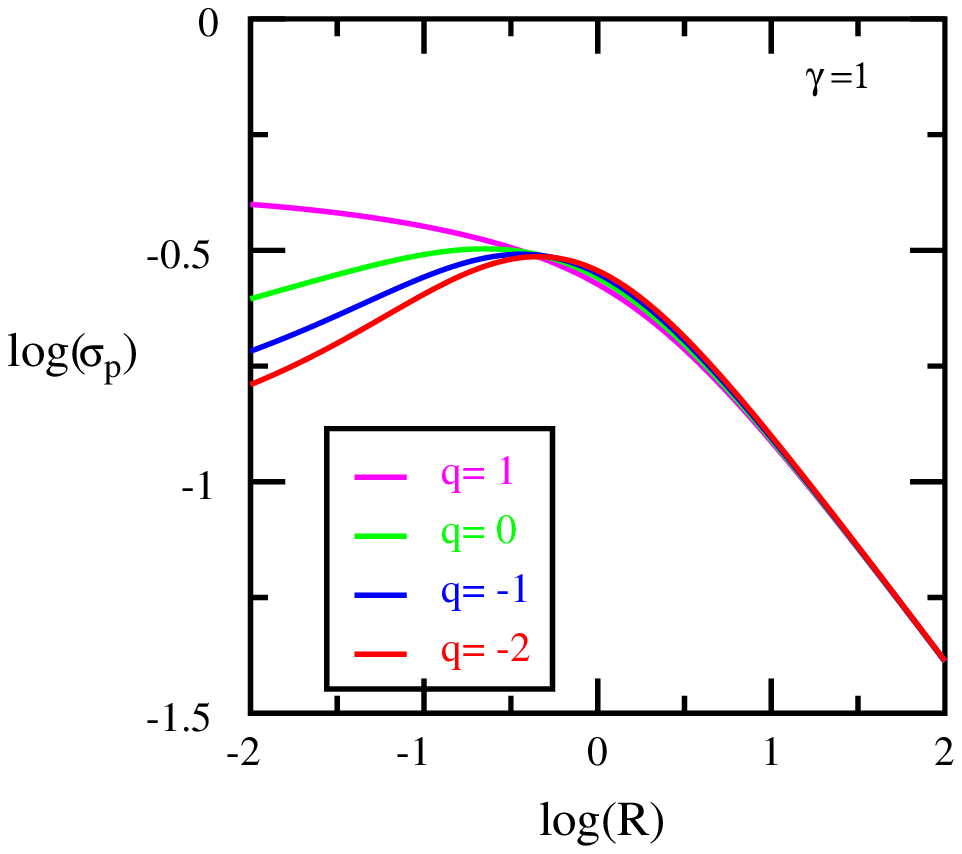}\hspace{1cm}
\includegraphics[width=5cm,angle=0,clip=true]{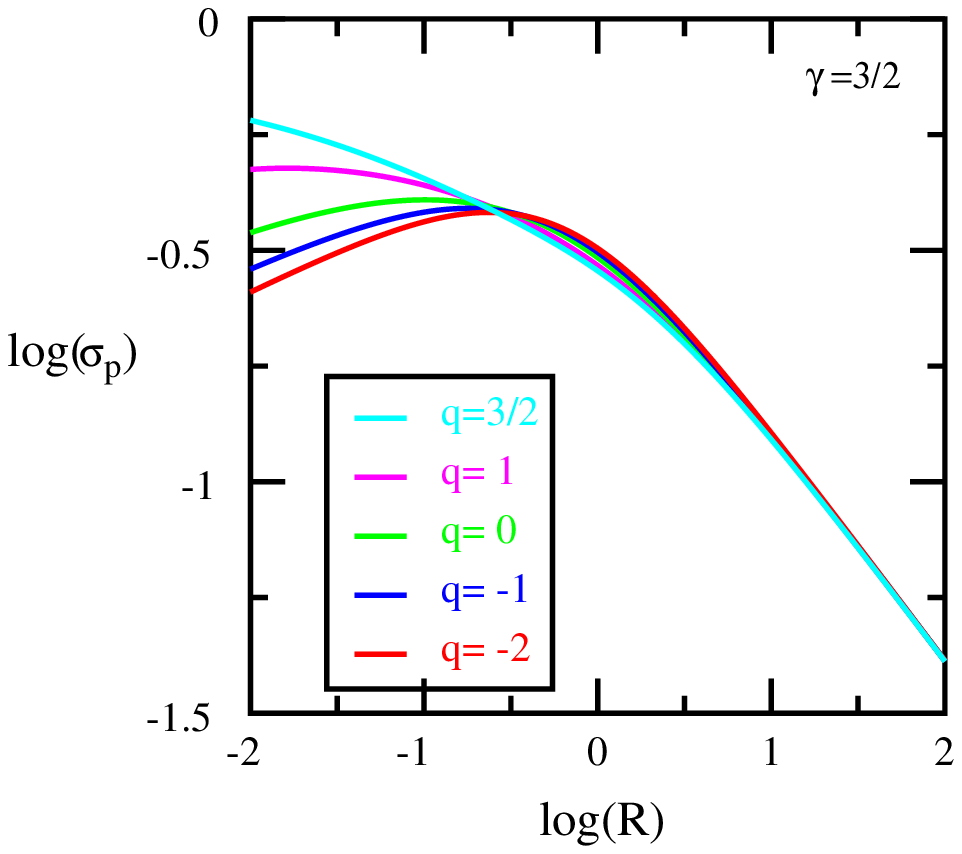}
\caption{The projected velocity dispersion profiles for the same set
of models as in Fig. \ref{fig:fig2}.\label{fig:dispp}}
\end{center}
\end{figure}

The projected density of a $\gamma$-model is found from the relation
\begin{equation}
\label{projden}
\Sigma(R)=2\int_{R}^{\infty}\frac{r\rho(r)dr}{\sqrt{r^2-R^2}},
\end{equation}
where we align the $z$-axis with the line-of-sight, and $R=\sqrt{x^2+y^2}$
is the usual plane polar coordinate, now in the plane of the sky.
The projected  velocity dispersion $\sigma_p(R)$ is given by
\citep{binney2,BT}
\begin{equation}
\label{projdisp}
\sigma_p^2(R)=\frac{2}{\Sigma(R)}\int_{R}^{\infty}
\left[1-\frac{R^2}{r^2}\beta(r)\right]
\frac{r\rho(r)\sigma_r^2(r)dr}{\sqrt{r^2-R^2}}.
\end{equation}
The integrations needed for both (\ref{projden}) and (\ref{projdisp})
must be done numerically for $\gamma$-models. Projected densities of
some $\gamma$-models are shown in Fig.~\ref{fig:rhop}; the dependence
of the central density slope on the parameter $\gamma$ is clearly
visible in projection. Projected velocity dispersions are shown in
Fig.~\ref{fig:dispp}. They decrease monotonically with the
projected radius $R$ in radial systems, but can have a central peak
for tangential systems when the tangential component of the velocity
dispersion becomes increasingly important with increasing $R$.

The normalised line-of-sight velocity profile $l(v_z,R)$ describes the
distribution of $v_z$ along the line-of-sight. It is obtained by
integrating the distribution function over the velocities $v_R$ and
$v_{\varphi}$ in the plane of the sky, followed by a spatial
integration along the line-of-sight. It is given by
\begin{equation}
\Sigma(R) l(v_z,R)=2\int_{R}^{\infty}\frac{rdr}{\sqrt{r^2-R^2}}
\int\!\!\!\int_{0 \le v_R^2+v^2_{\varphi} \le 2\psi-v_z^2} 
dv_{R} dv_{\varphi}{\cal F}(E,L).
\label{lp}
\end{equation}
It falls to zero at the extremities $v_z=\pm \sqrt{2\psi(R)}$. 
We express the arguments of the distribution function ${\cal F}$ as
\begin{equation}
E=\psi-\frac{1}{2}(v_z^2+v_R^2+v^2_{\varphi}), \quad
L=\sqrt{r^2v^2_{\varphi}+(zv_R-Rv_z)^2},
\end{equation}
and integrate using polar coordinates in $(v_R,v_{\varphi})$ velocity
space. The area under the normalised velocity profile is unity because
integrating ${\cal F}$ over all three velocity component $v_z$ gives
the density $\rho(r)$.  Hence integrating the right hand side of
equation (\ref{lp}) over $v_z$ gives the projected density $\Sigma$,
and hence integrating both sides of equation (\ref{lp}) over $v_z$
leads to the result $\int l(v_z,R) dv_z =1$.

\begin{figure}
\begin{center}
\includegraphics[width=5cm,angle=0,clip=true]{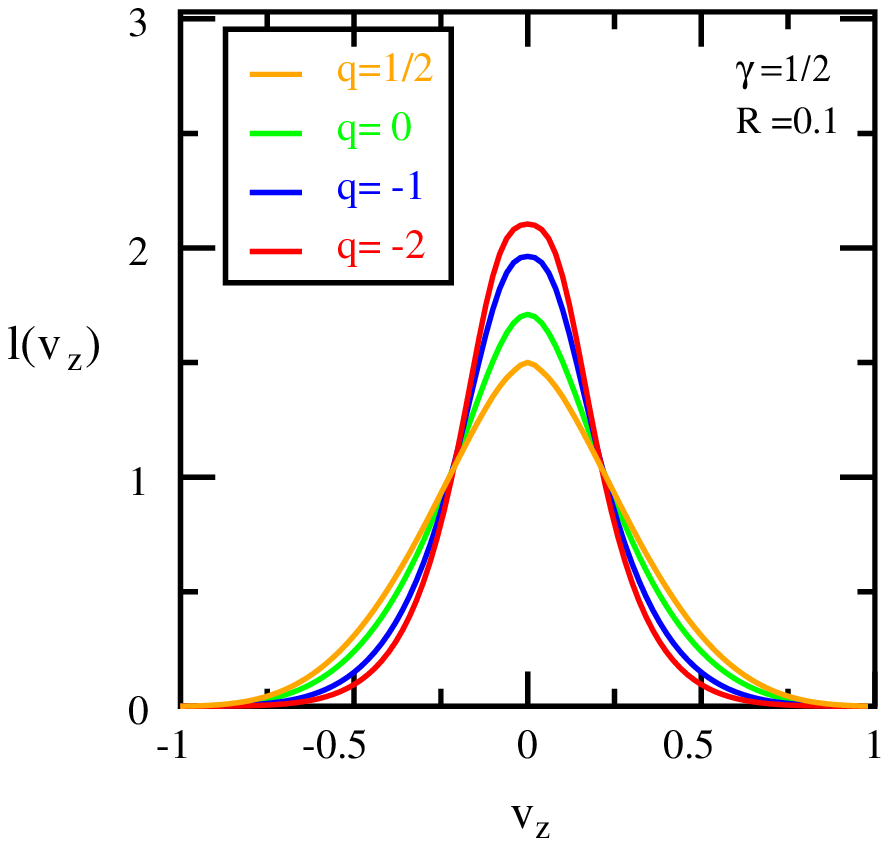}\hspace{1cm}
\includegraphics[width=5cm,angle=0,clip=true]{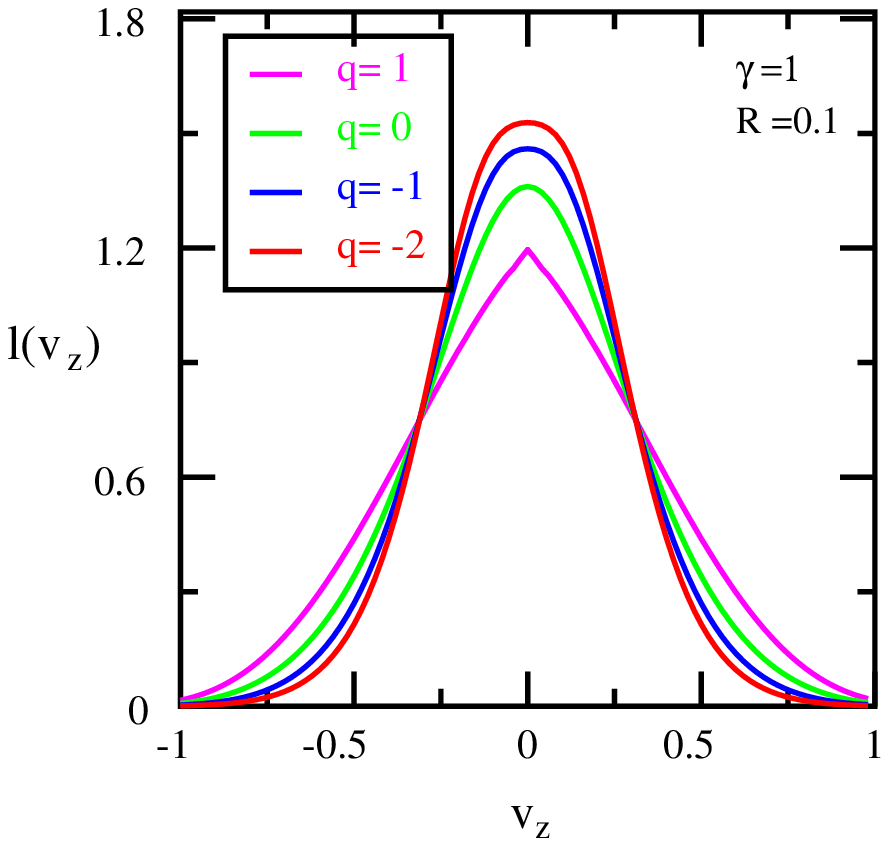}\hspace{1cm}
\includegraphics[width=5cm,angle=0,clip=true]{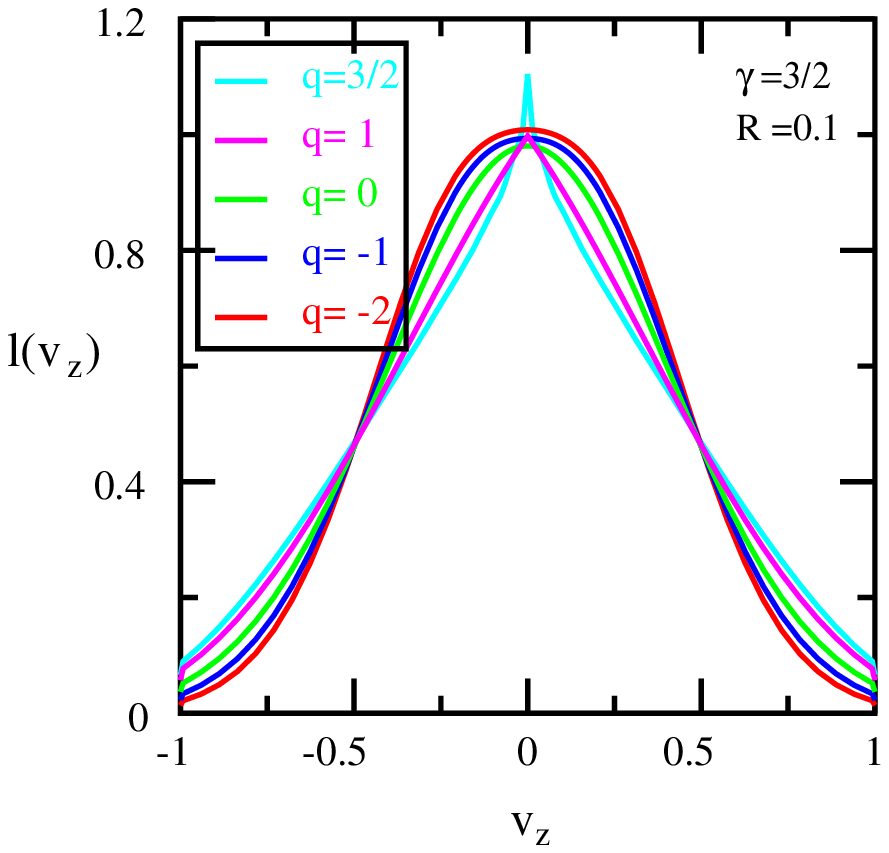}\vspace{1cm}
\includegraphics[width=5cm,angle=0,clip=true]{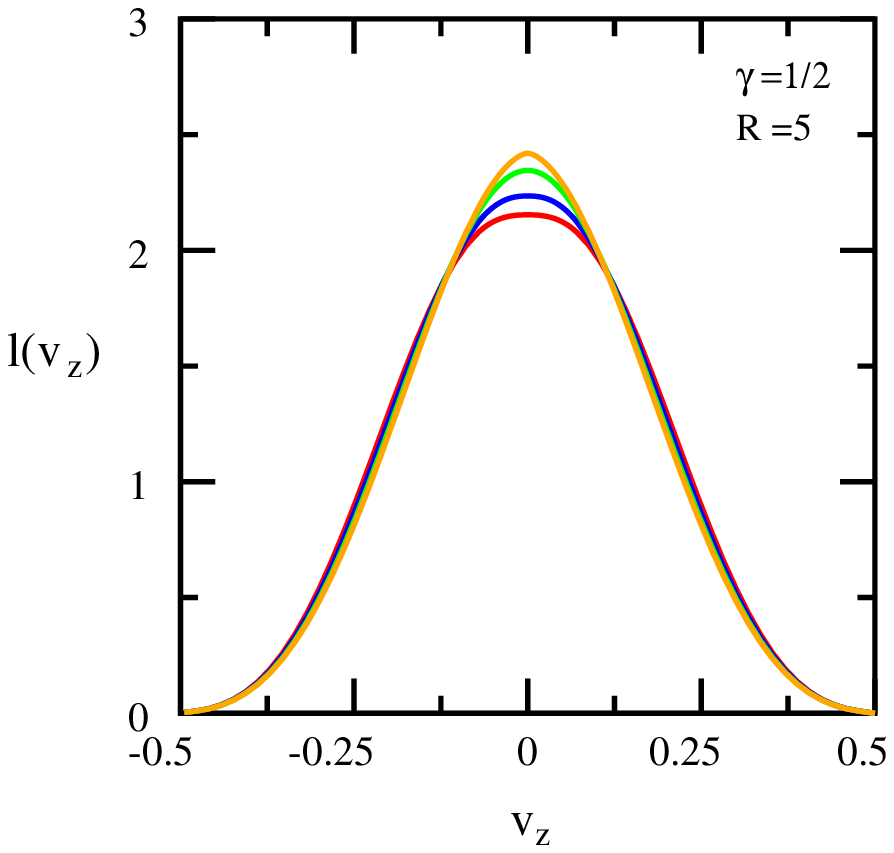}\hspace{1cm}
\includegraphics[width=5cm,angle=0,clip=true]{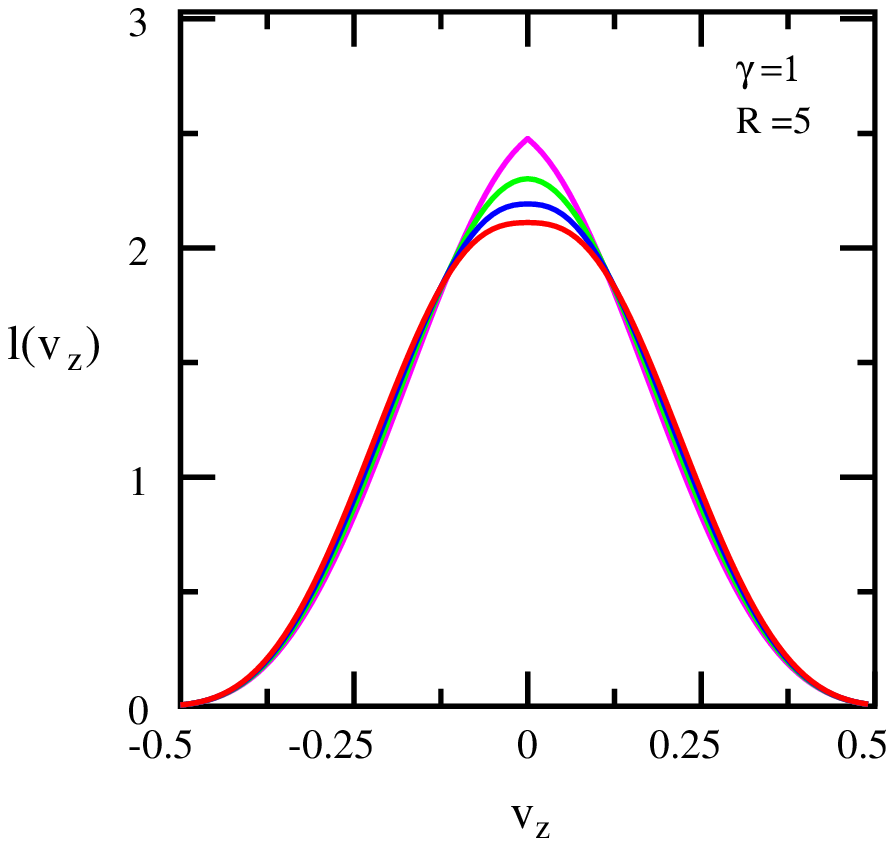}\hspace{1cm}
\includegraphics[width=5cm,angle=0,clip=true]{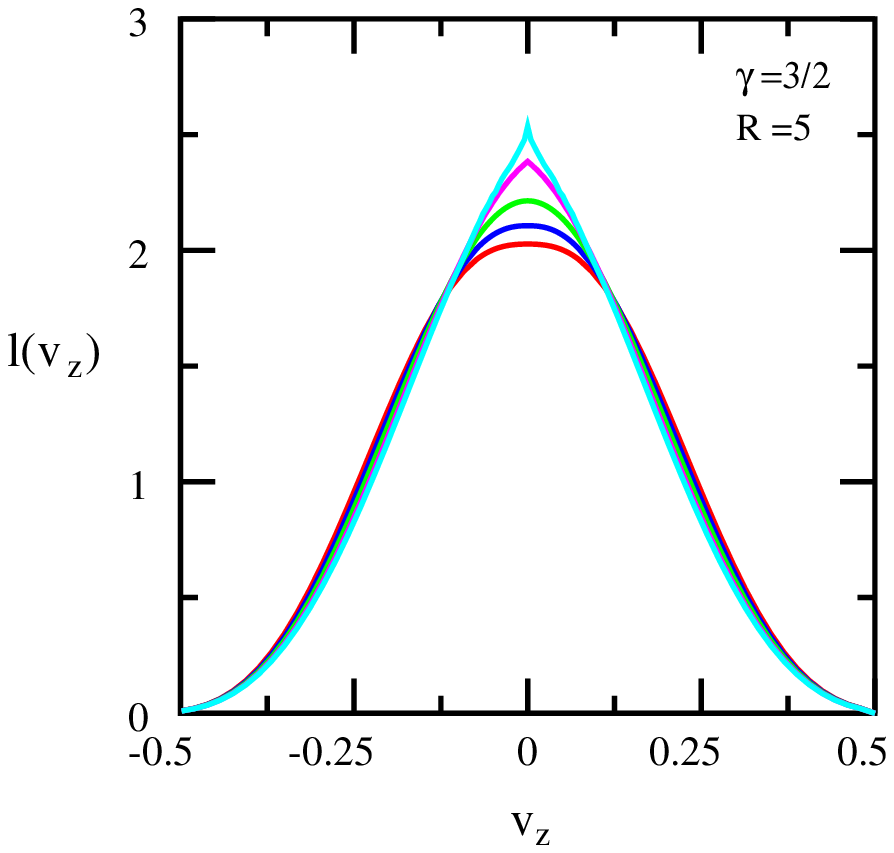}\hspace{1cm}
\caption{The normalised line-of-sight velocity profiles at projected radii
$R=0.1$ and $R=5$ for models with three different values of $\gamma$.
The same colour-codings are used in each column, i.e. for each $\gamma$.
\label{fig:losvdone}}
\end{center}
\end{figure}

Fig.~\ref{fig:losvdone} shows normalised line-of-sight velocity profiles for
models of the first family for three values of $\gamma$, all obtained
by evaluating equation (\ref{lp}) numerically.  The most striking
feature of these profiles is how much more they vary at small radii
with the central density slope $\gamma$ than with the orbital
composition. The profiles of the more tangential models are a little
narrower at small radii for all $\gamma$, and slightly broader at
large radii where they are almost isotropic.  Other features that
stand out are the discontinuous slopes at $v_z=0$ of $q=1$ profiles, 
and the sharp cusps there of $q=1.5$ profiles. They are further
visible effects of the $L^{-q}$ dependence of the distribution
functions for $L^2<2E$.


\section{More anisotropic models}
\label{sec:moremodels}

The models introduced in Section \ref{subsec:firstmodels} are all
isotropic at large radii. If isotropy is a consequence of mixing, then
it is more likely that galaxies are isotropic in their well-mixed
centres than in their outer regions. We now construct a second family
of models which are isotropic at their centres but anisotropic at larger
radii. We do this
by modifying the method of the previous section, in a way which
allows us to use similar mathematics. We derive the distribution 
functions of these models in \S\ref{subsec:secondmodels}. We give their
velocity dispersions, energy and transverse velocity distributions,
and projected line profiles in the next two sections.
Then in a brief final section \S\ref{subsec:radial models}, 
we construct some extreme $\gamma$-models for which all orbits are radial.

\subsection{Another two-parameter family of anisotropic models}
\label{subsec:secondmodels}

\begin{figure}
\begin{center}
\includegraphics[width=6cm,angle=0,clip=true]{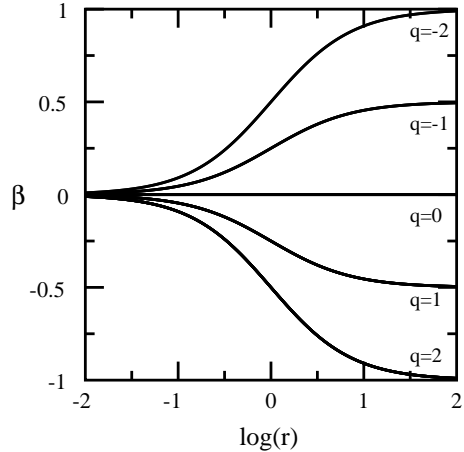}
\caption{The anisotropy parameter $\beta$ as a function of radius $r$ for
the second family of models of \S\ref{subsec:secondmodels}. It is again
independent of the model parameter $\gamma$.\label{fig:fig9}}
\end{center}
\end{figure}

We obtain these by writing the augmented density in the form of
$(1+r)^q$ times a function of $\psi$ as
\begin{equation}
\rho(\psi,r)=\frac{3-\gamma}{4\pi}
         (1+r)^q[1-(2-\gamma)\psi]^{\frac{-\gamma}{2-\gamma}}
\left\{1-[1-(2-\gamma)\psi]^{\frac{1}{2-\gamma}}\right\}^{4+q}.
\label{gammanirho2}
\end{equation}
As a result, Binney's anisotropy parameter is now
\begin{equation}
\beta(r)=-\frac{qr}{2(1+r)},
\label{betatwo}
\end{equation}
and has the opposite sign to $q$. It is zero at the isotropic centre,
and tends to $-q/2$ at large distances (see Fig. \ref{fig:fig9}).  Systems
are more strongly radial in the outer regions for negative $q$, and
more tangential for positive $q$. The parameter $q$ is restricted to 
$q \ge -2$ by the requirement that $\sigma_{\theta}^2 \ge 0$, but it is no
longer restricted by An \& Evans's (2006b) cusp slope-central anisotropy 
theorem because the centre is isotropic. Models with the extreme value
$q=-2$ are purely radial at large distances.

The augmented density (\ref{gammanirho2}) varies as $\psi^{4+q}$ 
for small $\psi$. Hence its expansion in powers of $\psi$ now 
ascends in integer steps from that initial term, and has the form
\begin{equation}
\label{newrhoexpand}
\rho(\psi,r)=\frac{3-\gamma}{4\pi}(1+r)^q \sum_{j = 0}^{\infty}
{\bar C}(\gamma,4+q+j,q)\psi^{4+q+j}, 
\end{equation}
for suitable coefficients ${\bar C}$. The distribution function is
\begin{equation}
\label{andistr2art}
{\cal F}(E,L,\gamma,q)=\frac{3-\gamma}{4\pi}\sum_{j = 0}^{\infty}
{\bar C}(\gamma,4+q+j,q){\bar F}_q^{4+q+j}(E,L).
\end{equation}
where ${\bar F}_q^p(E,L)$ is defined to be the component of the distribution 
function which corresponds to the component $(1+r)^q\psi^p$ of 
augmented density. Formulas for it, which again do not depend on the
parameter $\gamma$,  are given in Appendix \ref{app:formulas}.

\begin{figure}
\centering
\includegraphics[width=6cm,angle=0,clip=true]{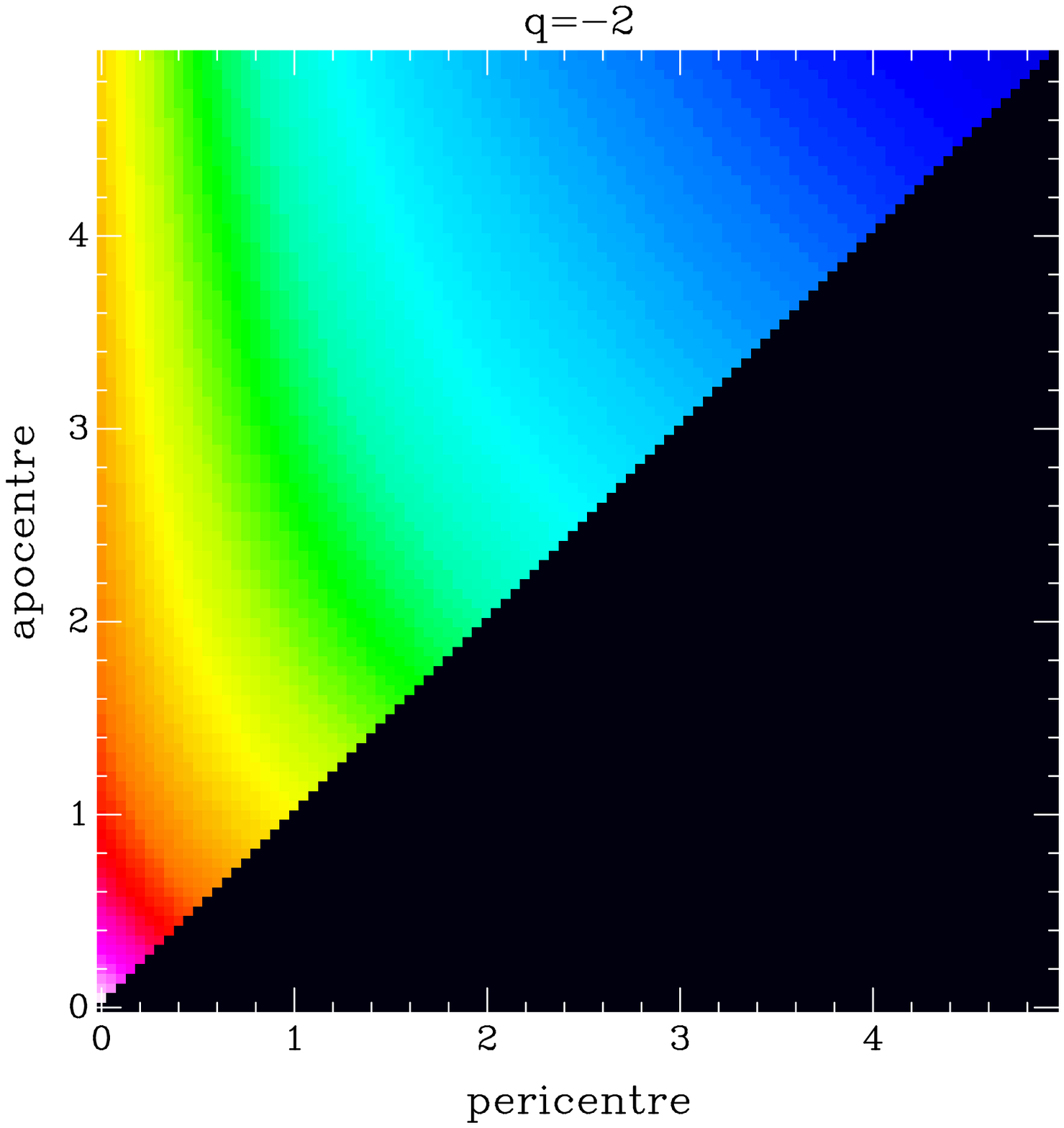}\hspace{1cm}
\includegraphics[width=6cm,angle=0,clip=true]{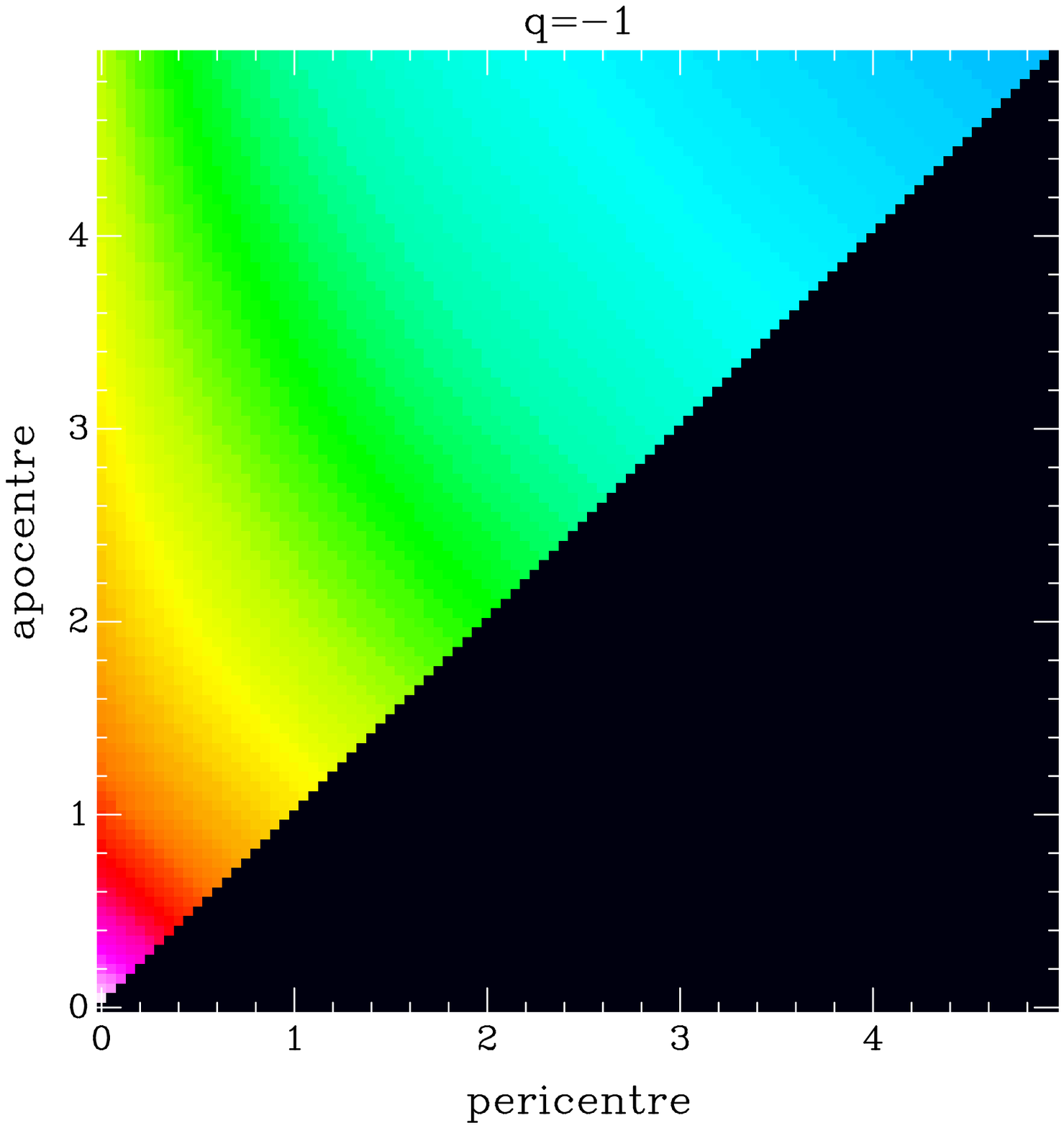}\vspace{1cm}
\includegraphics[width=6cm,angle=0,clip=true]{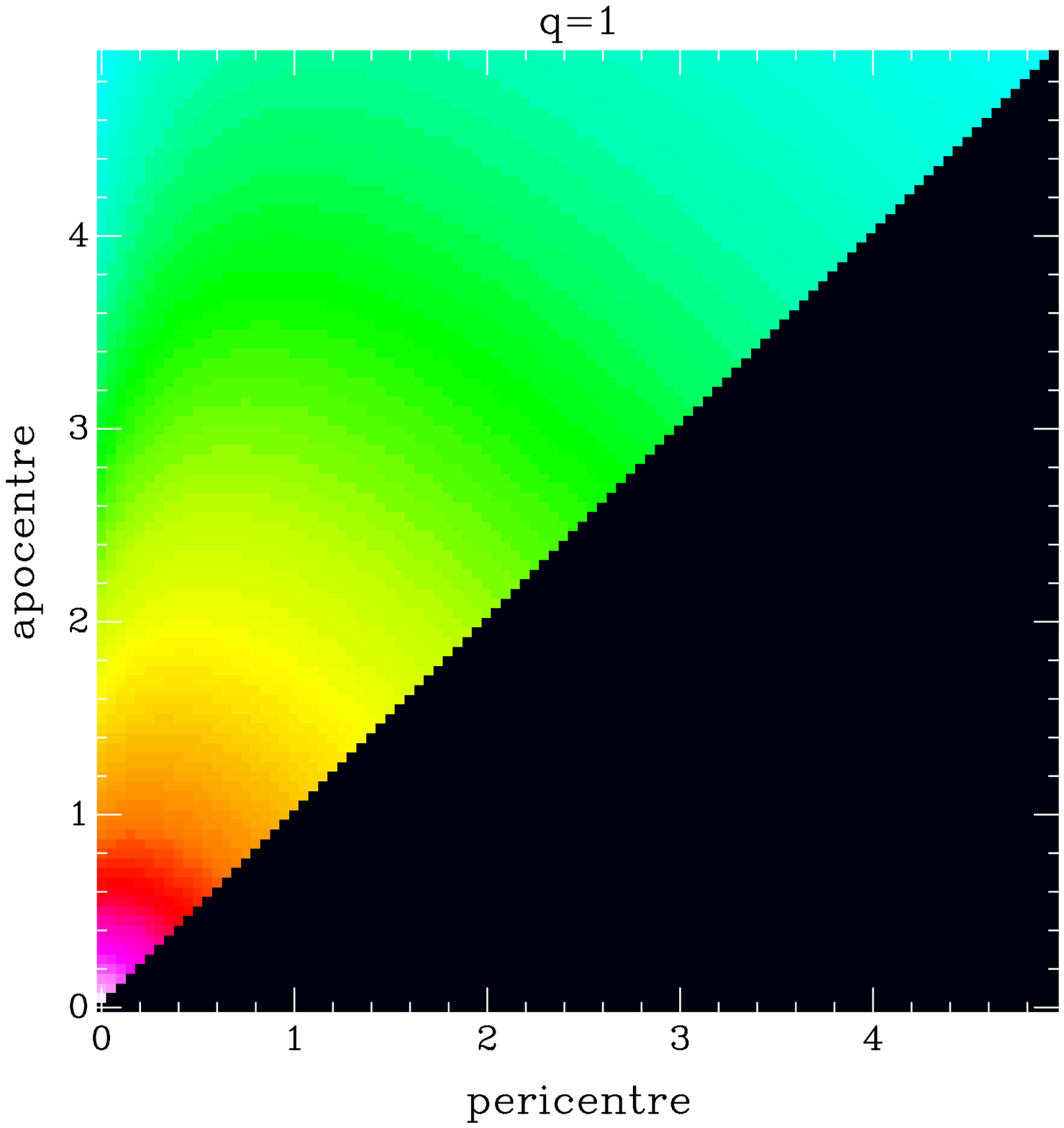}\hspace{1cm}
\includegraphics[width=6cm,angle=0,clip=true]{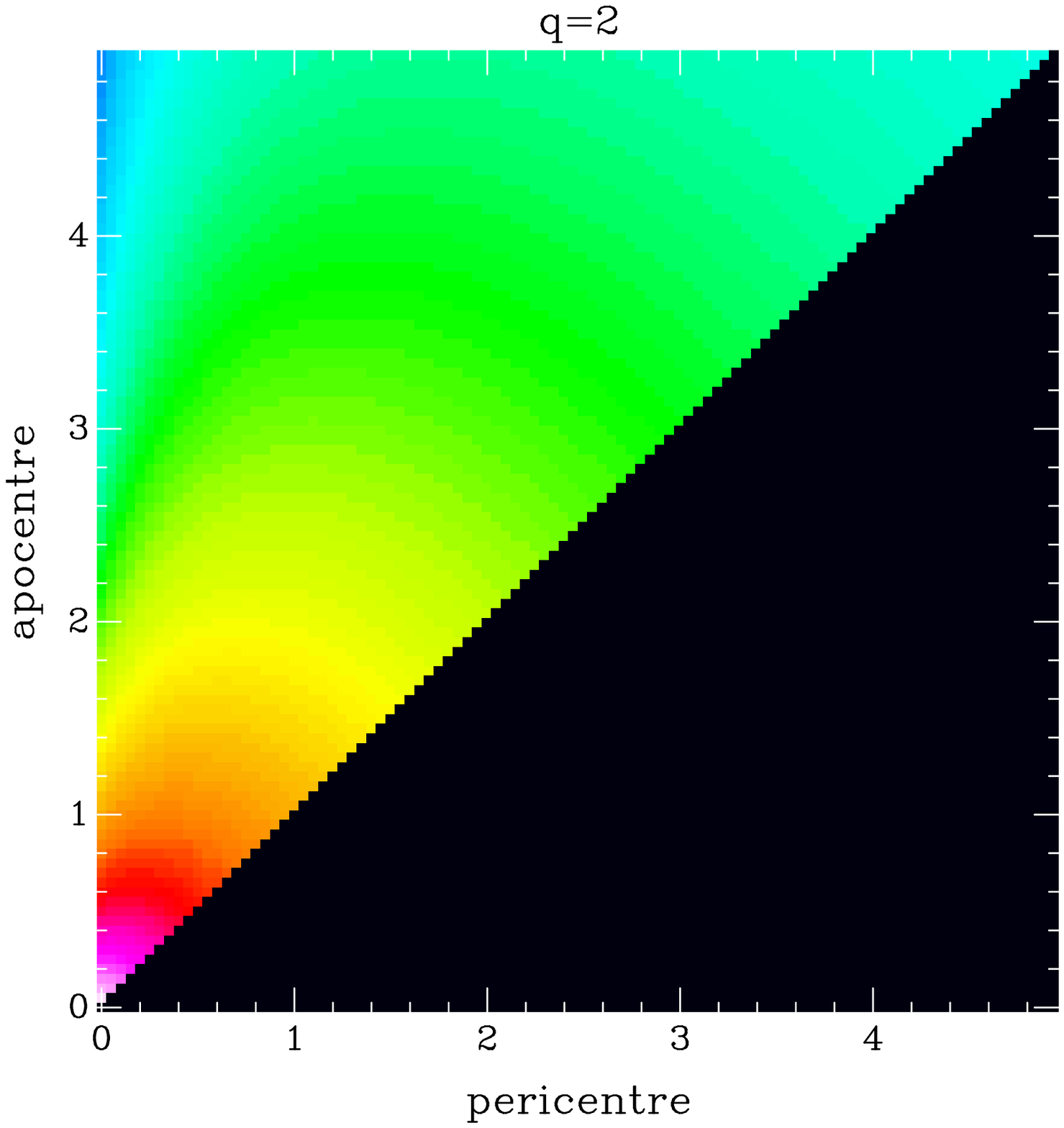}
\caption{Turning-point diagrams, coded as in Fig. \ref{figseruke1}, but
now for Hernquist models of the second family. The predominance of
near-radial orbits in the $q=-2$ model, and their decline as $q$
increases can be seen clearly seen. There is no $q=0$ plot because it
is the same as that in Fig. \ref{figseruke1}\label{fig:fig11}}
\end{figure}

The evaluation of the coefficients ${\bar C}$ is now more complicated.
The method of the previous section in which
we expand first in powers of $[1-(2-\gamma)\psi]$ and then in powers
of $\psi$, yields an expansion in integer powers of $\psi$, and so is
valid only when $(q+4)$ is a positive integer. For those cases
\begin{equation}
{\bar C}(\gamma,4+q+j,q) = (-1)^p(2-\gamma)^p
\sum_{k = 0}^{4+q}(-1)^k{4+q \choose k}
{\frac{k-\gamma}{2-\gamma} \choose 4+q+j}.
\end{equation}
As before, the closer $\gamma$ is to 2, the more
rapidly does the series (\ref{andistr2art}) converge.

More generally, one must expand the augmented density
(\ref{gammanirho2}) directly in powers of $\psi$ without any
intermediate expansion.  That leads to an infinite series in $\psi$
for $1-[1-(2-\gamma)\psi]^{1/(2-\gamma)}$ and the need to multiply the
$(4+q)$th power of that infinite series with another infinite series.
This procedure does simplify when $1/(2-\gamma)$ is a positive
integer. The simplest instances are the $\gamma=1$ Hernquist and the
$\gamma=\frac{3}{2}$ Dehnen models for which
\begin{equation}
\label{specialCbar}
{\bar C}(1,4+q+j,q)=1, \qquad {\bar C} \left(\frac{3}{2},4+q+j,q\right)=
\frac{1}{2^{j+1}} \sum_{k=0}^{j} \left(-\frac{1}{2}\right)^k
(j-k+1)(j-k+2) {4+q \choose k}.
\end{equation}

Although the infinite series expansions (\ref{andistr2art}) are
limited to the range $0 < \gamma < 2$ by the same convergence
conditions as in \S\ref{subsec:firstmodels}, that limitation
disappears when the sum in (\ref{newrhoexpand}) is finite. It is
finite for the most radial $q=-2$ case when $1/(\gamma-2)$ is a positive
integer, as for $\gamma=5/2$.  Equation (\ref{andistr2art}) then gives
exact distribution functions for strongly radial models with steep
cusps.

The elementary distribution functions are again simpler for some small
integer values of $q$.  Specifically, equation (\ref{barFsmallL})
gives
\begin{equation}
{\bar F}^p_q={\Gamma(p+1)E^p \over (2\pi E)^{3/2}}
  \left\{ {1 \over \Gamma(p-{1 \over 2})} +{2q \over \sqrt{\pi} \Gamma(p-1)}
 {L \over \sqrt{2E}} +{q(q-1)\over 2 \Gamma(p-{3\over 2})} {L^2 \over 2E}
\right\},
\end{equation}
for $q=0,1,2$, and equation (\ref{barFlargeL}) gives
\begin{equation}
{\bar F}^p_q={E^p \over \sqrt{\pi} (2\pi E)^{3/2}}
\left( {\sqrt{2E} \over L}\right)^{-2q-1}
        {\Gamma(p+1) \over \Gamma (p-1-q)}
        {}_2F_1\left(-q-{1\over 2},-q ; p-q-1;  -{2E \over L^2}\right),
\end{equation}
for $q=-1,-2$. Exact distribution functions of the forms ${\cal
F}(E,L,\gamma,q=1)=f_0(E) + Lf_1(E)$ and 
${\cal F}(E,L,\gamma,q=2)=f_0(E) + Lf_1(E)+L^2f_2(E)$ for $q=1$ and $q=2$
respectively can be found by the methods of An \& Evans
(2006a). The results, which are similar to those of
\S\ref{subsec:whenqisone}, are given in Appendix \ref{app:exactdfs}.

\subsubsection{Explicit distribution functions for Hernquist models}
\label{subsec:hernquist}

The fact that each ${\bar C}$ coefficient is 1 for Hernquist models
allows the summations needed for the full distribution function
(\ref{andistr2art}) to be performed explicitly for the simple ${\bar
F}^p_q$ functions listed above. Those for $q=-1$ and $q=-2$ can be
done using Euler's formula (15.3.1) of Abramowitz \& Stegun (1965) for
hypergeometric functions which gives
\begin{equation}
{}_2F_1\left(-q-{1\over 2},-q ; p-q-1;  -{2E \over L^2}\right) =
\frac{\Gamma (p-1-q)}{\Gamma(-q)\Gamma(p-1)} \int_0^1
t^{-q-1}(1-t)^{p-2}\left(1+\frac{2Et}{L^2}\right)^{q+\frac{1}{2}}dt.
\end{equation}
This integral converges for all positive $E$, regardless of the
magnitude of $L^2$. Using it, the series summation
$\sum_{j=0}^{\infty}(j+2)(j+1)x^j=2/(1-x)^3$ and a change of
integration variable to $y=Et$, we obtain the compact integrals
\begin{eqnarray}
\label{dftwointegrals}
{\cal F}(E,L,\gamma=1,q)&=&\left\{\begin{array}{lll}
\frac{1}{2\pi^3}\int_0^E\frac{dy}{(L^2+2y)^{1/2}}
\left[\frac{1}{(1-E+y)^3}-1\right], && q=-1, \\ 
\frac{1}{\pi^3}\int_0^E\frac{ydy}{(L^2+2y)^{3/2}(1-E+y)^3},
&& q=-2, \end{array}\right.
\end{eqnarray}
for the two distribution functions. They are evidently
non-negative everywhere, and so physically acceptable.  The integrals
(\ref{dftwointegrals}) can be evaluated by elementary methods as
\begin{equation}
{\cal F}(E,L,1,-1)=\frac{1}{2\pi^3} \left\{
\frac{3}{2\chi^2}\left[I_1+\sqrt{L^2+2E}-\frac{L}{(1-E)}\right]
+\frac{1}{2\chi}\left[\sqrt{L^2+2E}-\frac{L}{(1-E)^2}\right]
-\sqrt{L^2+2E} + L \right\},
\end{equation}
and
\begin{equation}
{\cal F}(E,L,1,-2)=\frac{2(1-E+2L^2){\cal F}(E,L,1,-1)}{\pi}
+\frac{1}{\pi^3\chi}\left[\frac{L^2}{\sqrt{L^2+2E}} -\frac{L}{(1-E)^2}
+(1-E+2L^2)(\sqrt{L^2+2E} - L)\right],
\end{equation}
where 
\begin{eqnarray}
\chi=2(1-E)-L^2, \qquad
I_1=\int_0^E\frac{dy}{(L^2+2y)^{1/2} (1-E+y)}
&=&\left\{\begin{array}{lll} \frac{2}{\sqrt{\chi}}\left[
\arctan \frac{\sqrt{L^2+2E}}{\sqrt{\chi}} - \arctan \frac{L}{\sqrt{\chi}}
\right], && \chi > 0, \\
\frac{1}{\sqrt{-\chi}} \ln \left[ \frac
{(\sqrt{L^2+2E}-\sqrt{-\chi})(L+\sqrt{-\chi})}
{(\sqrt{L^2+2E}+\sqrt{-\chi})(L-\sqrt{-\chi})} \right], && \chi < 0.
\end{array}\right.
\end{eqnarray}
Both distribution functions vary smoothly across the curve $2(1-E)=L^2$ where
\begin{eqnarray}
{\cal F}(E,\sqrt{2(1-E)},1,-1) &=& \frac{1}{\pi^3\sqrt{2}}\left\{
\frac{1}{5}\left[\frac{1}{(1-E)^{5/2}}-1\right]
+\sqrt{1-E} -1 \right\}, \\
{\cal F}(E,\sqrt{2(1-E)},1,-2) &=& \frac{1}{\pi^3\sqrt{2}}\left\{
\frac{1}{5}\left[\frac{1}{(1-E)^{5/2}}-1\right]
+\frac{(1-E)}{7}\left[1-\frac{1}{(1-E)^{7/2}}\right] \right\}.
\end{eqnarray}

We display distribution functions for these models for different $q$
values in Fig.\ref{fig:fig11}. They now show increasing radiality with
decreasing $q$, and can be compared with those of Fig.\ref{figseruke1}.
The $q=0$ case, missing from Fig.\ref{fig:fig11}, is the same as
that shown in Fig.\ref{figseruke1}.

\begin{figure}
\centering
\includegraphics[width=5cm,angle=0,clip=true]{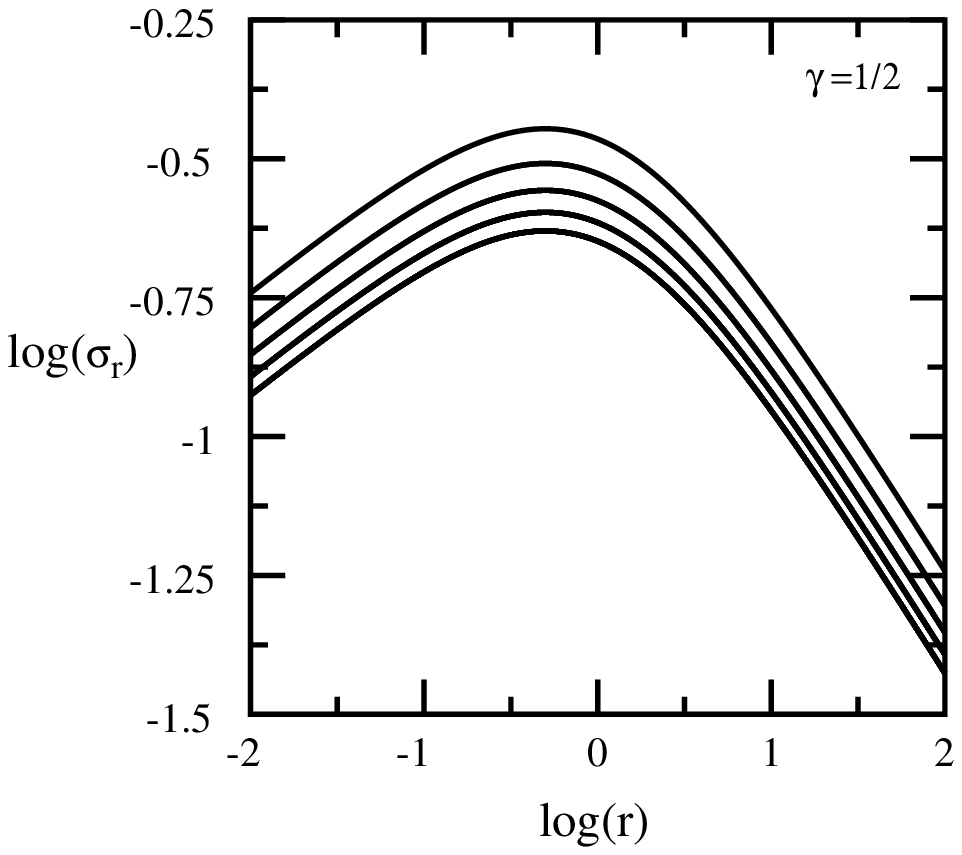}\hspace{1cm}
\includegraphics[width=5cm,angle=0,clip=true]{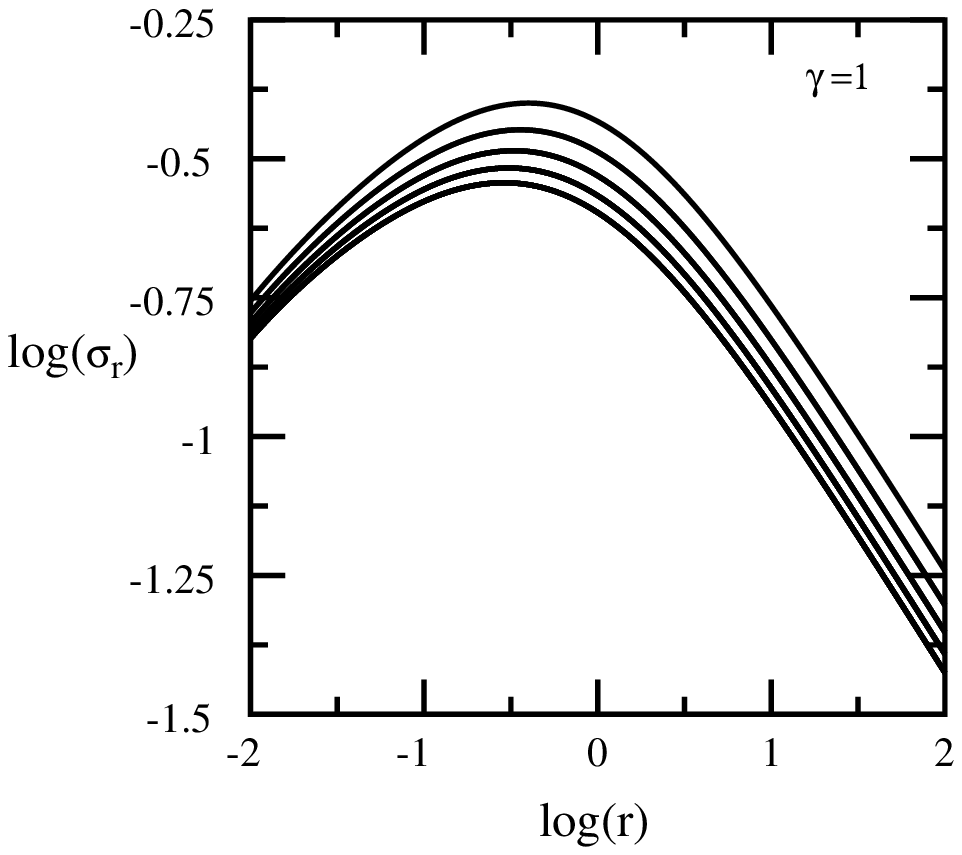}\hspace{1cm}
\includegraphics[width=5cm,angle=0,clip=true]{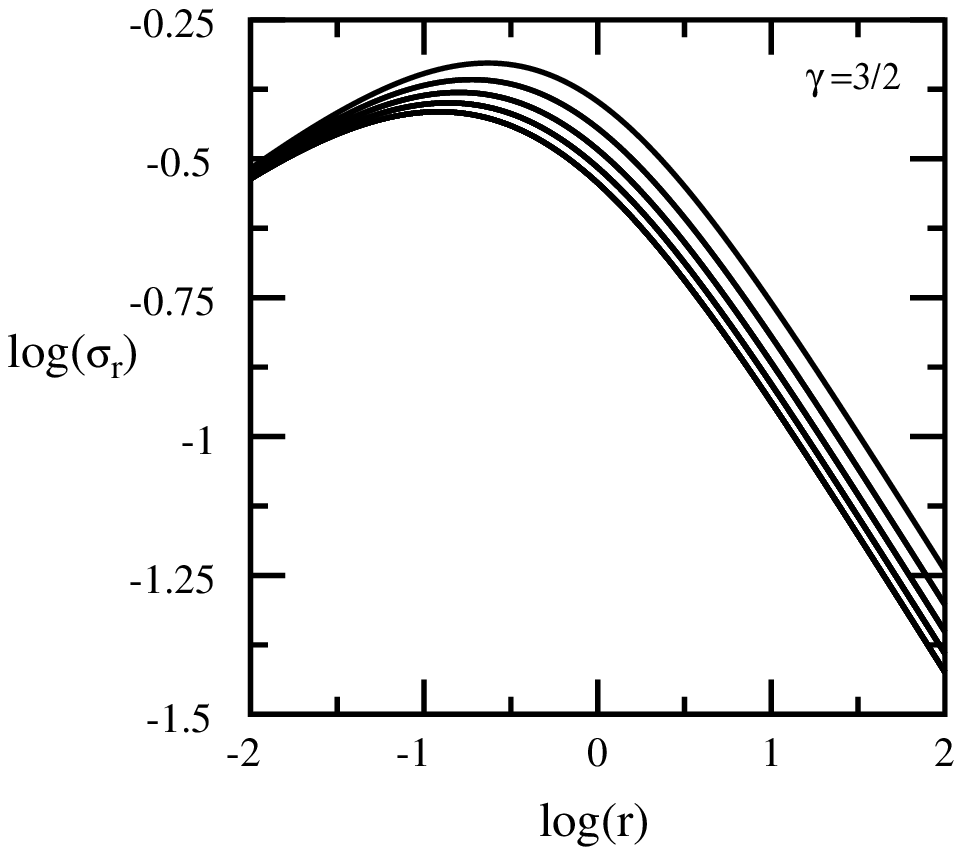}
\caption{The radial velocity dispersion profiles for three $\gamma$
values for the second family, and for $q=-2, -1, 0, 1, 2$.
The highest curve is for the lowest value of $q$ and the others
follow in sequence.\label{fig:fig12}}
\end{figure}

\subsection{Second-order moments}

The velocity dispersions are now related by 
\begin{equation}
\sigma_t^2(r,q)=\left[1+\frac{qr}{2(1+r)}\right]\sigma_r^2(r,q).
\label{dispttwo}
\end{equation}
and the radial velocity dispersions are
\begin{equation}
\sigma_r^2(r,q)=r^{\gamma}(1+r)^{4+q-\gamma}
B_{1/(1+r)} (5+q , 2-2\gamma).
\label{disprtwo}
\end{equation}
Equation (\ref{disprtwo}), when approximated for large $r$, shows that
the radial velocity dispersion $\sigma_r(r,q) \approx 1/\sqrt{(5+q)r}$
for large $r$ and increases as $q$ becomes more negative. The radial
velocity dispersion near the centre behaves differently according to
whether $\gamma$ is greater or less than $1$. It has the form
$\sigma_r(r,q) \approx \sqrt{r^{2-\gamma}/[2(\gamma-1)]}$, and
$\sigma_r(r,q) \approx \sqrt{(-r\ln r)}$ in the limit $\gamma = 1$,
which is independent of $q$ when $\gamma \ge 1$. It depends on $q$ 
for weaker cusps with $\gamma < 1$. Then $\sigma_r(r,q) \approx 
\sqrt{r^{\gamma}B(5+q,2-2\gamma)}$, and is larger for the more radial
models even near the centre as Fig. \ref{fig:fig12} confirms.
\begin{figure}
\begin{center}
\includegraphics[width=5cm,angle=0,clip=true]{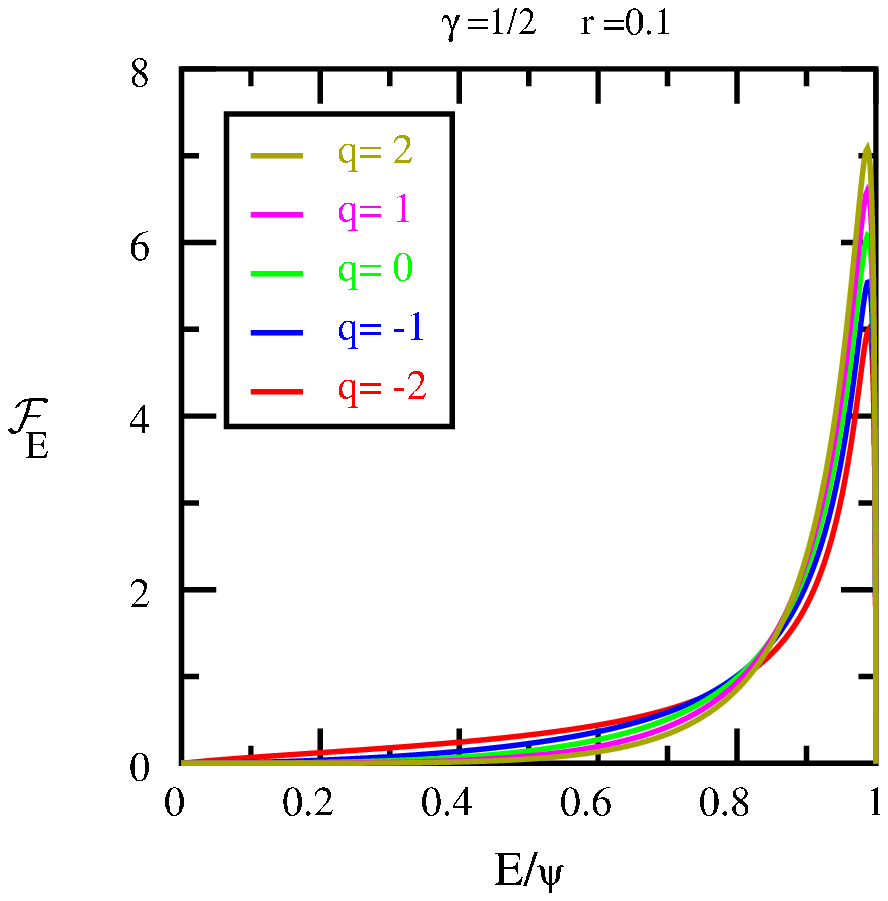}\hspace{1cm}
\includegraphics[width=5cm,angle=0,clip=true]{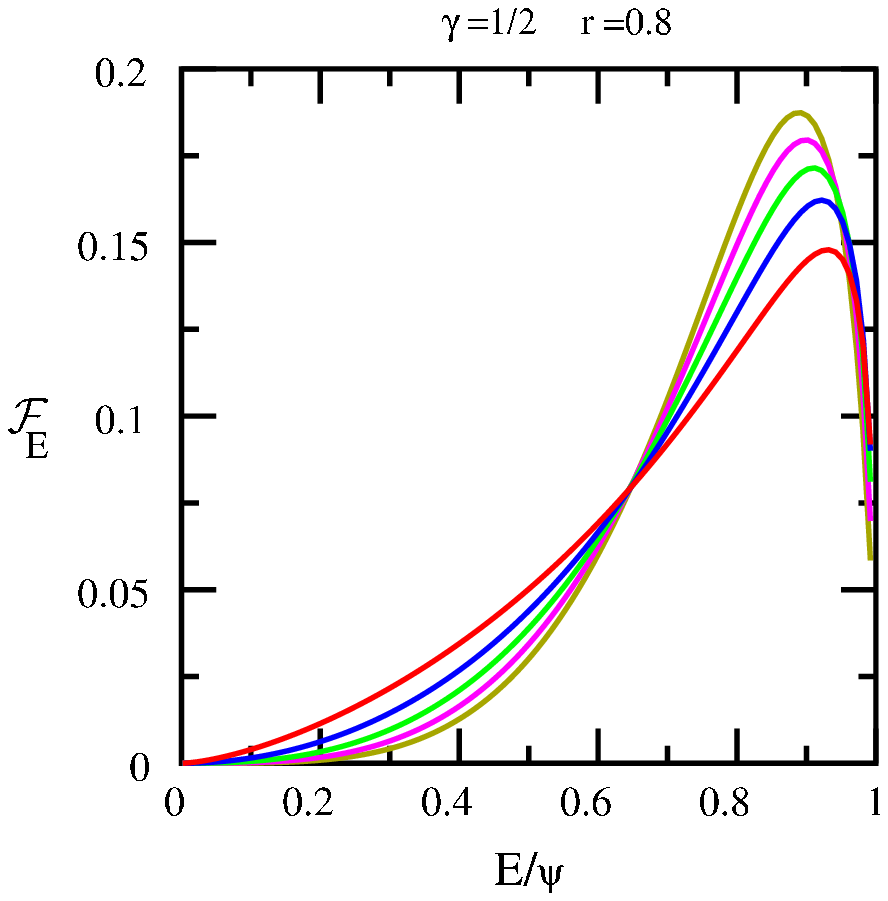}\hspace{1cm}
\includegraphics[width=5cm,angle=0,clip=true]{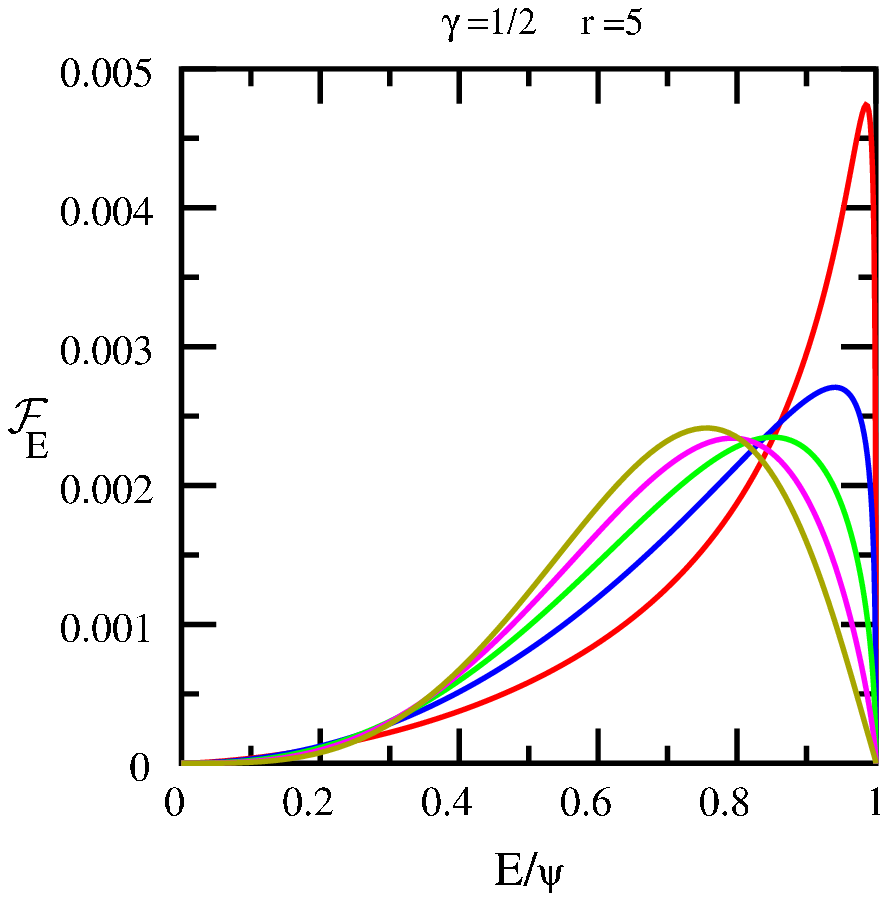}\vspace{1cm}
\includegraphics[width=5cm,angle=0,clip=true]{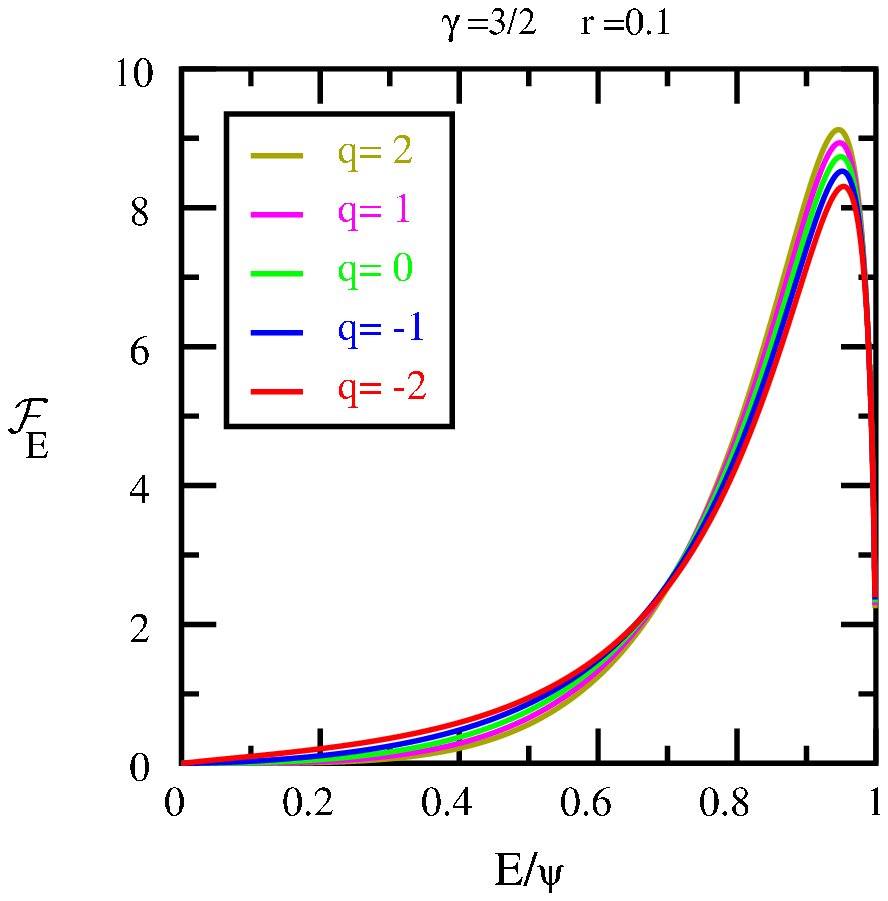}\hspace{1cm}
\includegraphics[width=5cm,angle=0,clip=true]{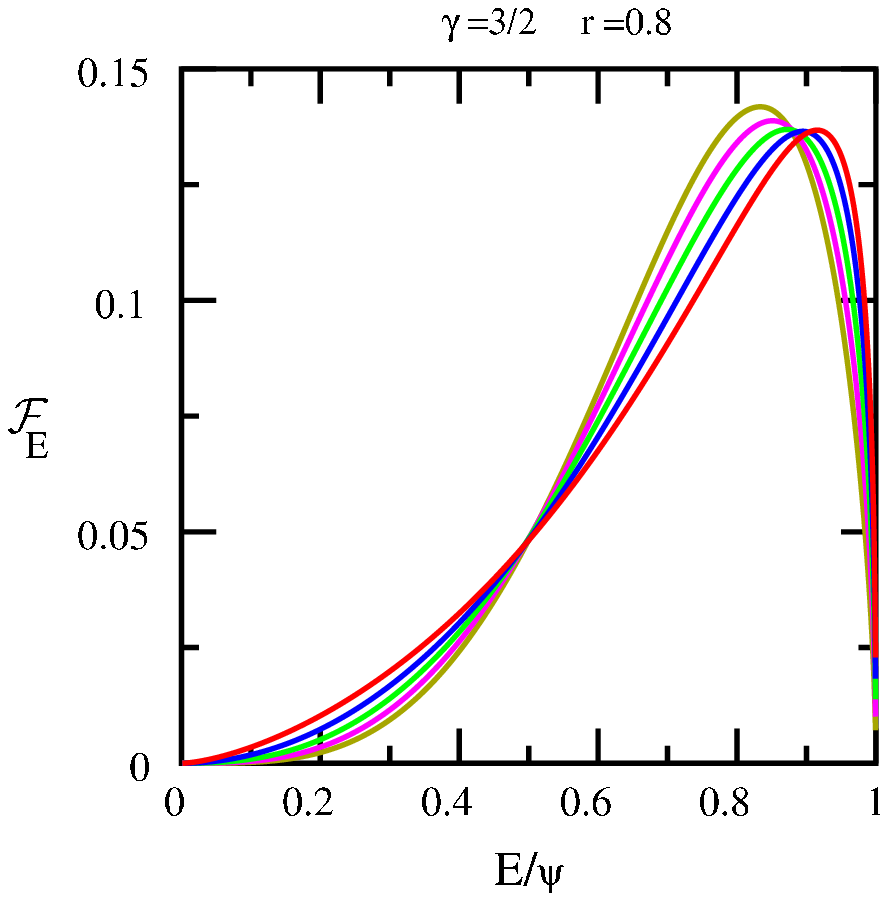}\hspace{1cm}
\includegraphics[width=5cm,angle=0,clip=true]{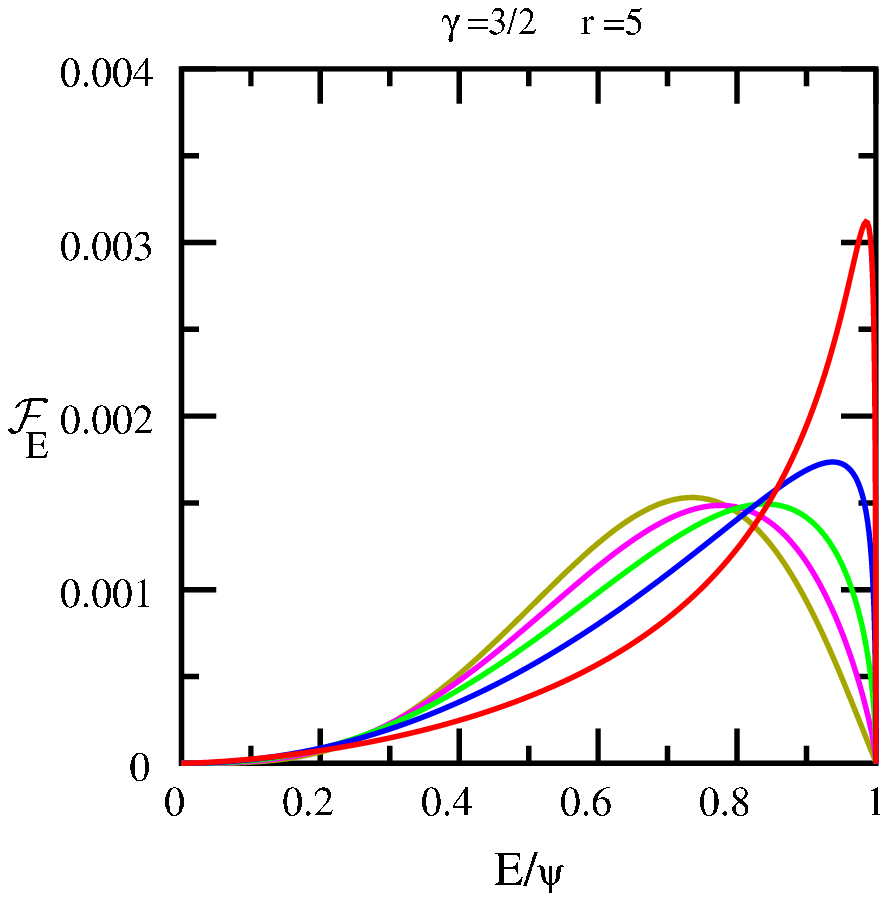}\hspace{1cm}
\caption{Energy densities, as in Fig. \ref{fig:fig4},
but for now for models of the second family. \label{fig:fig13}}
\end{center}
\end{figure}

\subsection{Other properties}

We define energy and transverse velocity densities ${\bar
F}_{E,q}^p(\psi,r,E)$ and ${\bar F}_{v_t,q}^p(\psi,r,v_t)$
corresponding to elementary distribution functions ${\bar F}_q^p(E,L)$
in the same way as we defined $F_{E,q}^p(\psi,r,E)$ and
$F_{v_t,q}^p(\psi,r,v_t)$ in \S\ref{subsec:energydist} and
\ref{subsec:transveldist}.  The energy and transverse velocity
densities are then given by sums of the same form as those of
equation (\ref{energysum}) and (\ref{transvelsum}) but now with barred
components.

Energy densities for models of the second family shown in
Fig.~\ref{fig:fig13} show the same relative differences between more
tangential (now larger $q$) and more radial (smaller $q$) as do the
models of the first family shown in Fig.~\ref{fig:fig4}.  The
differences are now small at small $r$ because models of the second
family are isotropic at their centres.  Differences appear as the
radius increases. The most radial $q=-2$ model remains strongly peaked
even at high $E/\psi$, while the peaks of the other models move to
lower relative energies with increasing radius and decreasing
radiality, as in Dejonghe's Fig. 1.  All ${\cal F}_{E}$ curves now
drop to zero at the right limit $E/\psi=1$ because the distribution
functions are not singular as $L \to 0$.  Our ${\cal F}_{E}$ are again
more concentrated towards high energies than those of Dejonghe (1987)
because our densities are more centrally concentrated than his.

\begin{figure}
\begin{center}
\includegraphics[width=5cm,angle=0,clip=true]{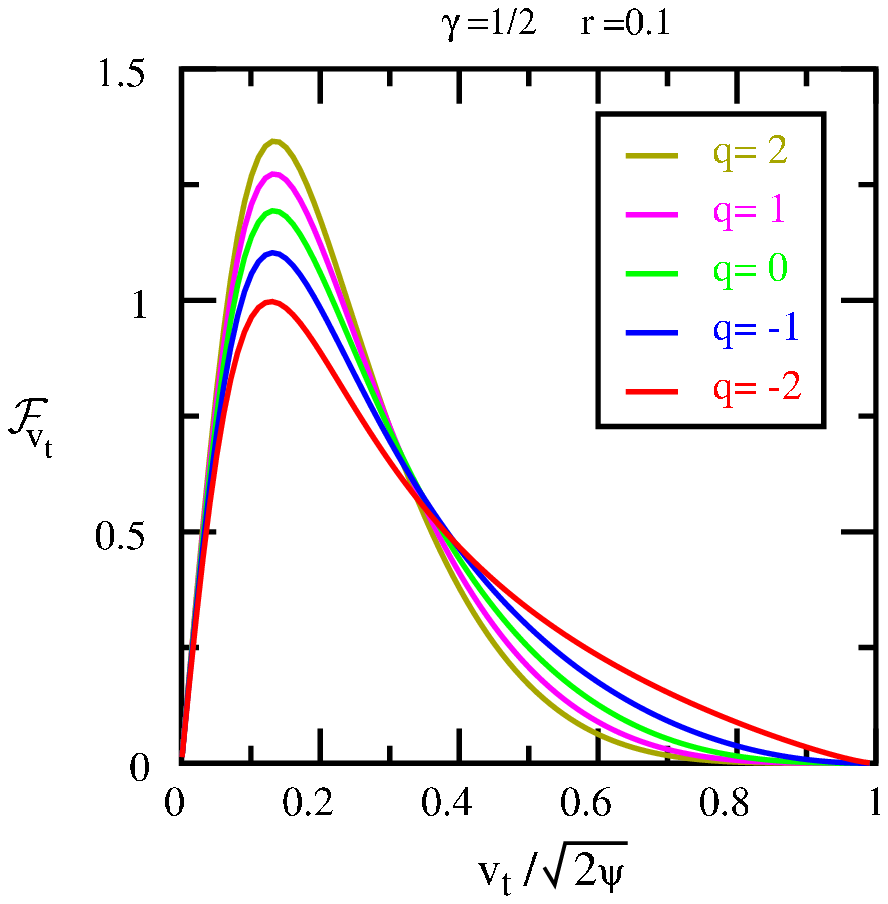}\hspace{1cm}
\includegraphics[width=5cm,angle=0,clip=true]{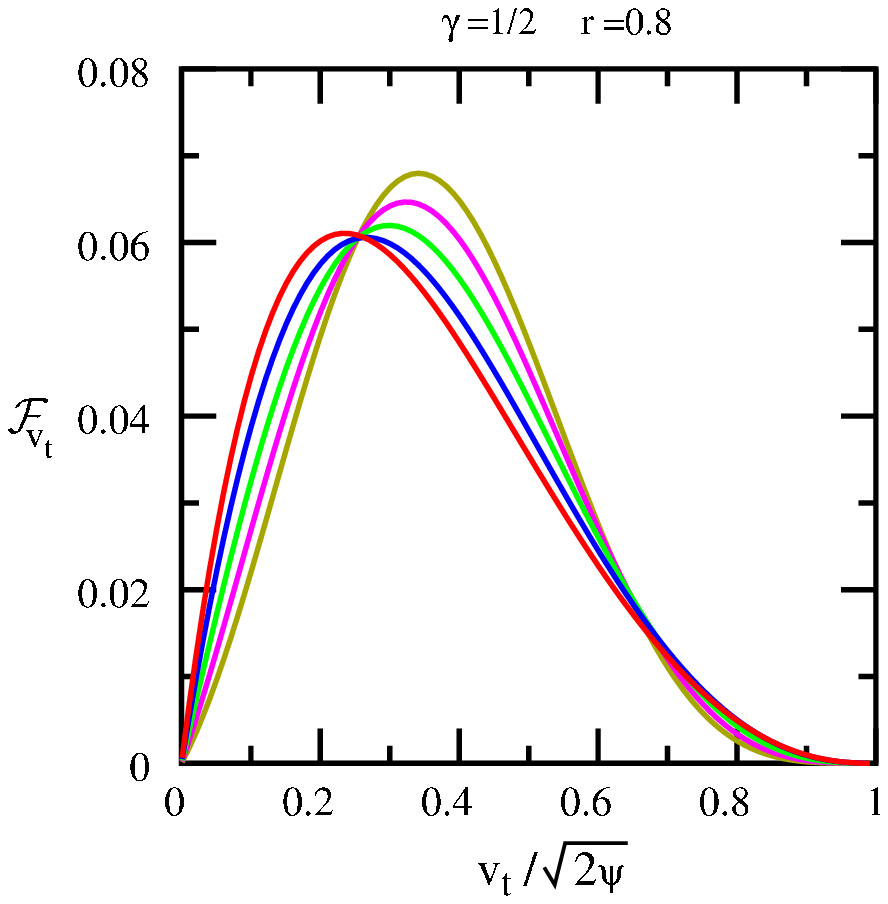}\hspace{1cm}
\includegraphics[width=5cm,angle=0,clip=true]{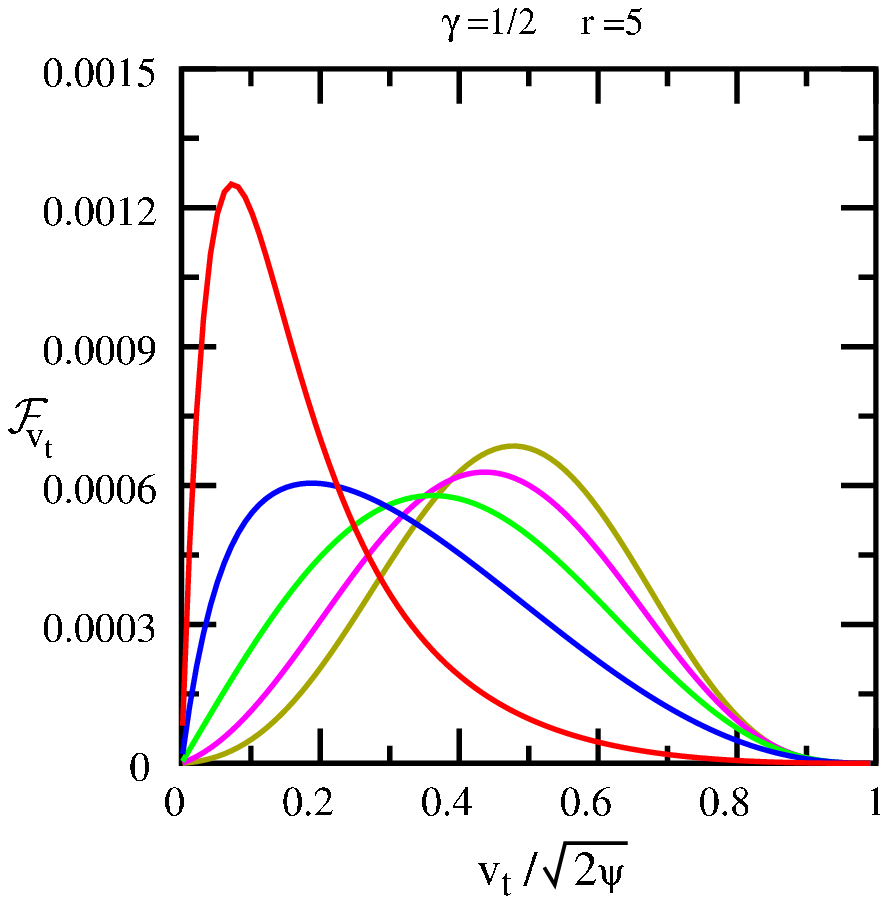}\vspace{1cm}
\includegraphics[width=5cm,angle=0,clip=true]{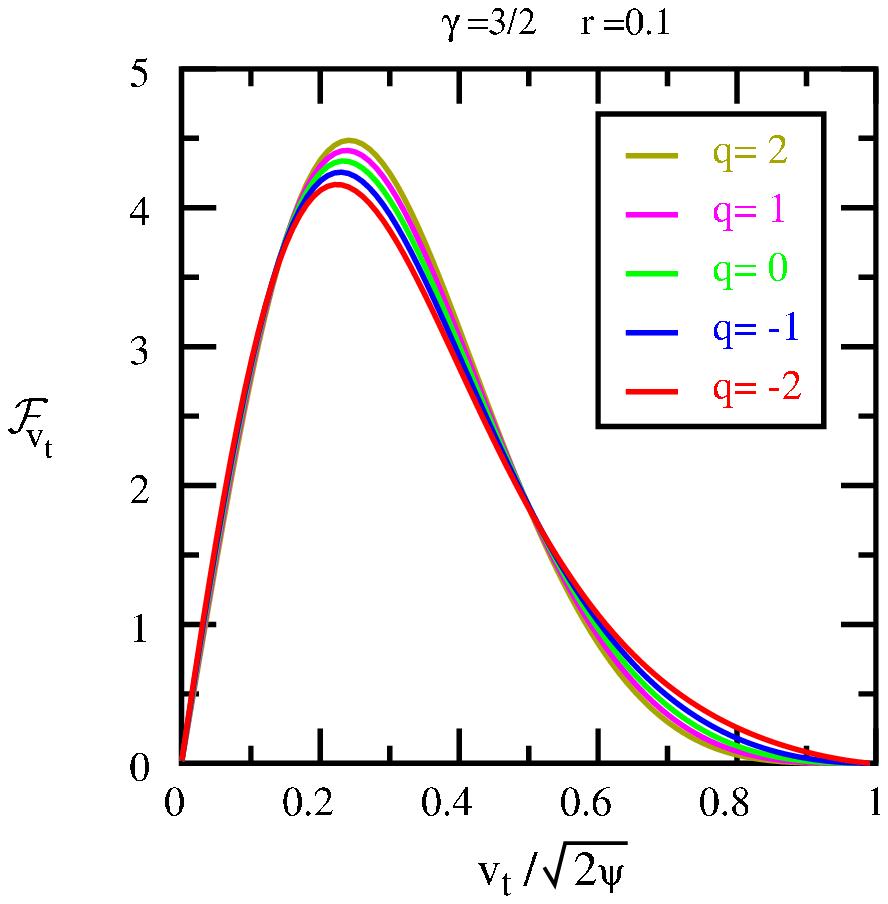}\hspace{1cm}
\includegraphics[width=5cm,angle=0,clip=true]{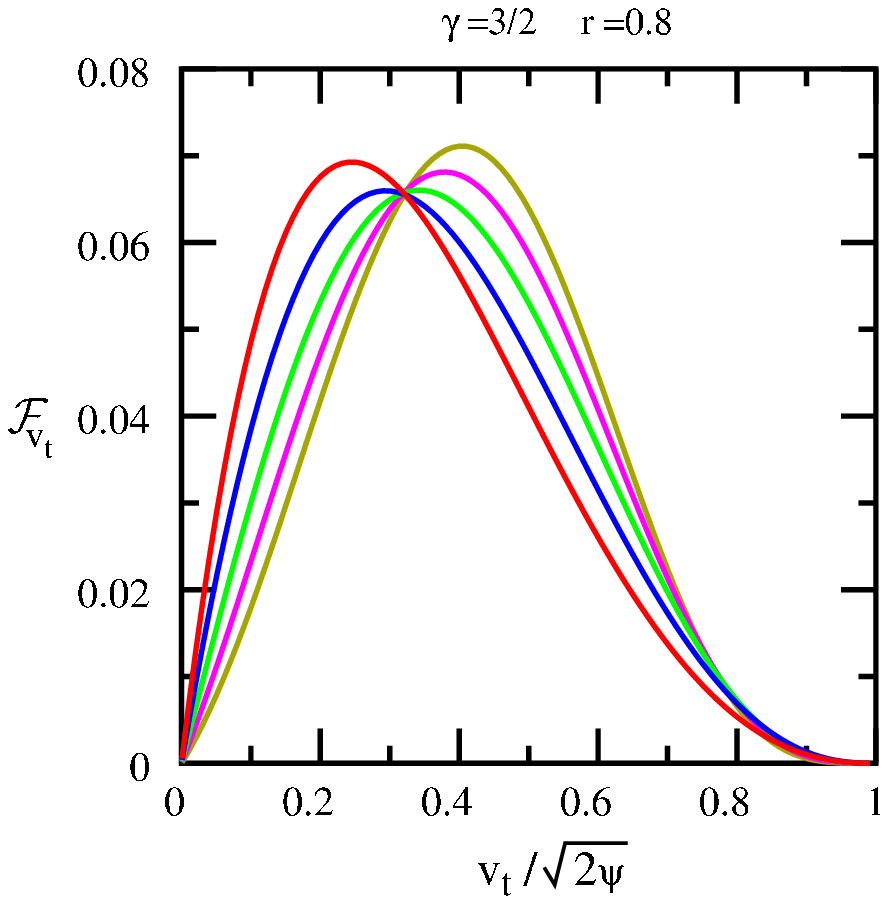}\hspace{1cm}
\includegraphics[width=5cm,angle=0,clip=true]{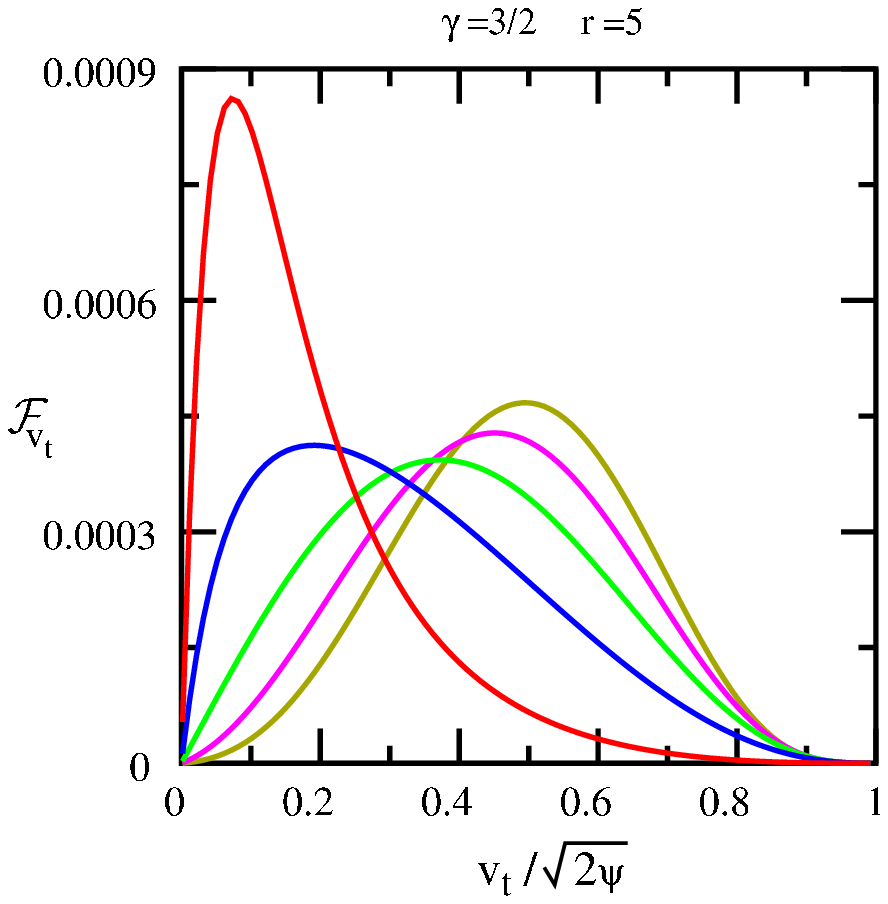}\hspace{1cm}
\caption{The transverse velocity densities, as in Fig. \ref{fig:fig5},
but for now for models of the second family. \label{fig:fig14}}
\end{center}
\end{figure}

The transverse velocity densities in Fig.\ref{fig:fig14} are also
similar at small radii. This distribution remains strongly peaked at
low relative $v_t$ for the most radial $q=-2$ model at large radii,
while the peaks of the other distributions become increasingly central
with increasing radius and increasing tangentiality. Dejonghe's
Fig. 4 shows similar behaviour at large radii.

\begin{figure}
\begin{center}
\includegraphics[width=5cm,angle=0,clip=true]{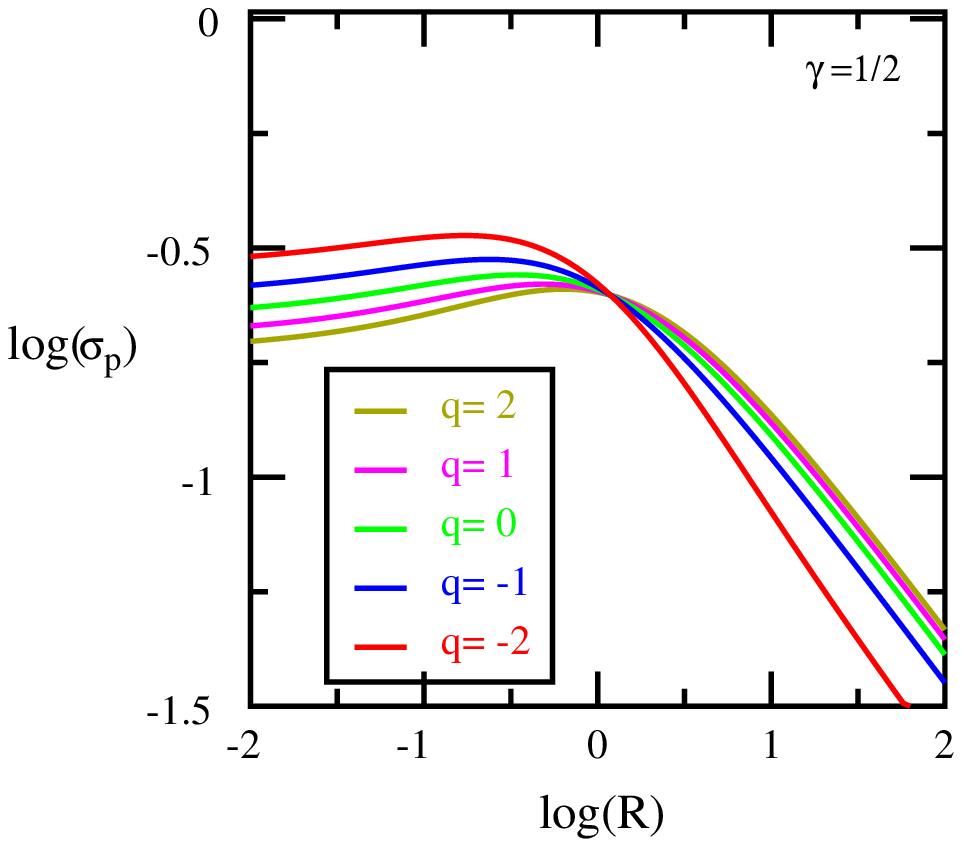}\hspace{1cm}
\includegraphics[width=5cm,angle=0,clip=true]{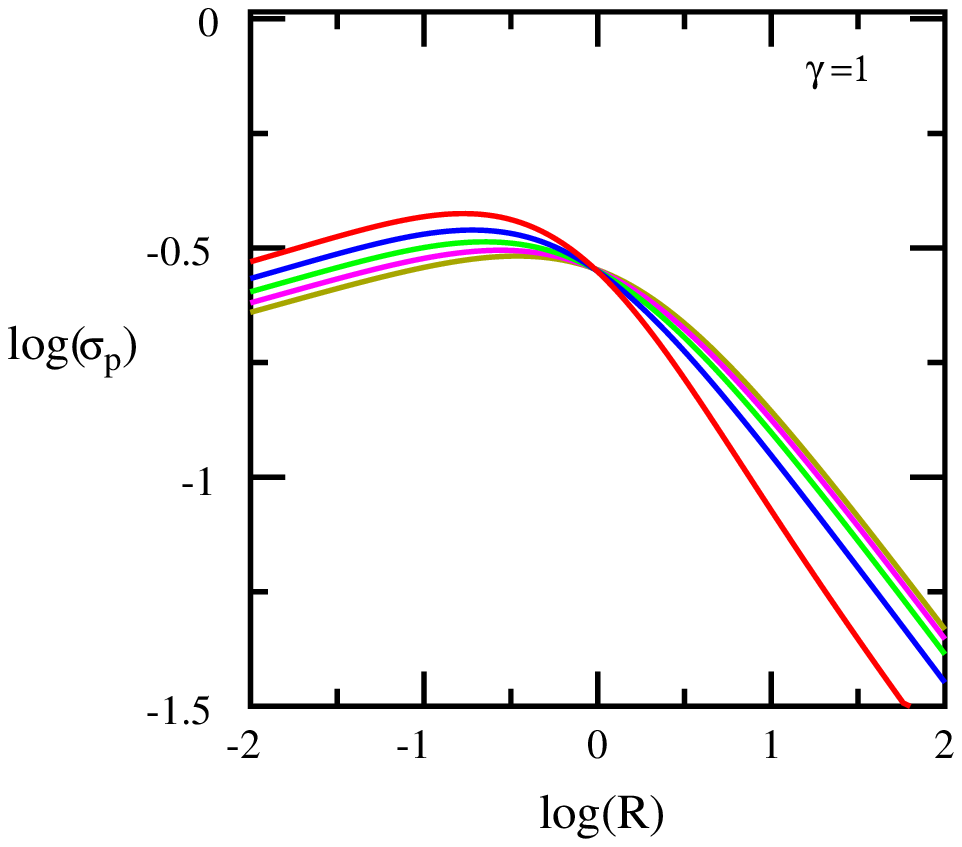}\hspace{1cm}
\includegraphics[width=5cm,angle=0,clip=true]{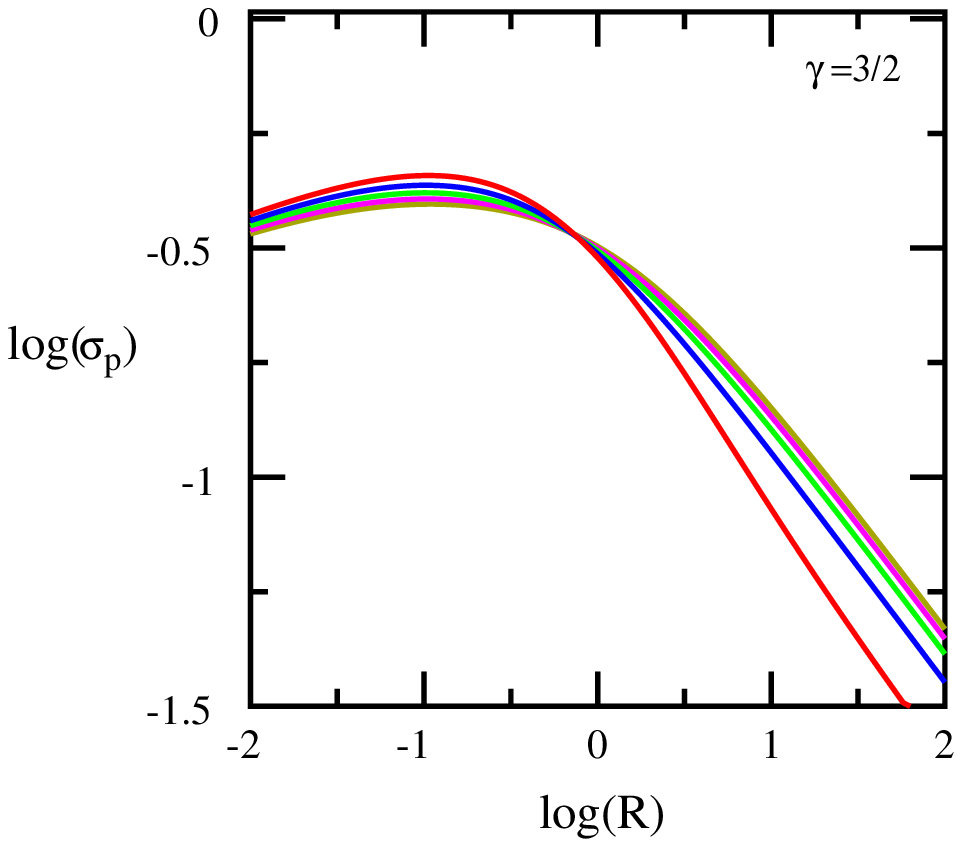}
\caption{The projected velocity dispersion profiles for the same models
as in Fig. \ref{fig:fig12}.
\label{fig:fig15}}
\end{center}
\end{figure}

Projected velocity dispersions for the second family are shown in
Fig.~\ref{fig:fig15}. All now peak at an intermediate projected radius
$R$. Unlike the unprojected radial velocity dispersions shown in
Fig.~\ref{fig:fig12}, the projected ones become larger near the centre
and smaller in the outer regions as the radiality increases.

\begin{figure}
\begin{center}
\includegraphics[width=5cm,angle=0,clip=true]{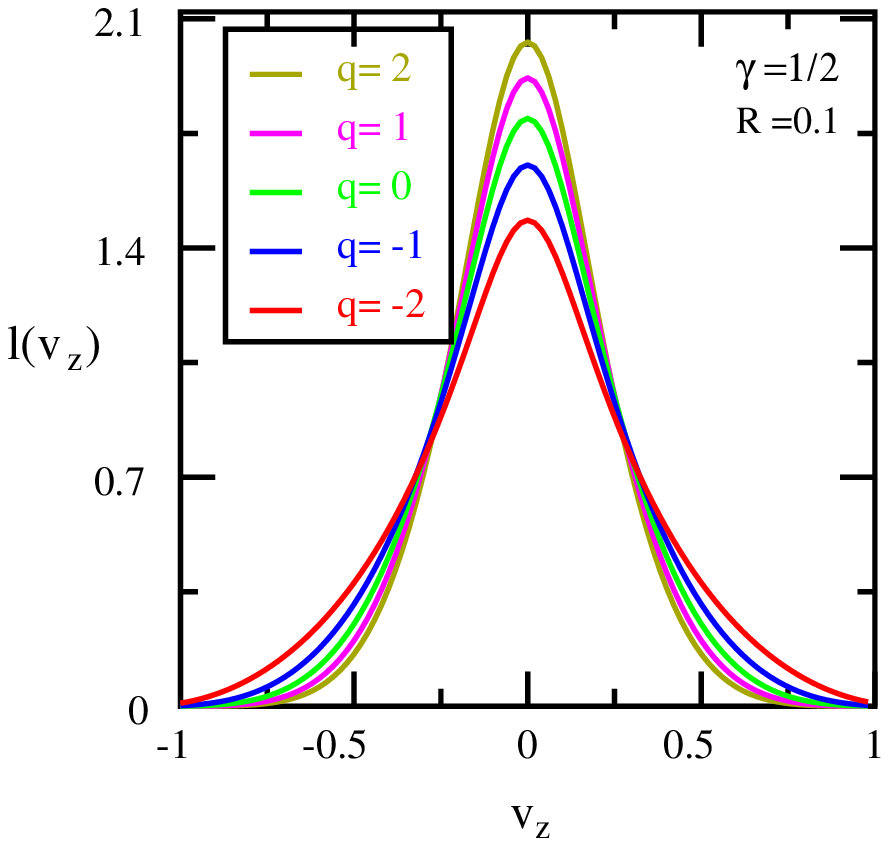}\hspace{1cm}
\includegraphics[width=5cm,angle=0,clip=true]{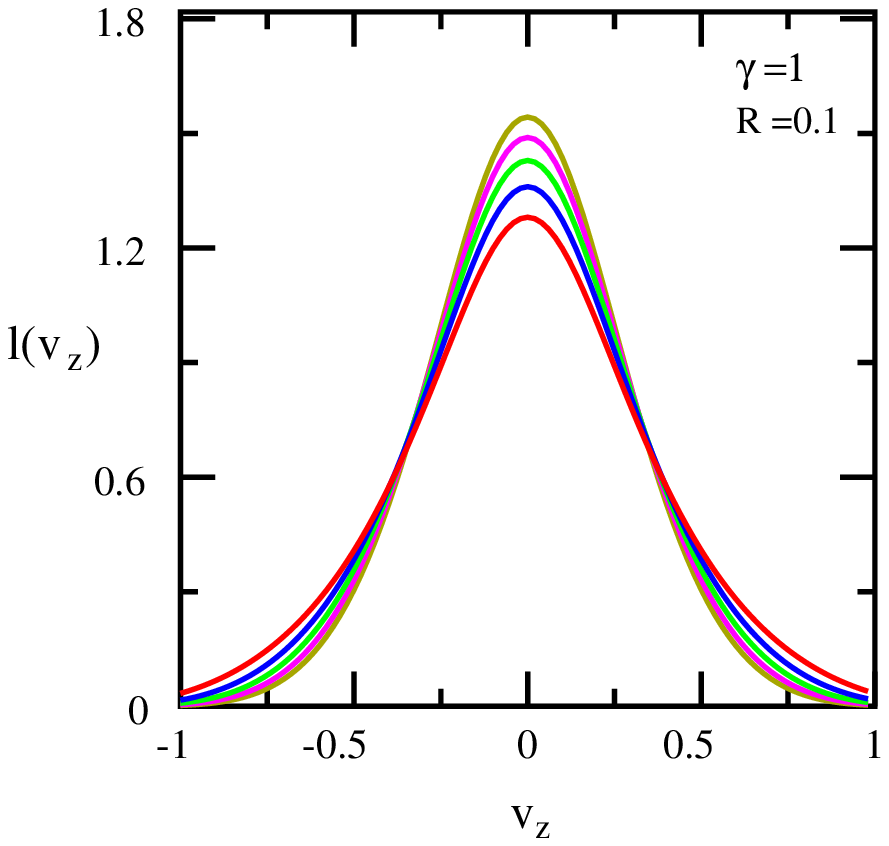}\hspace{1cm}
\includegraphics[width=5cm,angle=0,clip=true]{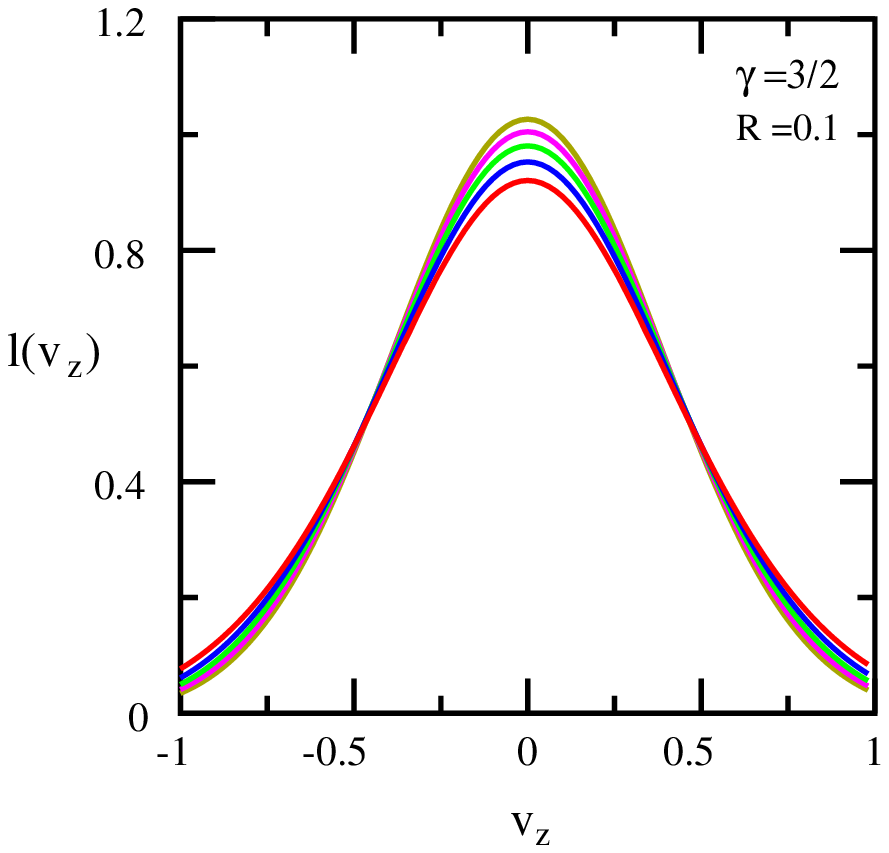}\vspace{1cm}
\includegraphics[width=5cm,angle=0,clip=true]{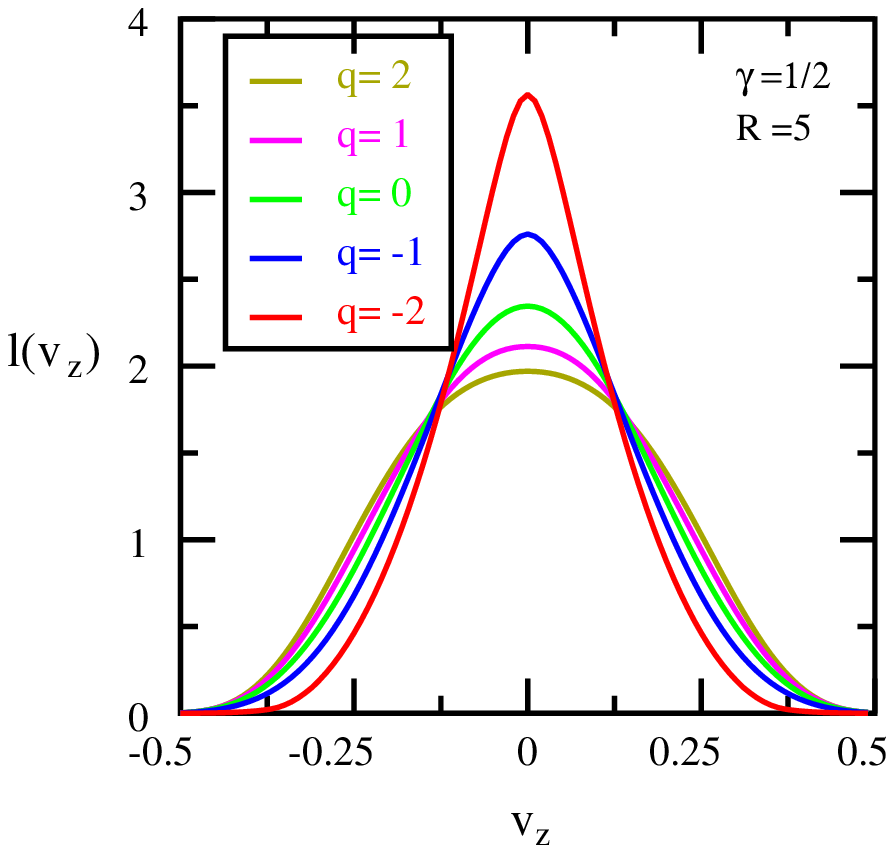}\hspace{1cm}
\includegraphics[width=5cm,angle=0,clip=true]{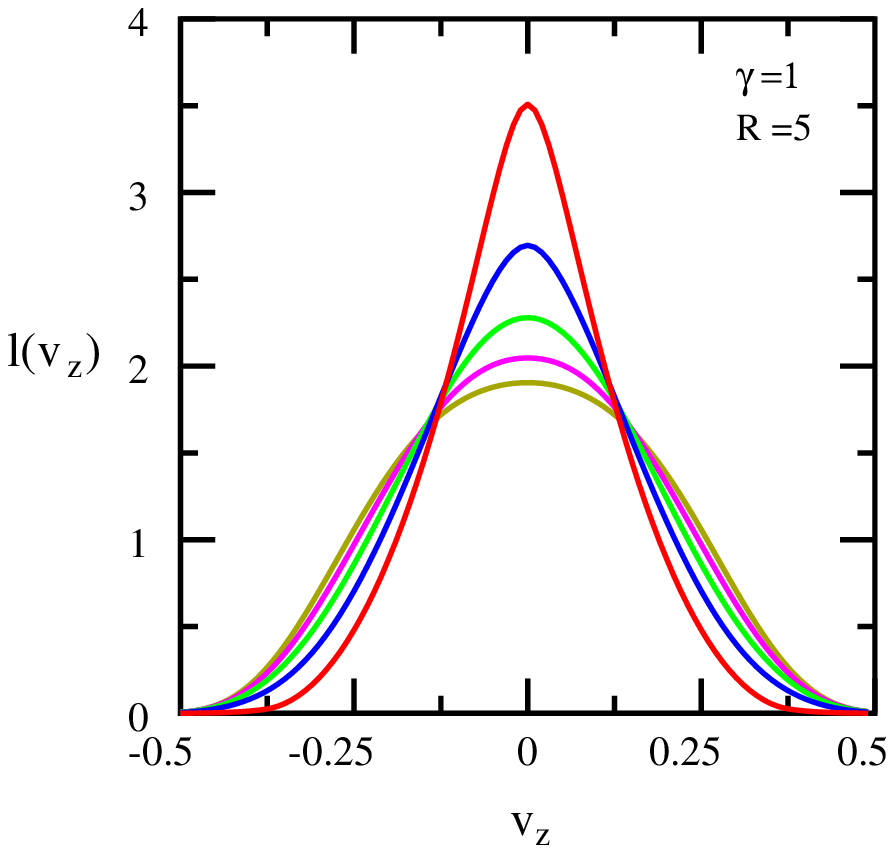}\hspace{1cm}
\includegraphics[width=5cm,angle=0,clip=true]{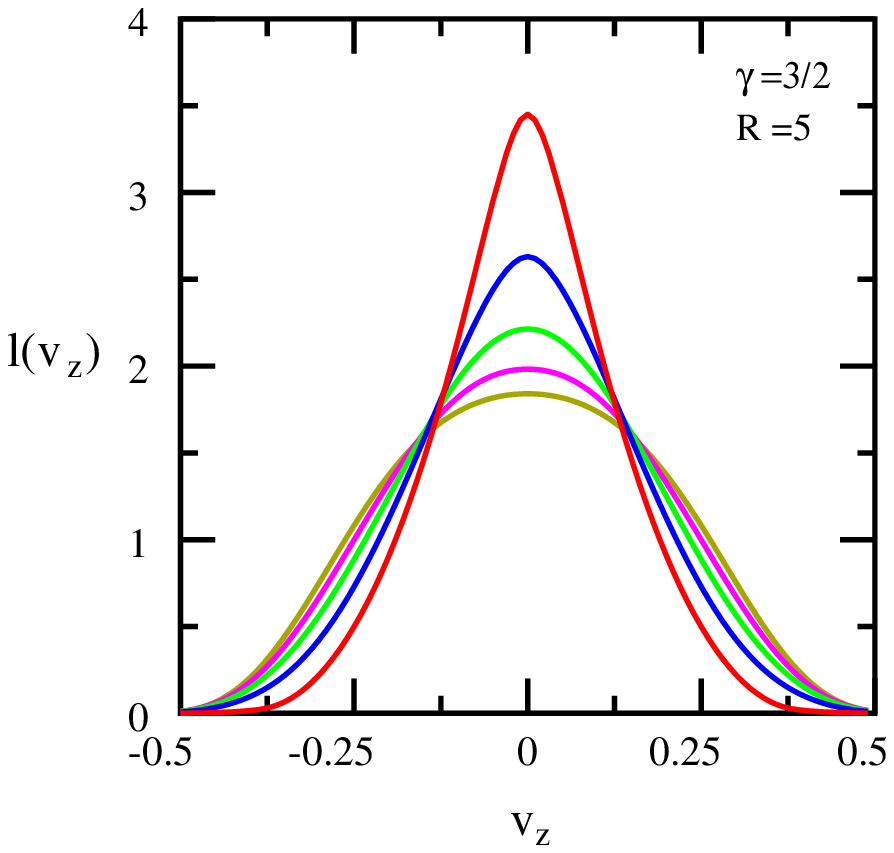}\hspace{1cm}
\caption{The normalised line-of-sight velocity profiles for models of the
second family for the same $R$ and $\gamma$ values as Fig.~\ref{fig:losvdone}.
\label{fig:losvdtwo}}
\end{center}
\end{figure}

Fig.~\ref{fig:losvdtwo} shows normalised line-of-sight velocity profiles for
models of the second family for three values of $\gamma$ and at
the same radii as those of Fig.~\ref{fig:losvdone}. They too vary more
at small radii with the central density slope $\gamma$ than with the orbital
composition. The profiles of the more tangential models are now more
noticeably narrower at small radii and broader at large radii than
the models of the first family. Now that the distribution functions
are isotropic at their centres, their velocity profiles peak smoothly
at $v_z=0$.

\subsection{Models with all radial orbits}
\label{subsec:radial models}

Models with all radial orbits have distribution functions of the form
\begin{equation}
{\cal F}(E,L,\gamma)=f(E)\delta(L^2).
\end{equation}
With this form, equation (\ref{rhoani}) becomes the following
Abel integral equation for $f(E)$ \citep{doug}:
\begin{equation}
\label{radialIE}
r^2\rho=2\pi \int_0^{\psi} \frac{f(E)dE}{\sqrt{2(\psi-E)}}.
\end{equation}
Physical solutions are possible only if $\lim_{r \to 0}r^2\rho >0$,
because (\ref{radialIE}) gives $r^2\rho$ as a weighted integral of
$f(E)$ over the whole physical range of $E$, and so a zero value would
imply that there are unphysical negative values of $f(E)$
\citep{doug}.  The solution of equation (\ref{radialIE}) is obtained
using the same $t$-substitution as in Baes et al. (2005) \S 5.2. It is
positive everywhere, and hence always physical, because it is an
integral with a positive integrand.  It can be expressed in
hypergeometric functions as
\begin{equation}
f(E)=\frac{3-\gamma}{4\pi^3}\sqrt{2E}
\left[\gamma-2 -2(\gamma-3)
\ _2F_1\left(1,\frac{1}{\gamma-2};\frac{3}{2};(2-\gamma)E\right)
+(\gamma-4)\ _2F_1\left(1,\frac{2}{\gamma-2};
\frac{3}{2};(2-\gamma)E\right)\right].
\end{equation}
This expression provides analytical $\gamma$-models for the range $2
\le \gamma < 3$, and adds to the only previously-known analytical
model with all radial orbits (Fridman \& Polyachenko 1984, Binney \&
Tremaine 1987, problem 4-19).  The hypergeometric functions are
elementary for the strong cusp case of $\gamma=\frac{5}{2}$, as they
are for that case in Tremaine et al. (1994), and give
\begin{eqnarray}
{\cal F}(E,L,\frac{5}{2})=\frac{\sqrt{2E}\delta(L^2)}{8\pi^3}
\left\{\frac{1}{2} \right. &+& \left.
\frac{1}{2(2+E)}-\frac{5}{4(2+E)^2}-\frac{15}{4(2+E)^3}
\right. \nonumber \\
&+& \left.\frac{2\ln\left[\sqrt{1+\frac{E}{2}}+\sqrt{\frac{E}{2}}\right]}
{\sqrt{E}(2+E)^{3/2}}\left[1-\frac{15}{4(2+E)^2}\right] \right\}.
\end{eqnarray}

\section{Conclusions}
\label{sec:conclusions}

We have provided a wide range of analytical anisotropic stellar
dynamic distribution functions for the widely used $\gamma$ density
profile. The simplest are two one-parameter families with constant
anisotropy, those of \S\ref{subsec:simpleaniso} for $\gamma \ge 1$ for
which Binney's parameter $\beta=0.5$, and the all-radial models of
\S\ref{subsec:radial models} for $\gamma \ge 2$ for which $\beta=1$.
Only the $\gamma=1$ member of the first family was known
previously \citep{baes2}.

The two two-parameter families of \S\ref{subsec:firstmodels} and 
\S\ref{subsec:secondmodels} provide a much wider range. They have a
second parameter $q$ in addition to $\gamma$. It controls their
anisotropy. The first family is isotropic at large distances, and $q$
gives the value of $2\beta$ at its centre. The second family is
isotropic at its centre, and $q$ gives the value of $-2\beta$ at
large distances.  A greater variety of behaviour can be obtained by
combining members of the families, as in Dejonghe (1989) and Baes
\& Dejonghe (2002), though we have not done this.

Finite expressions are given for some of the distribution functions.
These include those in \S\ref{subsec:whenqisone} for models of the
first family with $q=1$, and in Appendix \ref{app:exactdfs} for
tangentially biased models of the second family with $q=1$ and $q=2$.
Other more restricted cases are the radially biased Hernquist models
of the second family of \S\ref{subsec:hernquist} with $q=-1$ and
$q=-2$ (most radial), and the strongly cusped $q=-2$, $\gamma=2+1/N$,
$N$ an integer, cases identified, though not studied, in
\S\ref{subsec:secondmodels}.  More generally, distribution functions
must be found by summing series of ordinary or generalised
hypergeometric functions using formulas given in Appendix
\ref{app:formulas}. The coefficients of those series depend on powers
of $(2-\gamma)$, and the series converge rapidly for $\gamma$ close to
2 when few terms are needed. Conversely, long series are needed to
generate the $\gamma=0.5$ plots shown in the figures. It is not
difficult to compute hypergeometric functions; some advice on how to do so
is given at the end of Appendix \ref{app:evaluations}.

The first family of models, which are radially biased at their centres
when $q>0$, provide an interesting example of the cusp slope-central
anisotropy theorem of An \& Evans (2006b).  That theorem states that
the central value of $\beta$ can not exceed half the density slope
$\gamma$. In fact our models show that their theorem is somewhat more
widely applicable than their discussion of it allows. Their proof
assumes that the distribution function has a Laurent expansion of a
specific form in the whole neighbourhood of $L=0$.  Our equations
(\ref{andistrart}) and (\ref{FsmallL}) show that our distribution
functions have this form but not in the whole neighbourhood of $L=0$,
only in the region $L^2<2E$. A different form, given by equations
(\ref{andistrart}) and (\ref{FlargeL}) applies in $0<2E<L^2$, which
includes a part of the neighbourhood of $L=0$ near $E=0$.
Nevertheless their theorem is still valid, and their proof of it
requires no essential change, because a Laurent expansion of the form
they assume is valid near the cusp and its component that is proportional to
$L^{-q}$ dominates the behaviour there.  The wholly radial models of
\S\ref{subsec:radial models} with $\beta=1$ everywhere respect
the singular $q \to 2$ limit of their theorem because they exist only
for $\gamma \geq 2$.

Besides distribution functions, we have given and displayed velocity
dispersions, energy and transverse velocity densities, and observable
properties, including line-of-sight velocity profiles.  The singular
central behaviour of the radially biased models of the first family is
evident in the $q \ge 1$ curves in Fig.~\ref{fig:fig4},
Fig.~\ref{fig:fig5}, and Fig.~\ref{fig:losvdone}.  The latter are
particularly significant since they plot a directly observable
quantity.  The $\gamma$-models are commonly used to describe the
kinematics at the cores of giant elliptical galaxies, both to model
observations and to derive initial conditions for either N-body or
Monte-Carlo simulations. Because of the past lack of analytical
models, many workers have used Hernquist's (1993) simplified procedure
to generate approximate distribution functions.  Kazantzidis,
Magorrian \& Moore (2004) have shown that this procedure can be
hazardous when applied to galaxies that are strongly non-Gaussian, as
it can then lead to an initial state that is far from equilibrium.  We
have identified cusped galaxies with $\beta \ge 0.5$ at their centres
and whose line-of-sight velocities are cusped as instances of
non-Gaussian behaviour. The wide variety of analytic distribution
functions of this paper provides alternatives to the use of
Hernquist's method.  Our distribution functions are in exact equilibrium.
They can be coupled with a three-dimensional Monte-Carlo simulator to
provide valid initial conditions for N-body and Monte-Carlo
simulations \citep{buyle}, and so avoid the subsequent development
being influenced by artifacts of the start.

\section*{Acknowledgments}
PB acknowledges the Fund for Scientific Research Flanders (FWO) for
financial support.

\appendix

\section{Evaluation of distribution and related functions}
\label{app:evaluations}

We defined $F_q^p(E,L)$ in section \ref{subsec:firstmodels} to be the
distribution function which corresponds to the elementary augmented
density $r^{-q}(1+r)^q\psi^p$. We find it by taking a Laplace-Mellin
transform in $E$ and $L=rv_t$ as in \citep{herwig1}, and obtain
\begin{equation}
\mathcal{L}_{E\to\alpha}\mathcal{M}_{L\to\beta}\{{\cal F}(E,L)\}=
\frac{2^{\beta/2}}{(2\pi)^{3/2}}\frac{\alpha^{(3-\beta)/2}}{\Gamma(1-\beta/2)}
\mathcal{L}_{\psi\to\alpha}\mathcal{M}_{r\to\beta}\{\rho(\psi,r)\}
=\frac{\Gamma(p+1)}{\alpha^{p+1}}
\frac{\Gamma(\beta-q)\Gamma(-\beta)}{\Gamma(-q)}.
\label{lapmel}
\end{equation}
Although the Mellin transform of $r^{-q}(1+r)^q$ does not
exist for non-integer $q>0$, the final distribution function which 
we obtain for $q<0$ is valid in a continous interval of $q$, and remains
valid for $q>0$ by the principle of analytical combination; it
suffices to check that the proper limits can be taken in 
Equation~(\ref{distrwatch}) in order to check the existence of our results. 
The Laplace transform with respect to $\alpha$ is easily inverted, 
and we are left with a Mellin inversion integral
\begin{equation}
\label{melF}
F_q^p(E,L)=\frac{\Gamma(p+1)E^{p-3/2}}{(2\pi)^{5/2}2^q\Gamma(-q)}
\frac{1}{2\pi i}\int_{\beta_0-i\infty}^{\beta_0+i\infty}
\frac{\Gamma(-\frac{\beta}{2})\Gamma(\frac{1}{2}-\frac{\beta}{2})
\Gamma(-\frac{q}{2}+\frac{\beta}{2})
\Gamma(\frac{1}{2}-\frac{q}{2}+\frac{\beta}{2})}
{\Gamma(1-\frac{\beta}{2})\Gamma(p-\frac{1}{2}+\frac{\beta}{2})}
\left(\frac{L^2}{2E}\right)^{-\beta/2}d\left(\frac{\beta}{2}\right).
\end{equation}
for $F_q^p(E,L)$. We have here used applied the duplication formula for
the Gamma function \citep{abramowitz} equation 6.1.18 to expand the product
$\Gamma(\beta-q)\Gamma(-\beta)$.
The inversion of the $\beta$-integral in equation (\ref{melF}) gives,
\begin{equation}
\label{firstdf}
F_q^p(E,L)=\frac{\Gamma(p+1)E^{p-3/2}}{(2\pi)^{5/2}2^q\Gamma(-q)}
G_{33}^{22} \left(\frac{L^2}{2E}\left|
\begin{array}{ccccc}1&,&\frac{1}{2}&,&p-\frac{1}{2}\\
-\frac{q}{2}&,&\frac{1}{2}-\frac{q}{2}&,&0\end{array}\right.\right),
\end{equation}
by matching to the definition 5.3.1 in
Erdelyi (1953) of the Meijer G function $G^{22}_{33}$.
A similar analysis for
the elementary augmented density $(1+r)^q\psi^p$ gives
\begin{equation}
{\bar F}_q^{p}(E,L)=\frac{\Gamma(p+1)E^{p-3/2}}{(2\pi)^{5/2}2^q\Gamma(-q)}
G_{33}^{22}\left(\frac{L^2}{2E}\left|
\begin{array}{ccccc}1+\frac{q}{2}&,&\frac{1}{2}+\frac{q}{2}&,&p-\frac{1}{2}
\\0&,&\frac{1}{2}&,&0\end{array}\right.\right),
\label{seconddf}
\end{equation}
with a different Meijer G function.
Yet other Meijer G function are needed for the energy and transverse
velocity densities ${\cal F}_E$ and ${\cal F}_{v_t}$. 
Applying the operator on the right hand side of equation (\ref{formfe})
to equation (\ref{melF}) and then inverting the $\beta$-integral gives
\begin{equation}
\label{firstfe}
F_{E,q}^p(E,L)=\frac{\Gamma(p+1)E^{p-3/2}\sqrt{\psi-E}}{2^{q+1}\pi\Gamma(-q)}
G_{33}^{22} \left(\frac{r^2(\psi-E)}{E}\left|
\begin{array}{ccccc}1&,&\frac{1}{2}&,&p-\frac{1}{2}\\
-\frac{q}{2}&,&\frac{1}{2}-\frac{q}{2}&,
& -\frac{1}{2}\end{array}\right.\right).
\end{equation}
Applying the operator on the right hand side of equation (\ref{formfvt})
to equation (\ref{melF}) and then inverting the $\beta$-integral gives
\begin{equation}
\label{firstfvt}
F_{v_t,q}^p(E,L)=\frac{\Gamma(p+1)v_t}{2^{q+1}\pi\Gamma(-q)}
\left(\psi-\frac{v^2_t}{2}\right)^{p-1}
G_{33}^{22} \left(\frac{r^2v_t^2}{2\psi-v_t^2}\left|
\begin{array}{ccccc}1&,&\frac{1}{2}&,&p\\
-\frac{q}{2}&,&\frac{1}{2}-\frac{q}{2}&,& 0 \end{array}\right.\right).
\end{equation}
The corresponding barred quantities differ from these quantities in
that the first two columns of coefficients, which are the same in
equations (\ref{firstdf}), (\ref{firstfe}), and (\ref{firstfvt}),
are replaced by those of equation (\ref{seconddf}).

All the Meijer G functions which arise in this paper can be
expressed as the sum of two generalised hypergeometric functions,
using formulas 5.3.5 and 5.3.6 of Erdelyi (1953), as
\begin{eqnarray}
\label{firstG}
&&G_{33}^{22}\left(\chi \left|
\begin{array}{ccccc}a_1 &,& a_2 &,& a_3\\ b_1 &,& b_2 &, & b_3\end{array}
\right.\right) = 2\pi \Gamma(-q) 2^q\chi^{a_1-1} \times \nonumber\\
&&\left\{\frac{\ _3F_2\left(-\frac{q}{2},\frac{1}{2}-\frac{q}{2},1+b_3-a_1;
\frac{1}{2},1+a_3-a_1;-\frac{1}{\chi}\right)}
{\Gamma(a_3-a_1+1) \Gamma(a_1-b_3)}
+
\frac{q\ _3F_2\left(\frac{1}{2}-\frac{q}{2},1-\frac{q}{2},1+b_3-a_2;
\frac{3}{2},1+a_3-a_2;-\frac{1}{\chi}\right)}
{\sqrt{\chi}\Gamma(a_3-a_2+1) \Gamma(a_2-b_3)} \right\}
\end{eqnarray}
for $\chi>1$, and 
\begin{eqnarray}
\label{secondG}
&&G_{33}^{22}\left(\chi \left|
\begin{array}{ccccc}a_1 &,& a_2 &,& a_3\\ b_1 &,& b_2 &, & b_3\end{array}
\right.\right) = 2\pi \Gamma(-q) 2^q\chi^{b_1} \times \nonumber\\
&&\left\{\frac{\ _3F_2\left(-\frac{q}{2},\frac{1}{2}-\frac{q}{2},1+b_1-a_3;
\frac{1}{2},1+b_1-b_3;-\chi \right)}
{\Gamma(1+b_1-b_3) \Gamma(a_3-b_1)}
+
\frac{q\sqrt{\chi}\ _3F_2\left(\frac{1}{2}-\frac{q}{2},1-\frac{q}{2},1+b_2-a_3;
\frac{3}{2},1+b_2-b_3;-\chi \right)}
{\Gamma(1+b_2-b_3) \Gamma(a_3-b_2)} \right\}
\end{eqnarray}
for $\chi<1$. The coefficients $(a_1,a_2,b_1,b_2)$ here are
$\left(1, \frac{1}{2}, -\frac{q}{2}, \frac{1}{2} -\frac{q}{2} \right)$
for the models of \S \ref{subsec:firstmodels} and
$\left(1, \frac{1}{2}, -\frac{q}{2}, \frac{1}{2} -\frac{q}{2} \right)$
for the barred quantities for the models of \S \ref{subsec:secondmodels} .
Certain coefficient differences are the same for both sets, and those
differences have been used in equations (\ref{firstG}) and (\ref{secondG}).
The coefficients $(a_3,b_3)$ are $\left(p-\frac{1}{2},0 \right)$
for distribution functions, $\left(p-\frac{1}{2},-\frac{1}{2} \right)$
for energy densities, and $\left(p,0 \right)$ for transverse velocity
densities as in equations (\ref{firstdf}), (\ref{seconddf}),
(\ref{firstfe}), and (\ref{firstfvt}).
Equations (\ref{firstG}) and (\ref{secondG}) simplify
when the argument of one of the denominator Gamma functions is zero
or a negative integer. Then that Gamma function is infinite and
one term disappears.  

The ${}_3F_2$ generalised hypergeometric functions reduce to
polynomials if any of the first three coefficients are a negative
integer, and are 1 if any of those coefficients is zero. They reduce
to the simpler ${}_2F_1$ hypergeometric functions if any of the first
three coefficients is the same as the fourth or the fifth.  Otherwise
they can be computed by summing their series expansions when their
final argument is less than 1 in magnitude, as it is for the two
separate formulas (\ref{firstG}) and (\ref{secondG}), and in their
applications given below.  
There is no singularity at intermediate cases of $\chi=1$, because the
final arguments of the hypergeometric functions are then $-1$.
The programme given in \S 6.12 of Press et al. (1992)
for computing ordinary ${}_2F_1$ hypergeometric functions can easily
be extended to the ${}_3F_2$ generalised case. Series with large $a$
coefficients converge slowly, and Press et al.'s idea of combining
series summation with integration of a differential equation is then
helpful. Generalised hypergeometric functions are also available in
mathematical software packages such as Maple or Mathematica.

\section{Compendium of formulas}
\label{app:formulas}

\subsection{Distribution functions}
\begin{equation}
\label{FlargeL}
F_q^{p}(E,L)=\frac{\Gamma(p+1)E^{p-3/2}}{(2\pi)^{3/2}}
\left[ \frac{1}{\Gamma(p-\frac{1}{2})} 
+ \frac{q}{\Gamma(p)L} \sqrt{\frac{2E}{\pi}} 
\ _3F_2 \left( \frac{1}{2}-\frac{q}{2},1-\frac{q}{2},\frac{1}{2};
\frac{3}{2},p;-\frac{2E}{L^2}\right) \right],
\end{equation}
\begin{eqnarray}
\label{barFlargeL}
{\bar F}_q^p(E,L) &=& \frac{\Gamma(p+1)}{(2\pi)^{3/2}}E^{p-3/2}
\left(\frac{L}{\sqrt{2E}}\right)^q \times \nonumber\\
&&\left[\frac{\ _3F_2\left(-\frac{q}{2},\frac{1}{2}-\frac{q}{2},-\frac{q}{2};
\frac{1}{2},p-\frac{1}{2}-\frac{q}{2};-\frac{2E}{L^2}\right)} 
{\Gamma(p-\frac{1}{2}-\frac{q}{2})\Gamma(1+\frac{q}{2})}
+\frac{q\sqrt{2E}}{L} \frac{
\ _3F_2\left(\frac{1}{2}-\frac{q}{2},1-\frac{q}{2},\frac{1}{2}-\frac{q}{2};
\frac{3}{2},p-\frac{q}{2};-\frac{2E}{L^2}\right)}
{\Gamma(p-\frac{q}{2})\Gamma(\frac{1}{2}+\frac{q}{2})} \right],
\end{eqnarray}
when $L^2>2E>0$, and when $2E>L^2$
\begin{eqnarray}
\label{FsmallL}
F_q^{p}(E,L)&=&\frac{\Gamma(p+1)E^{p-3/2}}{(2\pi)^{3/2}}
\left( \frac{\sqrt{2E}}{L} \right)^q \times \nonumber\\
&&\left[\frac{\ _3F_2(-\frac{q}{2},\frac{1}{2}-\frac{q}{2},
\frac{3}{2}-p-\frac{q}{2};\frac{1}{2},1-\frac{q}{2};-\frac{L^2}{2E})}
{\Gamma(1-\frac{q}{2})\Gamma(p-\frac{1}{2}+\frac{q}{2})}
+\frac{qL}{\sqrt{2E}}
\frac{\ _3F_2(\frac{1}{2}-\frac{q}{2},1-\frac{q}{2},2-p-\frac{q}{2};
\frac{3}{2},\frac{3}{2}-\frac{q}{2};-\frac{L^2}{2E})}
{\Gamma(\frac{3}{2}-\frac{q}{2})\Gamma(p-1+\frac{q}{2})}\right],
\label{distrwatch}
\end{eqnarray}
\begin{eqnarray}
\label{barFsmallL}
{\bar F}_q^{p}(E,L)&=&\frac{\Gamma(p+1)}{(2\pi)^{3/2}}E^{p-3/2} 
\times \nonumber\\
&&\left[ \frac{_3F_2(-\frac{q}{2},\frac{1}{2}-\frac{q}{2},\frac{3}{2}-p;
\frac{1}{2},1;-\frac{L^2}{2E})}{\Gamma(p-\frac{1}{2})} 
+ \frac{2qL}{\sqrt{2\pi E}}
\frac{\ _3F_2(\frac{1}{2}-\frac{q}{2},1-\frac{q}{2},
2-p;\frac{3}{2},\frac{3}{2};-\frac{L^2}{2E})}{\Gamma(p-1)} \right].
\end{eqnarray}

\subsection{The energy distribution}

\begin{equation}
\label{FEsmallE}
F^p_{E,q} (\psi,r,E) = \Gamma(p+1)E^{p-3/2}\sqrt{\psi-E}
\left[ \frac{2}{\sqrt{\pi}\Gamma(p-\frac{1}{2})} 
{}_3F_2\left(-\frac{q}{2}, \frac{1}{2}-\frac{q}{2},-\frac{1}{2};
\frac{1}{2},p-\frac{1}{2};\frac {E} { r^2(E-\psi)} \right)
+\frac{q}{r\Gamma(p)} \sqrt{\frac{E}{\psi-E}} \right],
\end{equation}
\begin{eqnarray}
\label{barFEsmallE}
{\bar F}_{E,q}^{p}(\psi,r,E)&=&\Gamma(p+1)E^{p-3/2}\sqrt{\psi-E}
\left[\frac{r^2(\psi-E)}{E}\right]^{q/2} 
\left[ \frac{_3F_2(-\frac{q}{2}, \frac{1}{2}-\frac{q}{2},
-\frac{1}{2}-\frac{q}{2};\frac{1}{2},p-\frac{1}{2}-\frac{q}{2};
\frac{E}{r^2(E-\psi)})}
{\Gamma(p-\frac{1}{2}-\frac{q}{2})\Gamma(\frac{3}{2}+\frac{q}{2})} 
\right.\nonumber\\ &&\left.
+\frac{q}{r}\sqrt{\frac{E}{\psi-E}}
\frac{\ _3F_2(\frac{1}{2}-\frac{q}{2},1-\frac{q}{2},-\frac{q}{2};
\frac{3}{2},p-\frac{q}{2};\frac{E}{r^2(E-\psi)})}
{\Gamma(p-\frac{q}{2})\Gamma(1+\frac{q}{2})}\right],
\end{eqnarray}
when $r^2(\psi-E) > E$, and when $E > r^2(\psi-E)$
\begin{eqnarray}
\label{FElargeE}
F_{E,q}^{p}(\psi,r,E)&=&\Gamma(p+1)E^{p-3/2}\sqrt{\psi-E}
\left[\frac{E}{r^2(\psi-E)}\right]^{q/2} 
\left[\frac{_3F_2(-\frac{q}{2},\frac{1}{2}-\frac{q}{2},
\frac{3}{2}-p-\frac{q}{2};\frac{1}{2},\frac{3}{2}-\frac{q}{2};
\frac{r^2(E-\psi)}{E})}{\Gamma(\frac{3}{2}-\frac{q}{2})
\Gamma(p-\frac{1}{2}+\frac{q}{2})}
\right.\nonumber\\ &&\left.
+qr\sqrt{\frac{\psi-E}{E}}\frac{\ _3F_2(\frac{1}{2}-\frac{q}{2},1-\frac{q}{2},
2-p-\frac{q}{2};\frac{3}{2},2-\frac{q}{2};\frac{r^2(E-\psi)}{E})}
{\Gamma(2-\frac{q}{2})\Gamma(p-1+\frac{q}{2})}\right],
\end{eqnarray}
\begin{eqnarray}
\label{barFElargeE}
{\bar F}_{E,q}^{p}(\psi,r,E)&=&\Gamma(p+1)E^{p-3/2}\sqrt{\psi-E}
\left[ \frac{2}{\sqrt{\pi}\Gamma(p-\frac{1}{2})} 
\ _3F_2 \left(-\frac{q}{2},\frac{1}{2}-\frac{q}{2},\frac{3}{2}-p;
\frac{1}{2},\frac{3}{2};\frac{r^2(E-\psi)}{E}\right)
\right.\nonumber\\ &&\left.
+ \frac{qr}{\Gamma(p-1)}\sqrt{\frac{\psi-E}{E}}
\ _3F_2 \left(\frac{1}{2}-\frac{q}{2},1-\frac{q}{2},2-p;\frac{3}{2},2;
\frac{r^2(E-\psi)}{E} \right) \right].
\end{eqnarray}

\subsection{The distribution of the transverse motions}

\begin{equation}
\label{Fvtlargevt}
F_{v_t,q}^{p}(\psi,r,v_t)=\Gamma(p+1)v_t\left(\psi-\frac{v_t^2}{2}\right)^{p-1}
\left[\frac{1}{\Gamma(p)}+\frac{q\sqrt{2\psi-v_t^2}}
{\sqrt{\pi}\Gamma(p+\frac{1}{2})rv_t}
\ _3F_2\left(\frac{1}{2}-\frac{q}{2},1-\frac{q}{2},\frac{1}{2};
\frac{3}{2},p+\frac{1}{2};\frac{v_t^2-2\psi}{r^2v_t^2}\right) \right],
\end{equation}
\begin{eqnarray}
\label{barFvtlargevt}
{\bar F}_{v_t,q}^{p}(\psi,r,v_t)&=&\Gamma(p+1)v_t
\left(\psi-\frac{v_t^2}{2}\right)^{p-1}
\left(\frac{r^2v_t^2}{2\psi-v_t^2}\right)^{q/2}
\left[\frac{\ _3F_2(-\frac{q}{2},\frac{1}{2}-\frac{q}{2},-\frac{q}{2};
\frac{1}{2},p-\frac{q}{2};\frac{v_t^2-2\psi}{r^2v_t^2}) }
{\Gamma(p-\frac{q}{2})\Gamma(1+\frac{q}{2})}
\right.\nonumber\\
&&\left.+\frac{q\sqrt{2\psi-v_t^2}}{rv_t}
\frac{\ _3F_2(\frac{1}{2}-\frac{q}{2},1-\frac{q}{2},\frac{1}{2}-\frac{q}{2};
\frac{3}{2},p+\frac{1}{2}-\frac{q}{2};\frac{v_t^2-2\psi}{r^2v_t^2})}
{\Gamma(p+\frac{1}{2}-\frac{q}{2})\Gamma(\frac{1}{2}+\frac{q}{2})}\right],
\end{eqnarray}
when $r^2v_t^2>2\psi-v_t^2$, and when $2\psi-v_t^2 > r^2v_t^2$ 
\begin{eqnarray}
\label{Fvtsmallvt}
F_{v_t,q}^{p}(\psi,r,v_t)&=&\Gamma(p+1)v_t
\left(\psi-\frac{v_t^2}{2}\right)^{p-1}
\left(\frac{2\psi-v_t^2}{r^2v_t^2}\right)^{q/2}
\left[\frac{\ _3F_2(-\frac{q}{2},\frac{1}{2}-\frac{q}{2},1-p-\frac{q}{2};
\frac{1}{2},1-\frac{q}{2};\frac{r^2v_t^2}{v_t^2-2\psi})}
{\Gamma(1-\frac{q}{2})\Gamma(p+\frac{q}{2})}
\right.\nonumber\\
&&\left.+\frac{qrv_t}{\sqrt{2\psi-v_t^2}}
\frac{\ _3F_2(\frac{1}{2}-\frac{q}{2},1-\frac{q}{2},\frac{3}{2}-\frac{q}{2}-p;
\frac{3}{2},\frac{3}{2}-\frac{q}{2};\frac{r^2v_t^2}{v_t^2-2\psi})}
{\Gamma(\frac{3}{2}-\frac{q}{2})\Gamma(p-\frac{1}{2}+\frac{q}{2})}\right],
\end{eqnarray}
\begin{eqnarray}
\label{barFvtsmallvt}
{\bar F}_{v_t,q}^{p}(\psi,r,v_t)&=&\Gamma(p+1)v_t
\left(\psi-\frac{v_t^2}{2}\right)^{p-1}
\left[ \frac{1}{\Gamma(p)}
\ _3F_2 \left(-\frac{q}{2},\frac{1}{2}-\frac{q}{2},1-p;
\frac{1}{2},1;\frac{r^2v_t^2}{v_t^2-2\psi}\right)
\right.\nonumber\\
&&\left.+\frac{2qrv_t}{\Gamma(p-\frac{1}{2})\sqrt{\pi(2\psi-v_t^2)}}
\ _3F_2\left(\frac{1}{2}-\frac{q}{2},1-\frac{q}{2},\frac{3}{2}-p;
\frac{3}{2},\frac{3}{2};\frac{r^2v_t^2}{v_t^2-2\psi}\right)  \right].
\end{eqnarray}

\subsection{More exact distribution functions}
\label{app:exactdfs}

The $\gamma$-models have distribution functions of the form
${\cal F}(E,L)=f_0(E) + Lf_1(E)$ where
\begin{equation}
f_0(E)=\frac{3-\gamma}{8\pi^3}\sqrt{2E}
\left[3(\gamma-5)\Phi(-1-\gamma)-10(\gamma-4)\Phi(-\gamma)
+10(\gamma-3)\Phi(1-\gamma)-5(\gamma-1)\Phi(3-\gamma)
+2\gamma\Phi(4-\gamma) \right]
\end{equation}
and
\begin{equation}
f_1(E)=\frac{3-\gamma}{8\pi^3}(1-t)^3 t^{\gamma-4}
\left[20t^2+5(\gamma+1)t(1-t)+2\gamma(1-t)^2\right], 
\end{equation}
where
\begin{equation}
\label{tdef}
\Phi(a)=\ _2F_1\left(1,\frac{a}{2-\gamma};\frac{3}{2};(2-\gamma)E\right),
\qquad t=\left[1-(2-\gamma)E\right]^{\frac{1}{2-\gamma}}.
\end{equation}
They also have distribution functions of the form ${\cal F}(E,L)
=f_0(E) + Lf_1(E)+L^2f_2(E)$ where
\begin{eqnarray}
f_0(E)=\frac{3-\gamma}{4\pi^3}\sqrt{2E}\left[ \right. 
&-& \left.2(\gamma-6)\Phi(-2-\gamma) + 9(\gamma-5)\Phi(-1-\gamma)
-15(\gamma-4)\Phi(-\gamma)+10(\gamma-3)\Phi(1-\gamma) \right. \nonumber\\
&-& \left. 3(\gamma-1)\Phi(3-\gamma)+\gamma\Phi(4-\gamma)\right], 
\end{eqnarray}
\begin{equation}
f_1(E)=\frac{3-\gamma}{4\pi^3}(1-t)^4 t^{\gamma-4}
\left[15t^2+3(\gamma+1)t(1-t)+\gamma(1-t)^2\right], 
\end{equation}
\begin{eqnarray}
f_2(E)=\frac{3-\gamma}{8\pi^3}\sqrt{2E} \left[ \right.
&2& \left.(\gamma-6)(\gamma+2)\Phi(-2\gamma) -9(\gamma-5)(\gamma+1)
\Phi(1-2\gamma) +15\gamma(\gamma-4)\Phi(2-2\gamma) \right. \nonumber\\
&-& \left. 10(\gamma-1)(\gamma-3)\Phi(3-2\gamma)
+3(\gamma-1)(\gamma-3)\Phi(5-2\gamma)-\gamma(\gamma-4)\Phi(6-2\gamma) \right].
\end{eqnarray}

\label{lastpage}
\begin{figure}
\includegraphics[width=0.1mm,angle=0,clip=true]{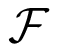}
\end{figure}
\end{document}